\documentclass[useAMS,usenatbib]{mn2e}
\usepackage{graphicx,hyperref,amssymb,wasysym}
\def\aj{AJ}%
\def\apj{ApJ}%
\def\apjl{ApJ}%
\def\apjs{ApJS}%
\def\aap{A\&A}%
\def\aaps{A\&AS}%
\def\mnras{MNRAS}%
\def\nat{Nature}%
\def\mrk501{Mrk\,501}
\newcommand{\eqb}{\begin{eqnarray}}
\newcommand{\eqe}{\end{eqnarray}}
\def\tql{\textquotedblleft}
\def\tqr{\textquotedblright~}
\title[Structure function: caveats and problems]{On the use of structure functions to study blazar variability: caveats and problems}
\author[Emmanoulopoulos et al.]{D.~Emmanoulopoulos$^{1}$\thanks{E-mail: D.Emmanoulopoulos@soton.ac.uk}, I.~M.~M\textsuperscript{c}Hardy$^{1}$ and P.~Uttley$^{1}$\\
$^{1}$ School of Physics and Astronomy, University of Southampton, Southampton SO17 1BJ}
\begin{document}
\date{Accepted 2009 Month \#. Received 2009 Month \#; in original form 2009 Month \#}
\pagerange{\pageref{firstpage}--\pageref{lastpage}} \pubyear{2222}

\maketitle
\label{firstpage}

\begin{abstract}
The extensive use of the structure function (SF) in the field of blazar variability across the electromagnetic spectrum suggests that characteristics time-scales are embedded in the light curves of these objects. We argue that for blazar variability studies, the SF results are sometimes erroneously interpreted leading to misconceptions about the actual source properties. Based on extensive simulations we caution that spurious breaks will appear in the SFs of almost all light curves, even though these light curves may contain no intrinsic characteristic time-scales. i.e.\ having a featureless underlying power-spectral-density (PSD). We show that the time-scales of the spurious SF-breaks depend mainly on the length of the artificial data set and also on the character of the variability i.e.\ the shape of the PSD.\par
The SF is often invoked in the framework of shot-noise models to determine the temporal properties of individual shots. We caution that although the SF may be fitted to infer the shot parameters, the resultant shot-noise model is usually inconsistent with the observed PSD. As any model should fit the data in both the time and the frequency domain the shot-noise model, in these particular cases, can not be valid.\par
Moreover, we show that the lack of statistical independence between adjacent SF points, in the standard SF formulation, means that it is not possible to perform robust statistical model fitting following the commonly used least-squares fitting methodology. The latter yields uncertainties in the fitting parameters (i.e.\ slopes, breaks) that are far too small with respect to their true statistical scatter. Finally, it is also commonly thought that SFs are immune to the sampling problems, such as data gaps, which affects the estimators of the PSDs. However we show that SFs are also troubled by gaps which can induce artefacts.   
\end{abstract}


\section{INTRODUCTION}
\label{sect:intro}
The flux variation of Active Galactic Nuclei (AGN) is a phenomenon that can provide us with significant information about the physical properties of these sources. The most variable AGN are the blazars which exhibit dramatic flux variations across the electromagnetic spectrum. The fastest variations are observed in the X-ray and $\gamma$-ray bands on time-scales of hours or even minutes \citep{catanese97,maraschi99,aharonian03,aharonian05_mrk421,aharonian_05B_pks2155} whereas in the optical \citep{tosti98,stalin05} and radio wavebands \citep{romero97,aller99,terasranta05} the temporal variations are seen on time-scales of days to weeks, up to years.\par
In the last fifty years several time-series analysis methods have been developed to study the variability properties of astronomical sources \citep[see for a review][]{rao97}. Linear and nonlinear analysis methods in the time and frequency domains, adjusted to the needs of astronomical data sets i.e.\ taking into account measurement errors and data-gaps, can describe adequately the time behaviour of AGN. Nevertheless, several authors, as we are going to discuss further, are convinced that more conventional, intuitive-tools, based on \tql running variance\tqr computations are able to give robust results with respect to the underlying variability properties of the observed source. Although these methods have value in some cases, they can often give misleading results as we shall show in this paper.\par
One of the most extensively used tools in the field of blazar variability is the {\it structure function} (SF) \citep[see e.g.][]{hughes92} which measures the mean value of the flux-variance for measurements, $x(t)$, that are separated by a given time interval, $\tau$, where $SF(\tau)=\left<\left[x(t)-x(t+\tau)\right]^2\right>$. The SF is commonly characterized in terms of its slope $\beta$, where $SF(\tau)\propto\tau^\beta$.\par
Historically the SF has been used in the study of turbulent plasmas \citep{kolmogorov41a,kolmogorov41b,yu03}. It was introduced to astrophysics by radio astronomers studying slow scintillations in the interstellar medium \citep{ricket84} following the methodologies of \citet{prokhorov75} and \citet{coles82} on the subjects of laser propagation in turbulent media and atmospheric scintillation respectively. The first systematic description of the SF methodology, adjusted to the needs of astronomical data sets, was made by \citet{simonetti85} using as a reference the work of \citet{rutman78} from the field of electrical and electronic engineering. \citeauthor{simonetti85} used the SF to demonstrate that the time-series of flat- and steep-spectrum radio sources differ qualitatively. During the same period, \citet{cordes85} estimate the SFs of 21 pulsars and \citet{hjellming86} derived the first quantitive SF results for the compact galactic radio source 1741-038. Then, \citet{fiedler87,heeschen87} also used the SF to study the variability properties of large extragalactic radio samples. Subsequently, the SF has been employed for the study of the timing properties of objects in higher energy bands. \citet{bregman88,neugebauer89,quirrenbach92,hufnagel92,smith93,lainela93,heidt96} whilst \citet{lewis96} and \citet{blandford97} applied the SF-analysis method to derive masses for microlenses. In the X-ray band, \citet{brinkmann00} applied the SF methodology to the ROSAT observations of PKS\,2155-304 and they attributed the derived shot-noise variability to either the accelerating inner jet model or to accelerating particles at shocks travelling down the relativistic jet. Thereafter many X-ray variability many X-ray observations have employed SF analysis \citep[e.g.][]{gliozzi01,brinkmann01,iyomoto01,zhang02a} in order to derive some short of characteristic time-scale.\par
SFs have been particularly commonly applied to observations of blazars. The {\it ASCA} \tql long-look\tqr observations of Mrk\,501, Mrk\,421, and PKS\,2155-304 \citep{takahashi00,tanihata01,tanihata03} are, even now, some of the most uniformly sampled light curves in the X-ray regime. The main outcome of the above SF analysis is an apparent characteristic time-scale of $\sim$1 day which appears to be in accordance with the duration of the shots existing in the internal shock particle acceleration scenario \citep{sikora01,spada01}.\par
In this work we show through extensive simulations that much of the existing literature on blazar SFs tends to misinterpret observed SF characteristics such as breaks and slopes as being real or physically meaningful, when often these are either artefacts intrinsic to SFs, or subject to much greater statistical variation than inferred from the commonly-used fitting procedures. In Section \ref{sect:lc_simul} we present the method that we are going to use in order to create artificial light curves lacking any sort of characteristic time-scales and in Section \ref{sec:xray_obs} we study the behaviour of the SF derived from the thoroughly studied {\it ASCA} data set of Mrk\,501 \citep[ hereafter TAN01]{tanihata01}. For the same source we employ the long-look light curve of {\it All-Sky Monitor} ({\it ASM}), onboard {\it Rossi X-ray Timing Explorer} ({\it RXTE}), in order to study the effects of the data length on the position of the SF-break. Then, in Section \ref{sect:shot_noise} we study the timing properties in the Fourier domain of the shot-noise model that TAN01 have used in order to study the SF properties. In Section \ref{sect:fitting_sf} we test the statistical robustness of the most commonly used fitting procedures which are employed in order to derive astrophysically interesting quantities from the SF. Finally, in Section \ref{sect:sf_gaps} we study the sensitivity of SF to the presence of data-gaps.
 
\section{LIGHT CURVE SIMULATIONS AND POWER SPECTRAL DENSITY}
\label{sect:lc_simul}
We simulate stationary light curves based on the procedure used by \citet{timmer95} which uses as an input the Power Spectral Density (PSD) function of the observed light curve and gives an ensemble of stochastic time-series produced from the same underlying PSD. This is the most robust method of mimicking the properties of a light curve since it takes into account correctly the intrinsic scatter in the power at a given frequency, described by a $\chi^2_2$, chi-squared distribution with two degrees of freedom \citep[e.g.][]{priestley81}, by randomizing both phases and amplitudes.\par
Throughout this paper we use the standard nomenclature, i.e.\ we refer to the periodogram as the modulus squared of the discrete Fourier transform of the particular data set under consideration whilst the PSD represents the mean of the true underlying distribution of variability power as a function of frequency \citep{priestley81,vaughan03}. Since the periodogram is highly scattered around the PSD having a standard deviation of 100 per cent \citep{press92}, we use the binned logarithmic periodogram \citep{papadakis93,vaughan05}. If there are more than 20 samples in each geometric-mean frequency bin then their distribution will be Gaussian, thus producing a statistically sensible standard deviation.\par
Concerning the length of the simulated light curves some special caution should be taken with respect to the effect of {\it red-noise leak} i.e.\ the transfer of power from low to high frequencies by the lobes of the window function which can produce features such as slowly increasing or falling trends across the generated light curve \citep[e.g.][]{priestley81}. To take this effect into account, we extend the PSD to very low frequencies (i.e.\ long time-scales), in order to generate artificial light curves 50 times longer than the observed one. We then truncate the simulated data trains to the desired length by selecting a random segment having equal length to the real light curve under study.\par
In the case of blazar variability the underlying PSD can be very-well described by simple or broken-power-laws with indices $1<\alpha<2.5$ flattening at long time-scales depicting the physical fact that the variability amplitude can not increase forever (stationarity). For these kinds of physically stationary processes the periodogram can correctly describe the underlying PSD even for steeper slopes of -4 (or even -5).\par
Apart from red-noise leak, spectral representations are also affected by {\it aliasing effects}. Frequency components outside the frequency range covered by the data set, are aliased into that range by the very act of discrete sampling \citep{press92}. Finally, irregular sampling produces a window function which smears the spectral estimates \citep{scargle89}, causing deviations from the true underlying PSD. Over the years a number of methods have been
derived to overcome these problems. In particular, we can estimate the underlying PSD by model fitting \citep[e.g.][]{done92,uttley02}. This process is now
commonly applied to observations of Seyfert galaxy X-ray variability \citep[e.g.][]{markowitz03,uttley05b,mchardy04,mchardy05,mchardy06,mchardy07} and produces reliable results. Similar simulation-based treatments of the SF results are missing from the blazar literature.

\section{CAVEATS REGARDING THE STRUCTURE FUNCTION BREAK-TIME-SCALES}
\label{sec:xray_obs}
\subsection{The {\it ASCA} data set of Mrk\,501}
\label{ssec:asca_sf}
\begin{figure*}
\hspace{-31em}\includegraphics[width=3.3in]{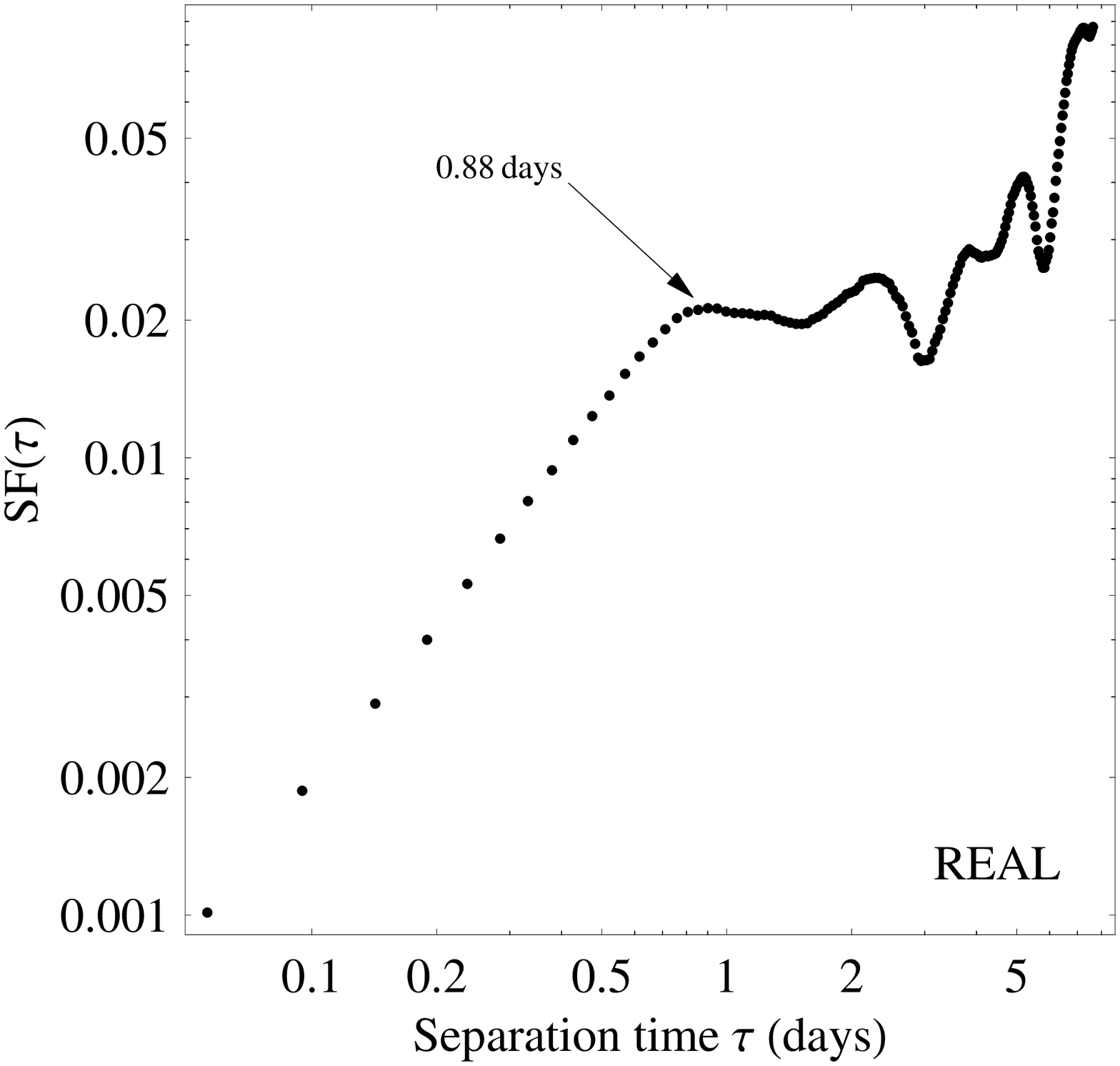}\\[-28.5em]\hspace{30em}
\includegraphics[width=3.3in]{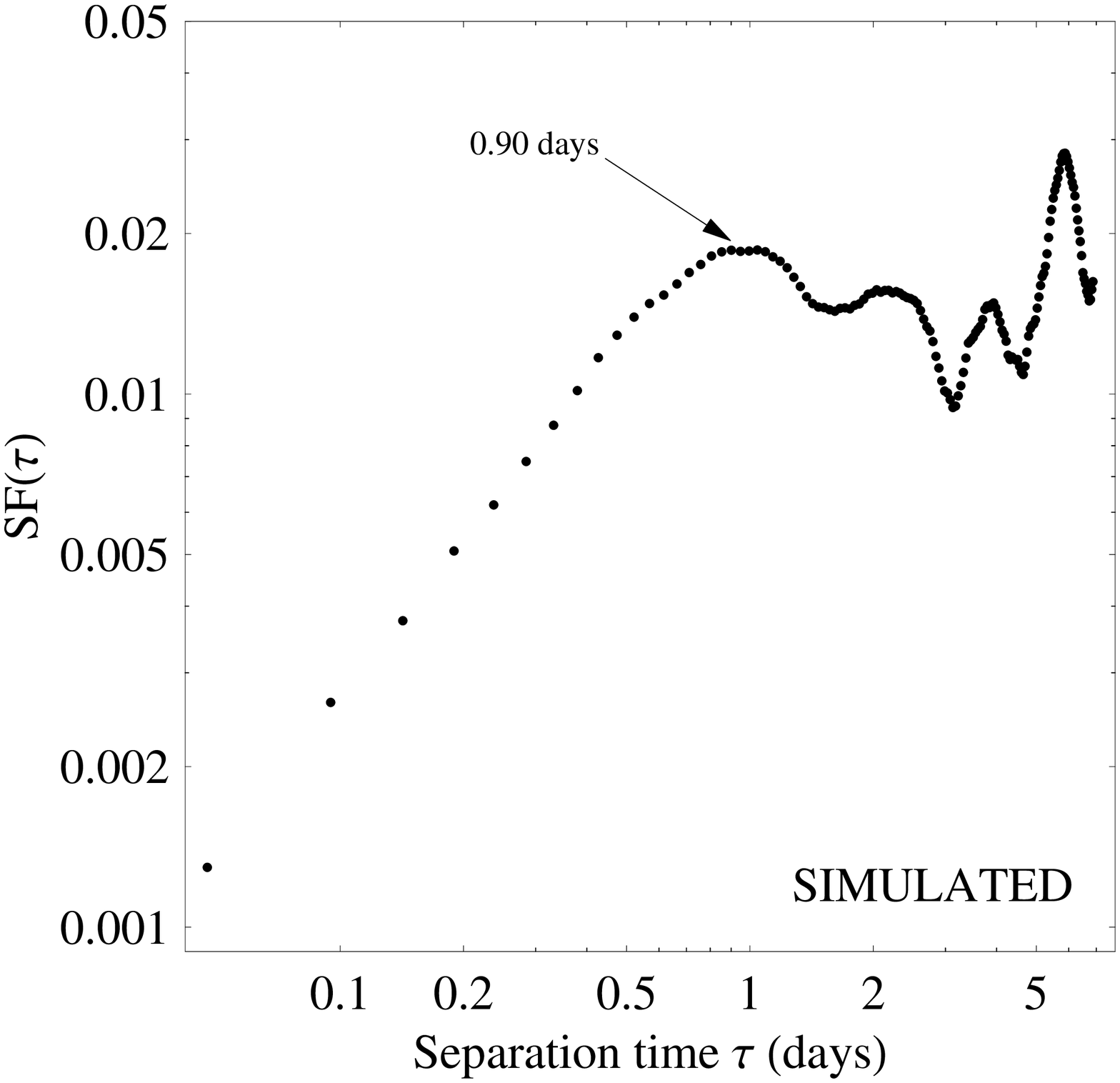}
\caption{[Left panel] The SF (in logarithmic scale) for the {\it ASCA} data set of \mrk501 as shown in panel (a) of figure 4 in TAN01. The arrow indicates the break time-scale which is around a day.\newline
[Right panel] A randomly chosen SF (in logarithmic scale) from the ensemble of the 2000 artificial light curves which are produced from a featureless power-law PSD of index -1.80 with no characteristic time-scale, having the same length as the {\it ASCA} data set. There is a clear break around one day, similar to the one derived from the {\it ASCA} SF.}
\label{fig:asca_arti_sf} 
\end{figure*}

To study the behaviour of the SF, we examine the {\it ASCA} data set of \mrk501 (TAN01) in 2--10 keV, with a sample interval of 5678.3 sec. This is one of the three data sets (the other two are Mrk\,421 and PKS\,2155-304) in which TAN01, based on the SF analysis method (Fig. \ref{fig:asca_arti_sf}, left panel), find characteristic time-scales and suggest that these are a signature of the minimum time-scales of individual shots. This is one of the longest and most continuous data sets ever obtained for a blazar in the X-rays, covering a time span of 10 days.\par
Initially, we estimate the binned logarithmic periodogram of the data set which can be very well fitted by a simple featureless power-law with index $\alpha=-1.80\pm0.09$ ($\chi^2=3.80$ for 6 degrees of freedom with a null hypothesis probability of 0.70). There is no evidence for a break in the PSD.\par
We next produce 2000 artificial light curves having the same power-law PSD of index $\alpha=-1.80$ and the same length as the studied {\it ASCA} data set. For each artificial light curve we estimate the SF by employing exactly the same functional form of the SF as TAN01  
\eqb
SF(\tau)=\frac{1}{N(\tau)}\sum w(i)w(i+\tau)[f(i+\tau)-f(i)]^2
\label{eq:sf_def_tanihata}
\eqe
where 
$N(\tau)=\sum w(i) w(i+\tau)$, $\tau$ is the separation time of the observations, and $w(i)$ is the weighting factor which is proportional to the data point $f(i)$ and inversely proportional to its error $\sigma_f(i)$. By an eye inspection, we note apparent breaks in almost all of the simulated SFs.\par
In our simulations we take into account the effect of the measurement errors in the SF-estimates by representing each measurement as a deviate drawn from a Gaussian distribution with mean and standard deviation equal to the measurement's value and error respectively (200 times for each artificial light curve). We have to note that for this particular {\it ASCA} data set the measurement errors do not play a major role in the SF-estimates. If we ignore these errors the values of the SF-breaks and SF-slopes change only by $\sim 1$ per cent and $\sim 0.5$ per cent respectively.\par
Next, we determine the position of the SF-break (Fig. \ref{fig:asca_arti_sf}, right panel). To localize the break we produce an interpolated version of the SF and we find the abscissa $\tau_{br}$ of the first local maximum, which indicates the position of the break. Even if a clear plateau is not formed after $\tau>\tau_{\rm br}$, several authors consider the first \tql hump\tqr as a signature of a genuine temporal property of the source \citep[e.g.][]{takahashi00,gliozzi01,kataoka01,agudo06,fuhrmann08}.\par
Finally we note that after the first break point, the SF usually exhibits \tql wiggly-patterns\tqr forming a plateau. These fluctuations indicate that the pairs $\{f(i+\tau),f(i)\}_{\tau>\tau_{\rm br}}$ which are averaged within each time bin $\tau>\tau_{\rm br}$, are linearly independent having zero linear correlation (Appendix \ref{app:sf_acf}) . In order to study the frequency of occurrence of these SF structures, we register the abscissas of all the local extrema occurring for $\tau>\tau_{\rm br}$.\par
\begin{figure*} 
\includegraphics[width=3.3in]{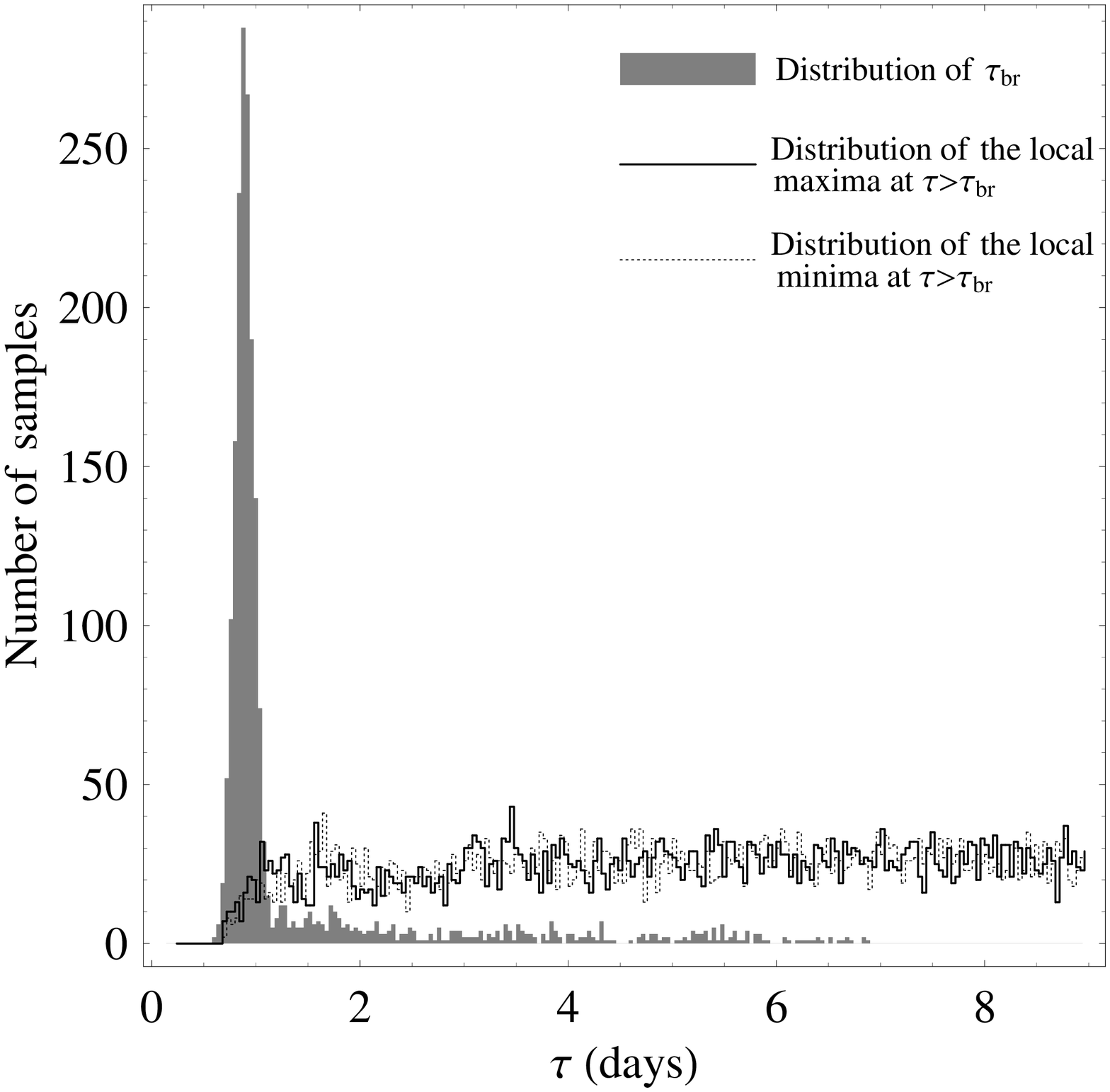}
\includegraphics[width=3.3in]{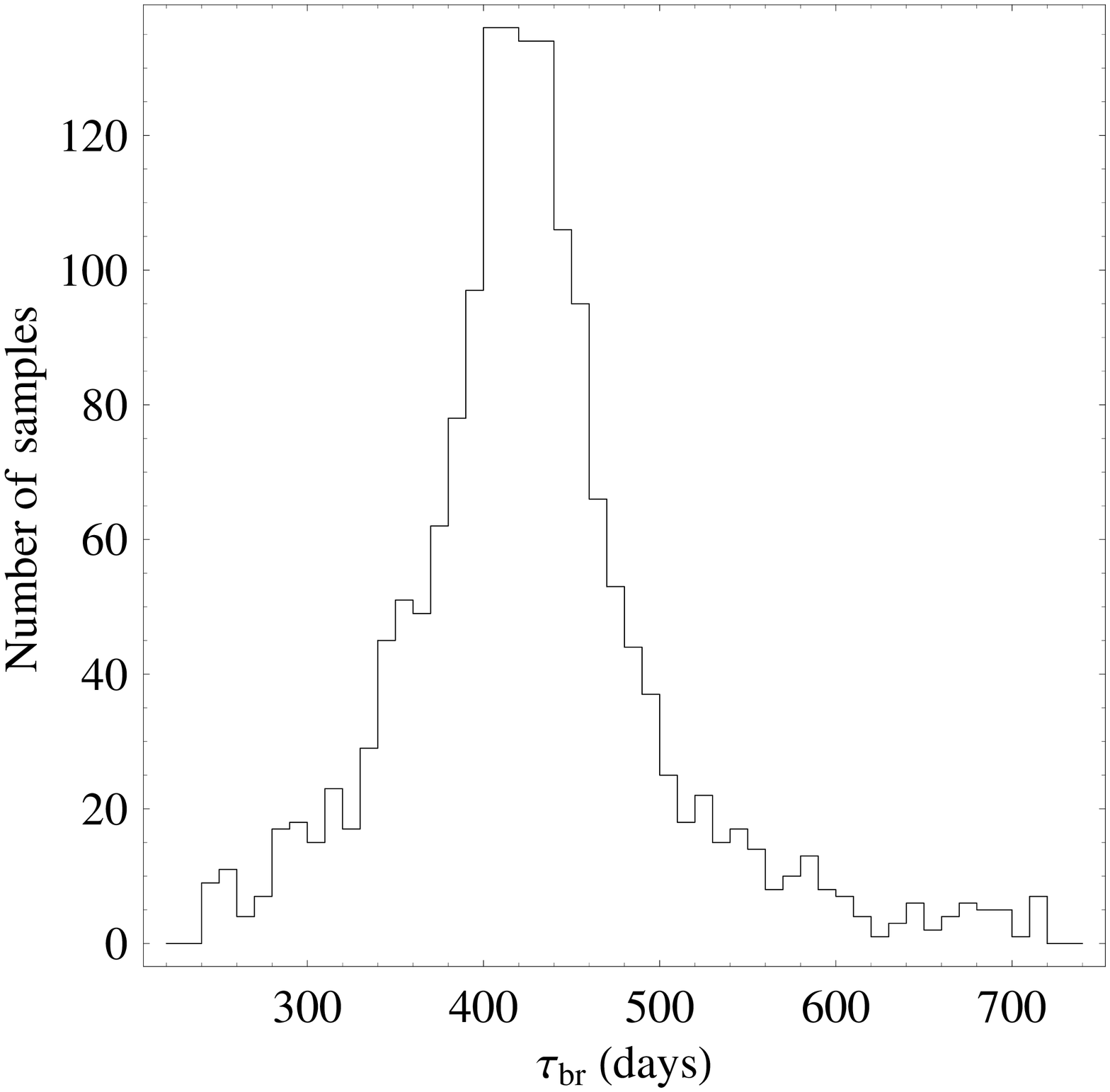}
\caption{The distribution of the SF's breaks and wiggling features.\newline
[Left panel] The grey area depicts the distribution of the SF-breaks coming from an ensemble of 2000 artificial light curves which are produced from a featureless power-law PSD of index -1.80 with no characteristic time-scale, for the case of the {\it ASCA} data of \mrk501 (the histogram bins have a length of 0.04 days). Based on a Gaussian fit the mean value and the standard deviation of the distribution are 0.92 and 0.09 days respectively. The solid and dashed lines represent the distribution of the local maxima and local minima for $\tau>\tau_{\rm br}$ mapping the positions of the wiggling features. \newline
[Right panel] The distribution of the SF-breaks for the case in which the simulated light curves, produced by the same PSD as above, extend to 2000 days (the histogram bins have a length of 10 days). It has a mean value of 399.35 days and a standard deviation of 74.73 days. }
\label{fig:sf_hist_break_and_long_dist}
\end{figure*}

In the left panel of Fig. \ref{fig:sf_hist_break_and_long_dist} we present with the filled grey area the distribution of the SF-breaks coming from our simulated light curves that have the same length and the same featureless PSD as the original {\it ASCA} light curve of \mrk501. From an ensemble of 2000 simulated data sets, we get breaks from 1903 of them. To specify the position of the maximum in our histogram we fit the histogram entries with a Gaussian distribution. The results of the fit yield an amplitude of 279.55 samples, a mean value of 0.91 days and a standard deviation of 0.09 days\footnote{By fitting a log-normal distribution the results remain practically the same giving an amplitude of 254.33 samples, $\mu=-0.09$, and $\sigma=0.10$, yielding a mean value of 0.92 days and a standard deviation of 0.09 days.}. With the solid and dotted lines we present the distribution of the local extrema (maxima with solid line and minima with the dotted line) occurring after the first break. The occurrence times of the \tql wiggling\tqr patterns are purely randomly distributed for $\tau>\tau_{\rm br}$ following a uniform distribution.\par
These simulations show us clearly that stochastic data sets having the same length as the {\it ASCA} data set of \mrk501 and the same featureless PSD, exhibit breaks in the SF around a day. Thus, the apparent break seen in the SF for \mrk501 (Fig. \ref{fig:asca_arti_sf}, left panel) should not be associated with any sort of physically meaningful time-scale.
\subsection{Longer time-series: The {\it ASM} data set of Mrk\,501}
\label{ssec:longe_ts}
Following Section \ref{ssec:asca_sf}, we extend our simulations to even longer time-scales to check how SF-break time-scales might be related to the length of a data set. Assuming that for long time-scales the variability properties of MRK\,501 are well-represented by the same PSD describing the short-term variability (i.e.\ same slope and normalisation), we produce 2000 artificial light curves, each being 2000 days long. For every light curve we calculate the SF and we specify the position of the SF-breaks. In total 1893 light curves exhibit SF-breaks and the distribution of their position can be well represented by a Gaussian distribution having a mean 399.35 days and a standard deviation of 74.73 days (Fig. \ref{fig:sf_hist_break_and_long_dist}, right panel).\par
In order to compare our predicted break with the data we use the long-term {\it RXTE-ASM} light curve of \mrk501\footnote{We have obtained the light curve from the {\it ASM} light curve site of MIT: \url{http://xte.mit.edu/ASM_lc.html}} spanning from MJD~50100 to MJD~52100 and we estimate its SF. The SF-break occurs around 402 days, something which is absolutely in accordance with the results from our simulated light curves of the same length which did not include of any sort of characteristic time-scale.\par
The aforementioned example reflects in a clear way that the SF deals only with the properties of the observed light curve ignoring the properties of the true underlying variability process. During astronomical observations we observe a source for a given time and from the observed data series we have to extract in a statistically robust way the properties of the underlying variability process. If the result that we obtain is a strong function of the length of a given data set then this can lead to a serious misunderstanding of the variability properties of the source.\par
Similar SF effects to those derived for the {\it ASCA} and {\it ASM} light curves of \mrk501 should be expected for all similar light curves. By producing an artificial light curve 6000 time units (t.u.) long, from a featureless power-law PSD of slope -2, we estimate the normalized SF (NSF) (equation (\ref{eqe:normalized_sf})) (i.e.\ the SF divided by two times the value of the variance of the data set, see Appendix \ref{app:sf_acf}) for the overall data train and for a 500 t.u. subset of it (Fig. \ref{fig:long_short_ds}). In this case we use the NSF, instead of the classical SF, for an easy and direct comparison of the position of the break, since the formed plateaus in both SFs are distributed around 2. The top right panel of Fig. \ref{fig:long_short_ds} shows the behaviour of the NSF for these two data sets, coming from the same underlying variability process. The break in the NSF for the short data set occurs at around 40 days and for the overall light curve at 1600 days respectively. In this case, the fact that the break occurs at two different time points for two different durations of the data set which is fully described from the same and completely featureless PSD, indicates that the break does not reflect any physically interesting time property of the process itself.\par
For comparison purposes we also estimate the binned logarithmic periodograms for the two light curves discussed above. As we can see from the bottom left panel of (Fig. \ref{fig:long_short_ds}) both can very well reproduce the input PSD slope of -2.
\begin{figure*} 
\includegraphics[width=3.3in]{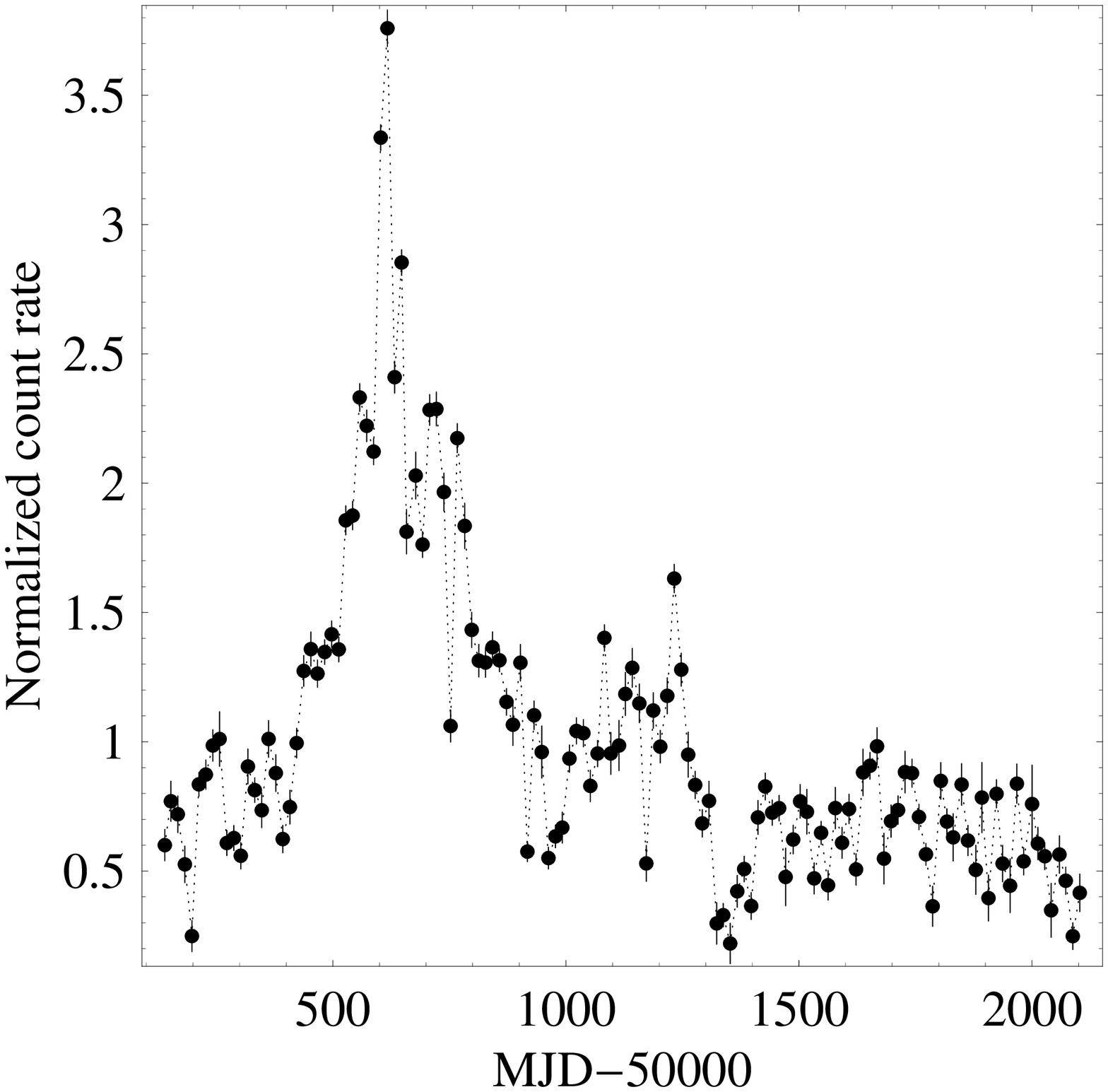}
\includegraphics[width=3.35in]{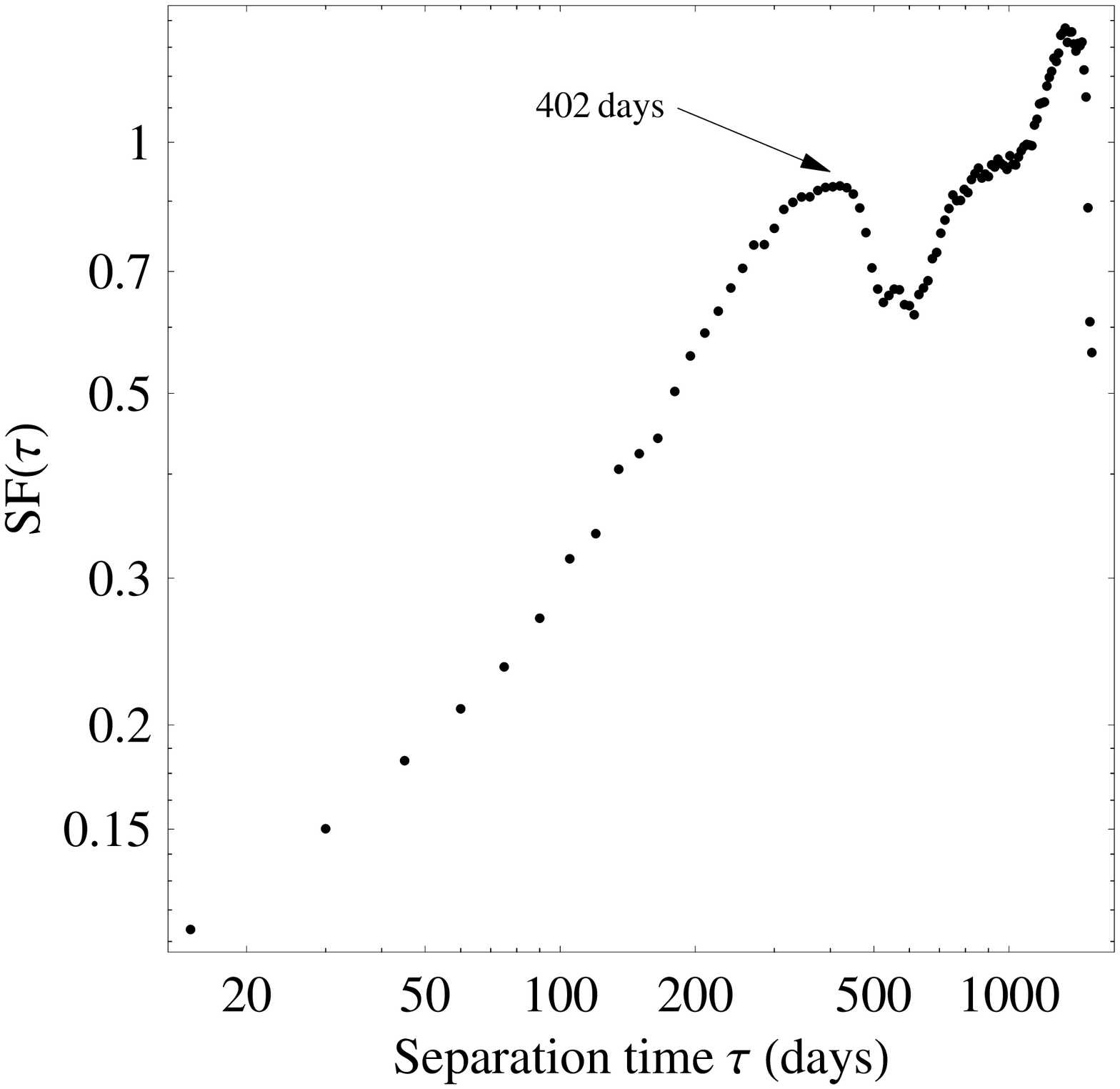}
\caption{[Left panel] The long-term 2--10 keV {\it ASM} light curve of \mrk501, between 50100 MJD--52100 MJD, in bins of 15-days. The dotted line is linear interpolation intended to guide the eye.\newline
[Right panel] The SF of the {\it ASM} light curve of \mrk501, in bins of 15-days (in logarithmic scale). The SF-break occurs around 402 days.}
\label{fig:long_sf} 
\end{figure*}
\begin{figure*} 
\parbox{0.5\linewidth}{\hspace*{-15em}\includegraphics[height=3in]{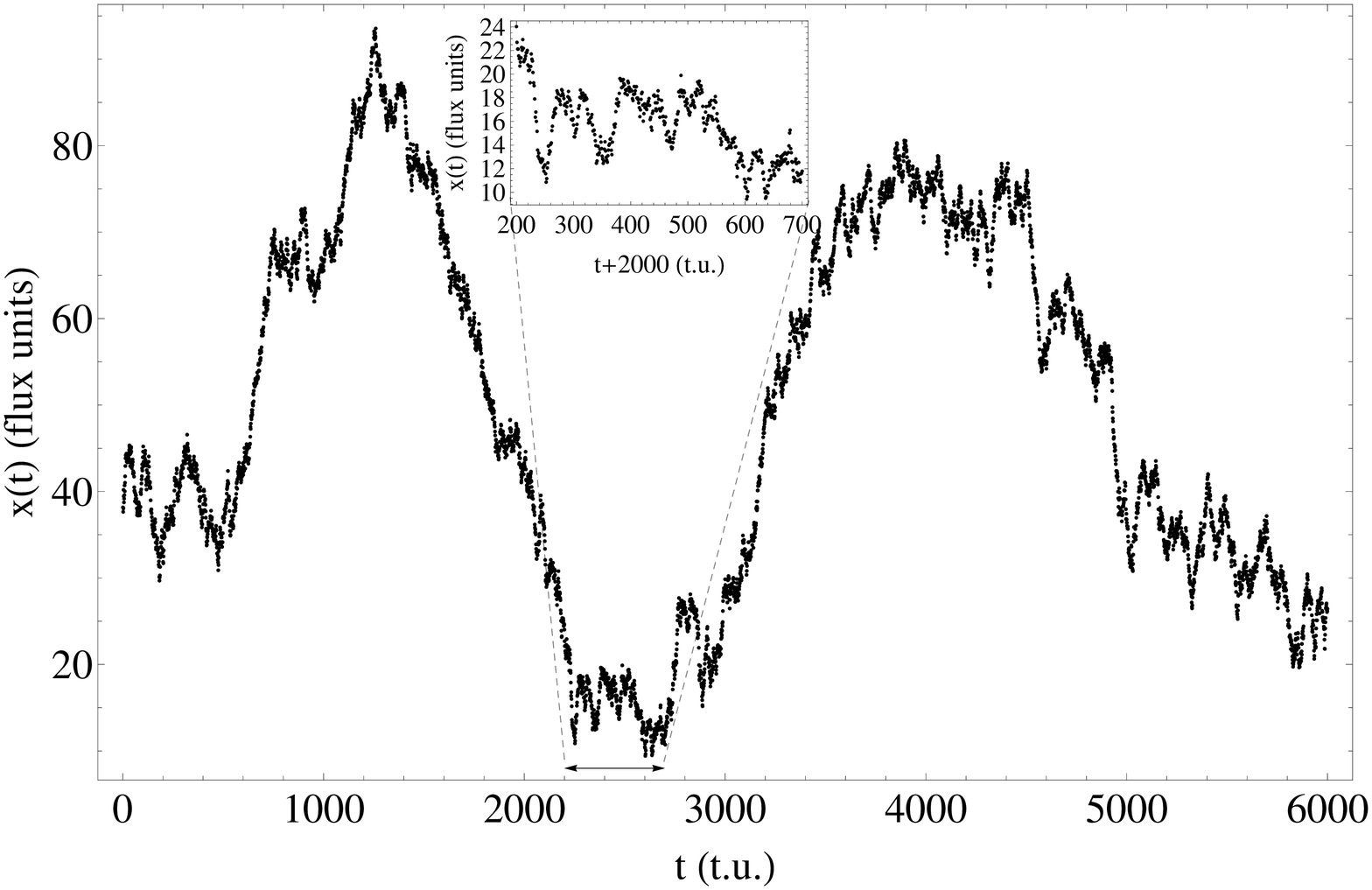}}\\
\parbox{0.5\linewidth}{\vspace*{-38.4em}\hspace{26em}\includegraphics[width=2.2in]{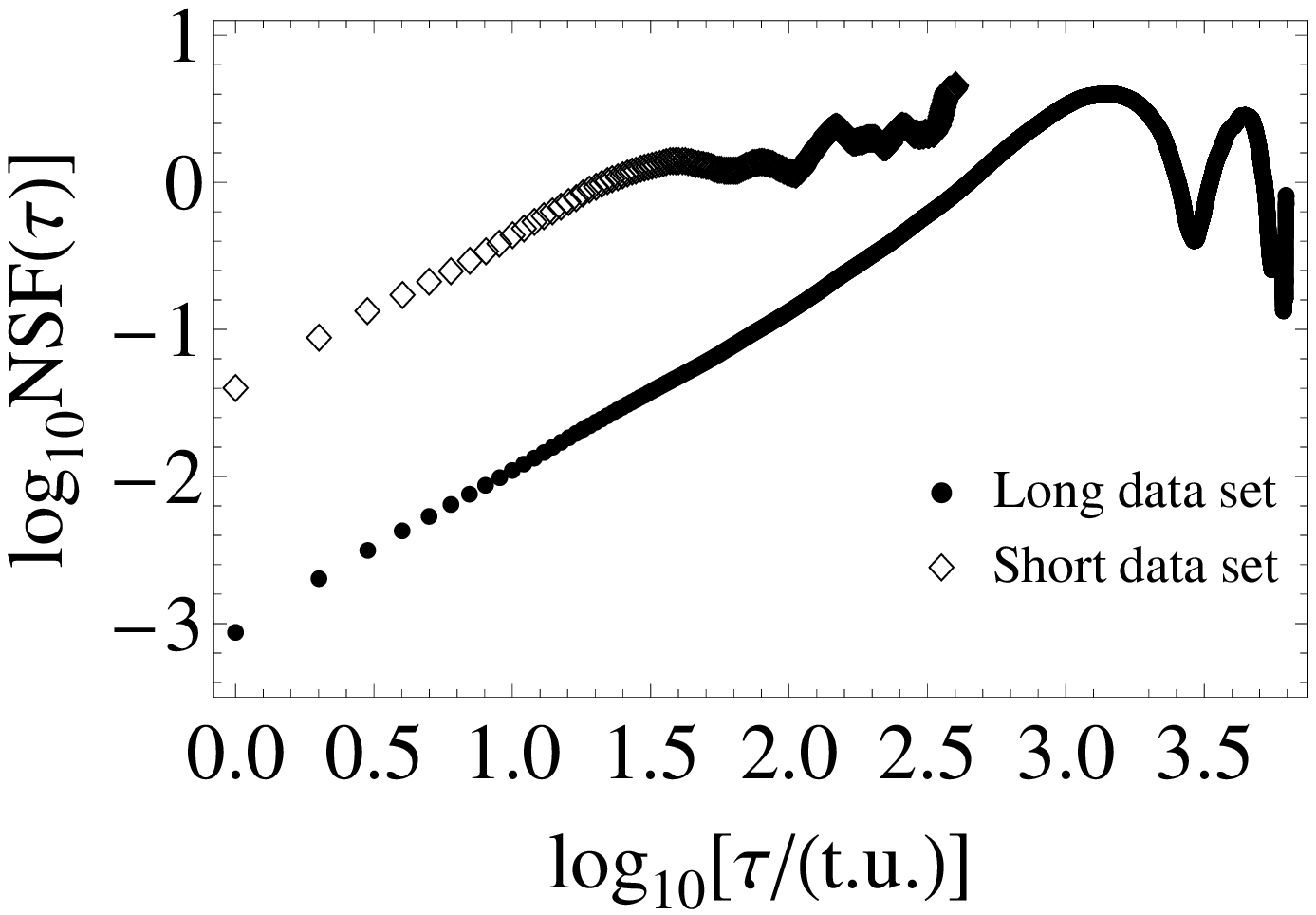}}\\
\parbox{0.5\linewidth}{\vspace*{-17em}\hspace{26.05em}\includegraphics[width=2.2in]{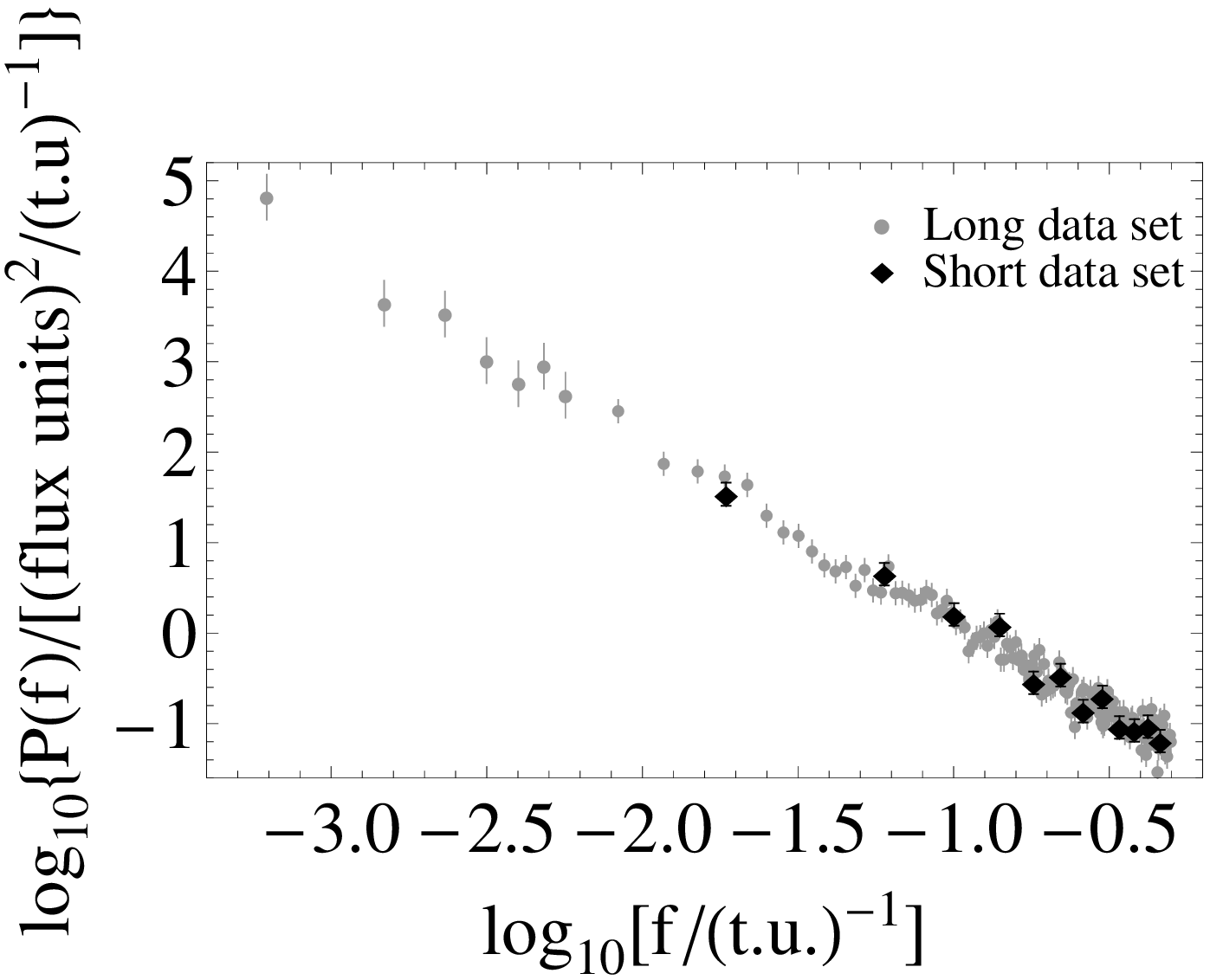}}
\caption{[Left panel] The overall simulated light curve 6000 t.u. long produced from an underlying PSD having a power-law form with index -2. The inlay shows a short segment of the light curve 500 t.u. long.\newline
[Top right panel] The NSF of the data in the inlay (empty diamonds) and the overall data set (black points). The break of the SF occurs at 40 and 1600 t.u. respectively.\newline
[Bottom right panel] The binned logarithmic periodograms of the short (filled diamonds) and the overall data sets (grey points). Both of them match the underlying PSD, being a featureless power-law of index -2.}
\label{fig:long_short_ds}
\end{figure*}

\subsection{SF-breaks and data set length}
\begin{figure} 
\includegraphics[width=3.3in]{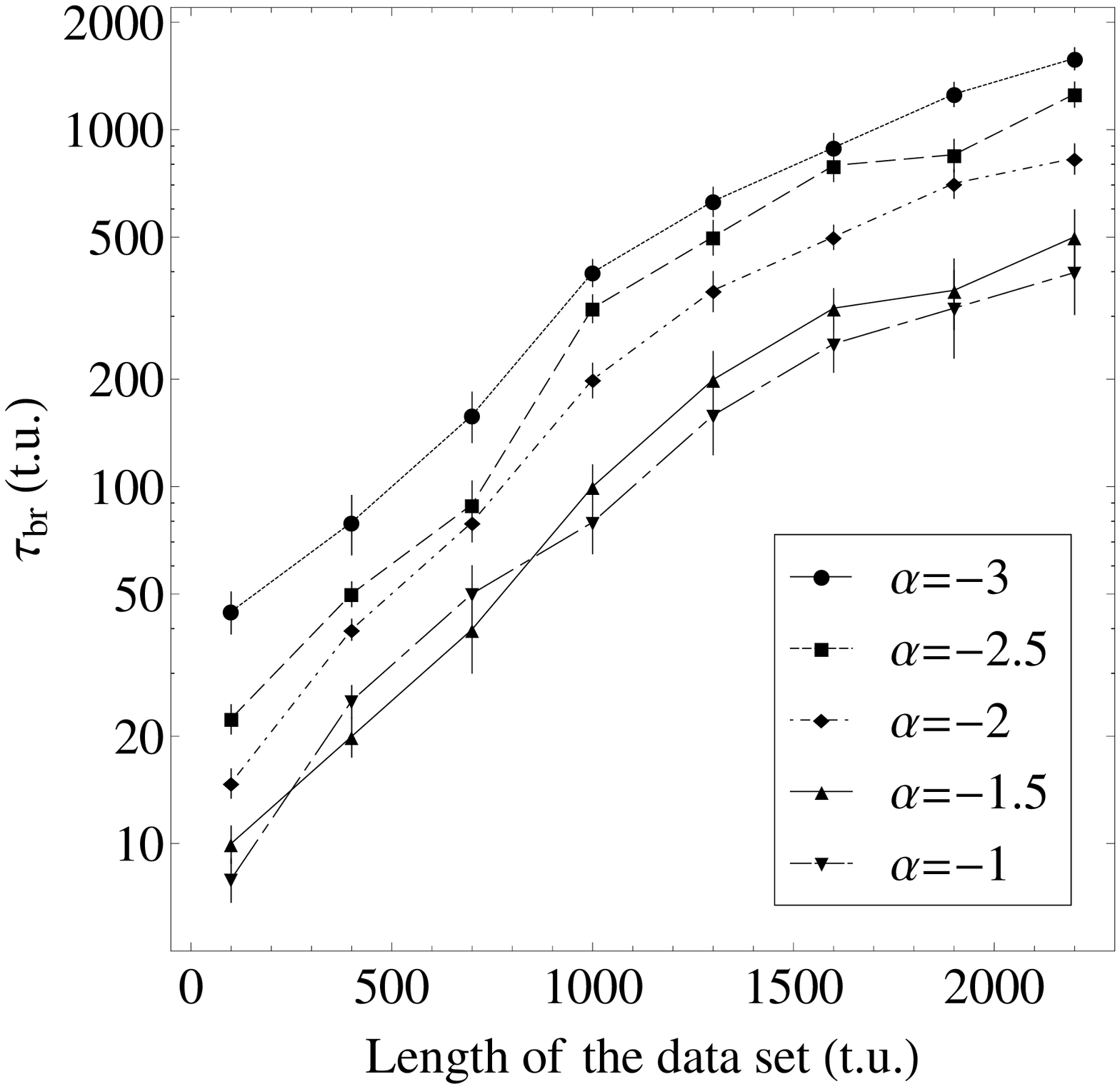}
\caption{The position of the SF-break, $\tau_{\rm br}$ (in logarithmic scale), as a function of the length of the data set assuming an underlying PSD of a power-law form of index $\alpha$. Each data point represents the mean value and the standard deviation of the distributions of the SF-breaks yielding from an ensemble of 2000 simulations having a power-law PSD of index -1 (filled circles), -1.5 (filled squares), -2 (filled diamonds), -2.5 (filled up-pointing triangles) and -3 (filled down-pointing triangles). The dashed lines are linear interpolations intended to guide the eye.}
\label{fig:sfbr_psd_length}
\end{figure}

To quantify the relationship between the spurious SF-breaks, $\tau_{\rm br}$, and both the length and the PSD-slope of the data set, we perform a set of simulations. For each featureless power-law PSD (with index $\alpha=-1$, $\alpha=-1.5$, $\alpha=-2$, $\alpha=-2.5$ and $\alpha=-3$) and for various data-set lengths  (100, 400, 700, 1000, 1300, 1600, 1900 and 2200 t.u.) we produce 2000 artificial light curves. In total we perform 40 simulations, each one consisting of 2000 artificial light curves. Then, for each PSD-index and light curve length we localise the position of the $\tau_{\rm br}$, as described in Section \ref{ssec:asca_sf}, and after forming its distribution we estimate its mean and its standard deviation\footnote{The distributions are very close to Gaussian (see e.g.\ Fig. \ref{fig:sf_hist_break_and_long_dist}).} (Fig. \ref{fig:sfbr_psd_length}).\par
For all the PSD-slopes we can readily see that as the length of the data set increases from $t$ t.u. to $t'$ t.u. $(t'>t)$, the SF-break gradually shifts to larger values. This can be understood for the case of a random walk process ($\alpha=-2$) where the mean distance after $T$ steps is $\sqrt{T}$. The mean distance in our case is the mean variance which is measured from the SF, hence for $t'>t$ the maximum variance and hence the position of the break is expected to shift towards larger time-scales.\par
The important point here is that the measured variance of a data set (and hence the maximum variance depicted by the break) does not reflect the true underlying variability properties of the source. As the data set increases, larger variations produced from the same variability process having the same PSD enter the data sets and this is mapped in the SF as a break even in the case of featureless PSD which are lacking characteristic time-scales.\par
In Fig. \ref{fig:sfbr_psd_length} we show the position of the SF-breaks, $\tau_{\rm br}$, induced by the different length of the data sets for various PSD-slopes. This diagram can be used only for continuously sampled light curves for which the underlying PSD is a pure power-law with the given indices. The existence of data-gaps in the data set can alter significantly this picture since sampling patters play a major role in the form of the SF and hence the position of the break (Section \ref{sect:sf_gaps}). In the case that our data set has a PSD of a broken-power-law form, similar simulations should be performed in order to specify the position of the fake SF-break. Note that the physically interesting PSD-breaks will be mapped in the SF but in statistically unsound way with respect to their position and their uncertainties (Section \ref{ssec:phys_meaningful_ts}).\par
Consider the case of the blazar PKS\,2155-304 as presented by \citet{zhang02a}. The 1997 ($\sim$100 points) and 1999 ($\sim$200 points) data sets are the most continuous ones \citep[figure 2 and figure 3 in][]{zhang02a} and they have featureless PSD with slopes $\alpha=-2$ and $\alpha=-3$ respectively \citep[Table 3 in][]{zhang02a}. The SF of the corresponding light curves \citep[figure 6, middle right panel and bottom right panel respectively, in][]{zhang02a} exhibit breaks of the order of $\tau_{\rm br}\approx12.8$ ksec and $\tau_{\rm br}\approx61.7$ ksec respectively (or 13 t.u. and 61.7 t.u respectively for SF-bins of 1 ksec). A SF-break for the 1997 data set is predicted (from Fig. \ref{fig:sfbr_psd_length}) to lie at around $14.8\pm1.4$ t.u. and for the 1998 data set, at around 60 t.u. with a simple linear interpolated estimation. This outcome readily tells us that these breaks can originate from a variability process with no characteristic time-scale simply described by a featureless PSD of a power-law form.     
\section{THE STRUCTURE FUNCTION AND THE BLAZAR SHOT-NOISE MODEL}
\label{sect:shot_noise}
Particle acceleration in shocks \citep{kirk98} is commonly invoked to explain the observed temporal and spectral variability of blazars. There may be a variety of intrinsic timescales, i.e.\ acceleration, escape and light crossing time-scales, associated with a single emitting region. TAN01 simulate light curves on the assumption of certain shot parameters and, from the similarity between the simulated and observed SFs, estimate some temporal shot properties. Here we caution against this approach as, in general, the light curves derived from such shots do not exhibit the same PSD as those observed from the blazars. A physically correct model should reproduce both the SF and the PSD.\par
In TAN01 a number of simulations were carried out in order to associate the SF-breaks with the properties of the observed flares. The light curve is regarded as a superposition of {\it triangular shots}\footnote{The last 40 years the commonly used shot-noise models consist of pulses with either a symmetric exponential rise and decay profile or only an exponential decay profile \citep[e.g.][]{lochner91,burderi97}.} with rise and decay time-scale $t_{\rm r}$ and $t_{\rm d}$ respectively, occurring randomly at $t_{\rm p}$ following a Poisson distribution, with an intensity $I_0$ at $t_{\rm p}$  
\eqb
\mathcal{I}_{\rm tr.sh.}(t)=\left\{
\begin{array}{c l}
  \frac{I_0}{t_{\rm r}}\left[t-(t_{\rm p}-t_{\rm r})\right]& t_{\rm p}-t_{\rm r}<t\leq t_{\rm p} \\ \\
  -\frac{I_0}{t_{\rm d}}\left[t-(t_{\rm p}-t_{\rm d})\right]& t_{\rm p}<t<t_{\rm p}+t_{\rm d}
\end{array}
\right.
\label{eqe:triang_shot}
\eqe
Then, the SF properties were studied with respect to the aforementioned time-scales and the main conclusion was that only the shortest time-scale between $t_{\rm r}$ and $t_{\rm d}$ determines the position of the break. Moreover, due to the fact that the {\it ASCA} data set of \mrk501 consists mainly of symmetric flares, they connect the shortest time-scale derived from the SF to the light crossing time of the emission region.\par
The apparent similarities between the observed SFs and those coming from the shot-noise simulations, presented by TAN01, give an erroneous impression that such correspondences hold. Apart from the fact that breaks in the SF can occur in data sets with no characteristic time-scales (Section \ref{sec:xray_obs}), we posit that the shot-noise model is not a physically realistic approach that can be used in order to associate the observed breaks of the SF to the smallest time-scales embedded in the data sets. In order for the shot-noise simulations to be realistic representations of the observed {\it ASCA} data set of \mrk501, they must be able to reproduce the observed PSD as well as the SF.\par
We repeat the simulations of TAN01 by creating shot-noise light curves 2000 t.u. long and we calculate the corresponding PSD. Since the general shape of the latter can be understood qualitatively by the PSD form of the individual triangular shots, used for the construction of each light curve, it is useful to look at the general analytical form of the PSD for a single triangular shot as derived directly from the Fourier transform of the continuous functions (Appendix \ref{app:psd_triang}). We should note here that since the simulated light curves are discretised, apart from their noisy form, due to the randomness of the occurrence times and of the intensities, we expect the PSD to flatten towards high frequencies due to aliasing effects \citep[e.g.][]{papadakis93}.\par
\begin{figure} 
\includegraphics[width=2.8in]{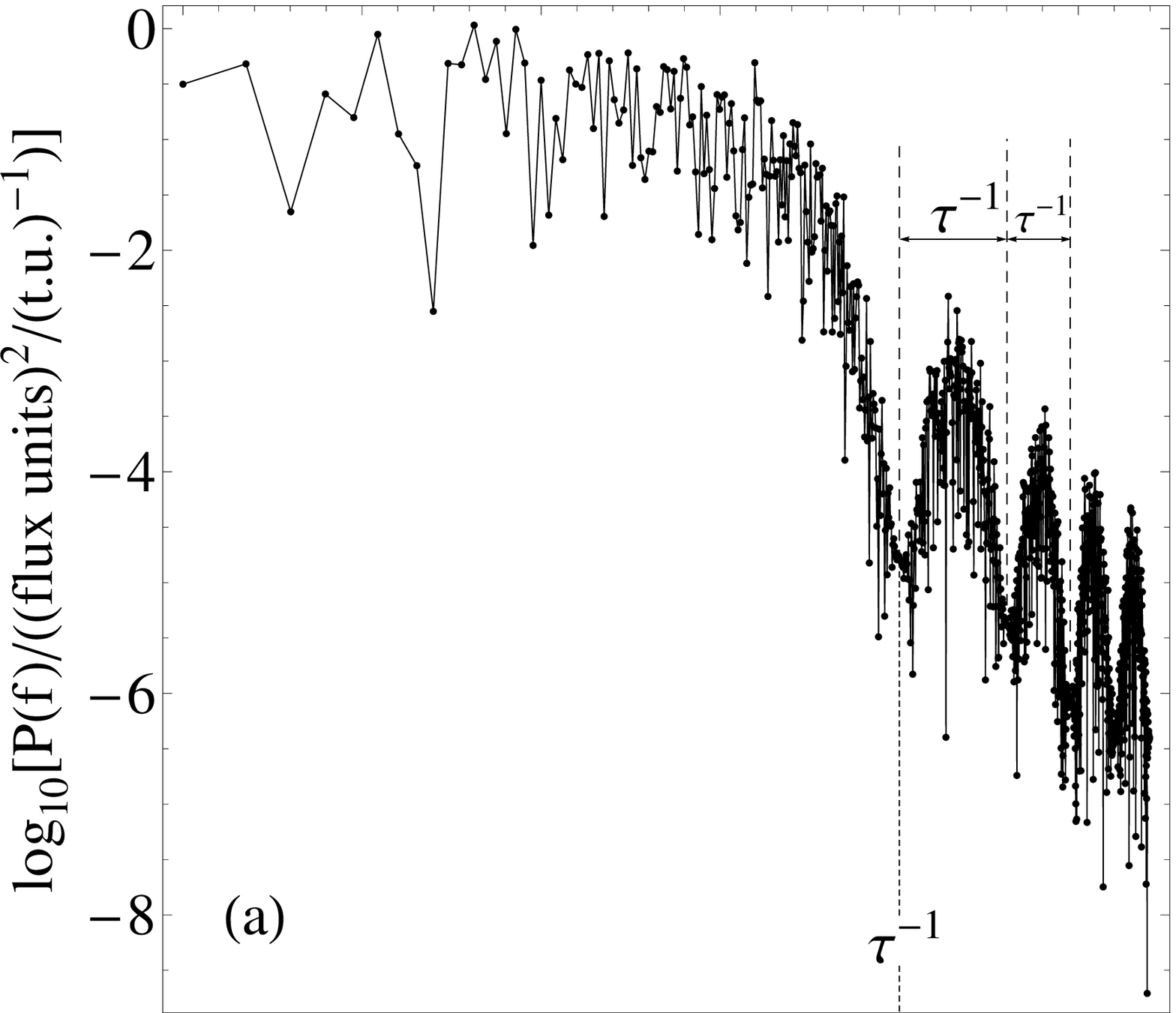}\\[-0.17em]
\includegraphics[width=2.8in]{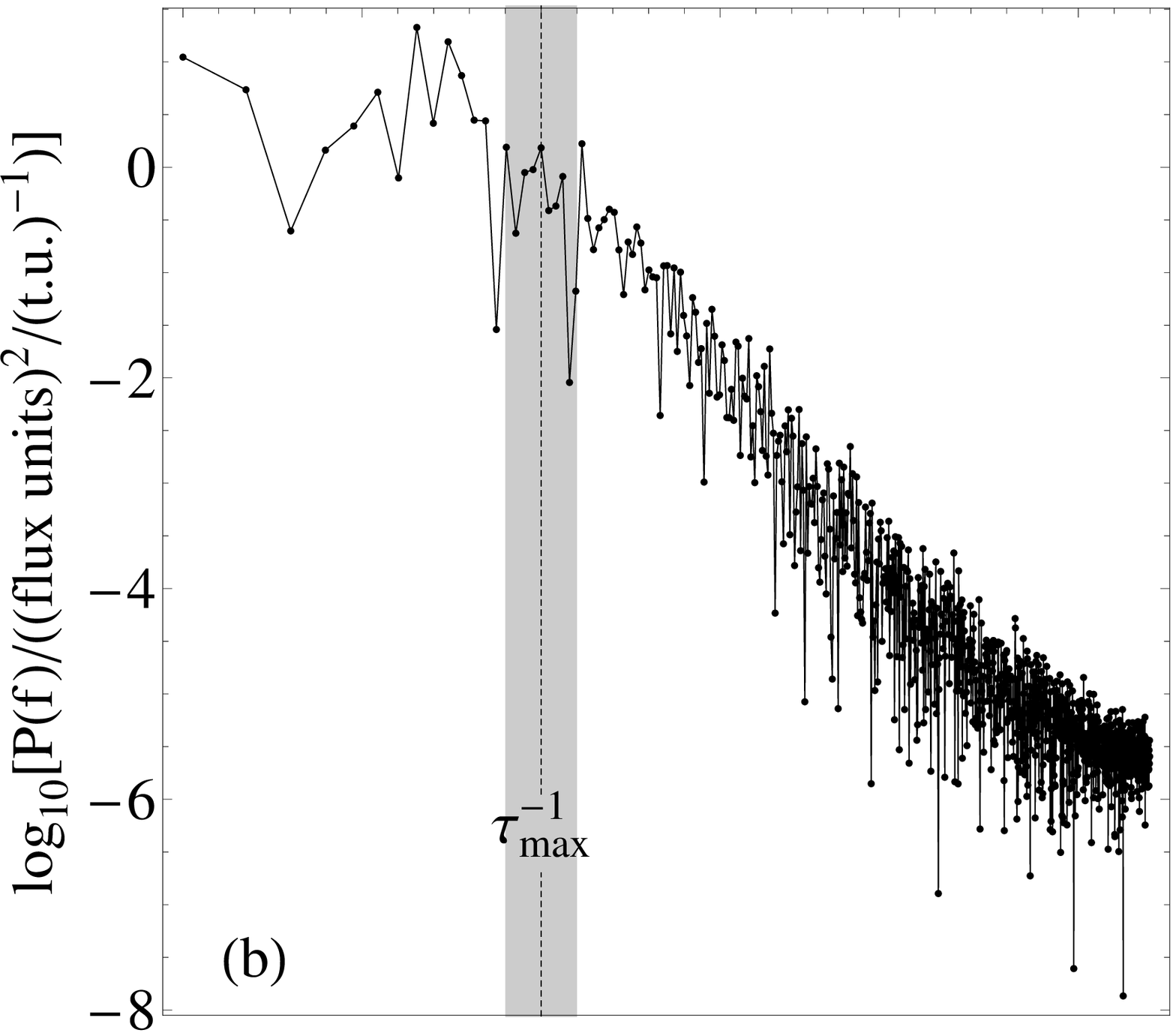}\\[-0.7em]
\includegraphics[width=2.8in]{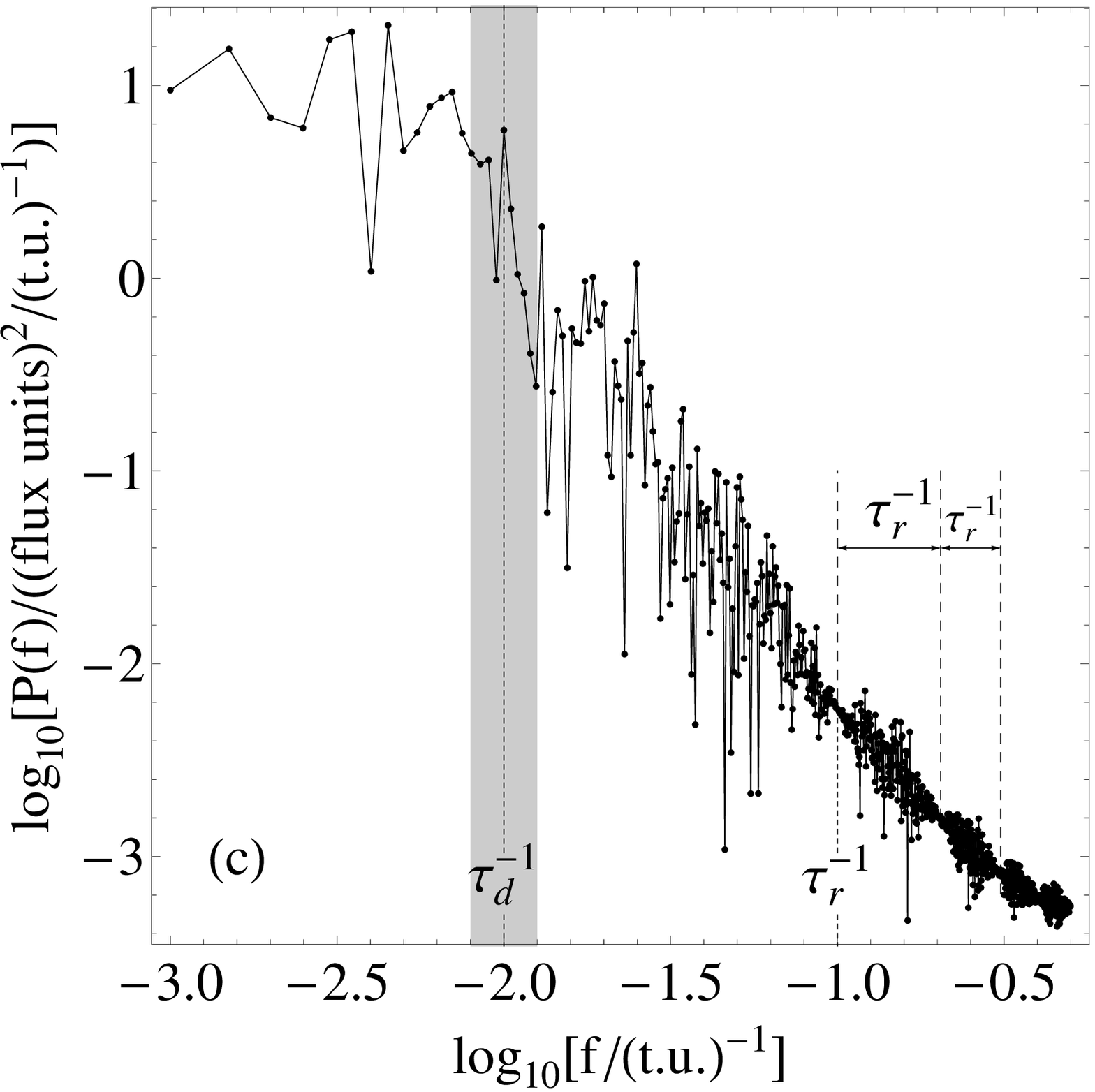}
\caption{\label{fig:psd_simul_triang_shots}Example PSDs for the shot-noise model consisting of triangular shots for three cases of shots.\newline
[Top panel (a)] Identical symmetric shots with rise and decay time, $\tau_{\rm r}$ and $\tau_{\rm d}$ respectively, equal to $10$ t.u.\newline
[Middle panel (b)] Symmetric shots with rise and decay time equal to $\tau$ ($\tau_{\rm r}=\tau_{\rm d}=\tau$), and $\tau$ distributed between $\tau_{\rm min}=10$ t.u. and $\tau_{\rm max}=100$ t.u.\newline
[Bottom panel (c)] Asymmetric identical shots with rise and decay time $\tau_{\rm r}=10$ t.u. and $\tau_{\rm d}=100$ t.u., respectively.}
\end{figure}

Initially TAN01 consider that the light curve consists of identical symmetric triangular shots i.e.,\ $\tau=t_{\rm r}=t_{\rm d}=10$ t.u.
The PSD of such a process breaks at $\tau^{-1}$ and becomes very steep at higher frequencies with power-law of index $\alpha\approx-4$ superimposed on a wave with peaks separated by $\tau^{-1}$ (Fig. \ref{fig:psd_simul_triang_shots}, panel a). Qualitatively we can justify this behaviour by considering the PSD of a single symmetric triangular shot which is proportional to $f^{-4}\sin^4(\pi\tau f) $  (equation (\ref{eqe:psd_symm_triang})). We expect that in the case of randomly occurring symmetric triangular shots, having different intensity but the same duration, the overall shape of their PSD should be similar to the sum of the PSDs of the individual symmetric triangular shots.\par
The second case considered by TAN01 is that of non-identical symmetric shots ($\tau=t_{\rm r}=t_{\rm d}$) whose time-scale $\tau$ varies randomly between $\tau_{\rm min}=10$ t.u. and $\tau_{\rm max}=100$ t.u. In this case the PSD is flat until $\tau_{\rm max}^{-1}$ and then it breaks following a simple steep power-law of index $\alpha\approx-4$ (Fig. \ref{fig:psd_simul_triang_shots}, panel b). The oscillatory behaviour, caused by the sine term of equation (\ref{eqe:psd_symm_triang}), is smoothed out due to the random distribution of $\tau$s coming from every symmetric shot.\par
The third and the last case considered by TAN01 is that of asymmetric identical shots with $\tau_{\rm r}=10$ t.u. and $\tau_{\rm d}=100$ t.u. This is actually a generalization of the first case. The PSD breaks at $\tau_{\rm d}^{-1}$, then it becomes steep with a power-law index of $\alpha\approx-2$ and after $\tau_{\rm r}^{-1}$ beats separated by $\tau_{\rm r}^{-1}$ are superimposed on the underlying power-law (Fig. \ref{fig:psd_simul_triang_shots}, panel c). Qualitatively, based on the PSD of a single asymmetric shot (Fig. \ref{fig:anal_psd_triang}) we can readily distinguish very similar spectral features such as the position of the first break, the slope between $\tau_{\rm d}^{-1}$ and $\tau_{\rm r}^{-1}$, and the separations in frequency of the beating frequencies. \par
Each of the aforementioned simulated PSDs differs significantly from the PSD derived from the {\it ASCA} observations i.e.\ a simple power-law with index $\alpha=-1.80\pm0.09$. Neither steep slopes nor sinusoidal features in oscillating or beating forms appear in the featureless PSD of \mrk501. The fact that the PSD functions differ significantly between simulations and observations clarifies in an unambiguous way that the variability properties of \mrk501 can not be described in a physically correct way from the shot-noise model. The latter can very well reproduce the various SF-features i.e.\ slopes and breaks observed in the actual SFs of the observed in blazars, but it can not reproduce the featureless PSD. This means that the shot-noise simulations are not representative of the true underlying variability properties of \mrk501 and hence should not be used in order to associate the observed SF-breaks with characteristic physical time-scales.\par

\section{FITTING MODELS TO THE STRUCTURE FUNCTION: PARAMETER AND ERROR ESTIMATION}
\label{sect:fitting_sf}
\subsection{Lack of Gaussianity and statistical independence}
\label{ssect:non_indepen_gauss}
One of the major problems that affects the use of SF is that the various estimates, $SF(\tau_i)$, are not independent of each other. This problem severely affects the fitting routines e.g.\ least-squares, maximum likelihood, that are commonly used in the published blazar-SF-literature to derive the SF-breaks and the -slopes. These routines all require that the data points should be statistically independent so that the probability of obtaining the full ensemble, in our case $SF(\tau_i)$ for all $\tau_i$, is equal to the product of the probabilities of obtaining the individual $SF(\tau_i)$ \citep[e.g.][]{bevington92}.\par
That means that if we want to fit a broken-power-law of the form \citep[e.g.][]{zhang02a}
\eqb
SF_{\rm M}(\tau;C,\tau_{\rm max},\beta_1,\beta_2) = \left\{
\begin{array}{c l}
C\left(\frac{\tau}{\tau_{\rm max}}\right)^{\beta_1} & \tau\leq\tau_{\rm max} \\
C\left(\frac{\tau}{\tau_{\rm max}}\right)^{\beta_2} & \tau>\tau_{\rm max}
\end{array}
\right.
\label{eqe:sf_fit_model}
\eqe
to a data set of the form $$\bigl\{\bigl(\tau_1,SF(\tau_1)\bigr),\bigl(\tau_2,SF(\tau_2)\bigr),\ldots,\bigl(\tau_N,SF(\tau_N)\bigr)\bigr\}$$ we have to maximize the probability of obtaining the ensemble of estimates $\{SF(\tau_1),SF(\tau_2),\ldots,SF(\tau_N)\}$, known as the {\it joint probability}
\eqb
P(C,\tau_{\rm max},\beta_1,\beta_2)&=&\prod_{i=1}^N P_i\left(SF_{\rm M}(\tau_i;C,\tau_{\rm max},\beta_1,\beta_2)\right)
\label{eqe:independ_prob}\eqe
where $P_i\bigl(SF_{\rm M}(\tau_i;C,\tau_{\rm max},\beta_1,\beta_2)\bigr)$ is the probability of obtaining the estimate $SF(\tau_i)$. Equation (\ref{eqe:independ_prob}) is valid (i.e.\ the joint probability equals to the product of individual probabilities) {\it iff} the estimates $\{SF(\tau_1),SF(\tau_2),\ldots,SF(\tau_N)\}$ are independent of each other.\par

\begin{figure} 
\includegraphics[width=3.1in]{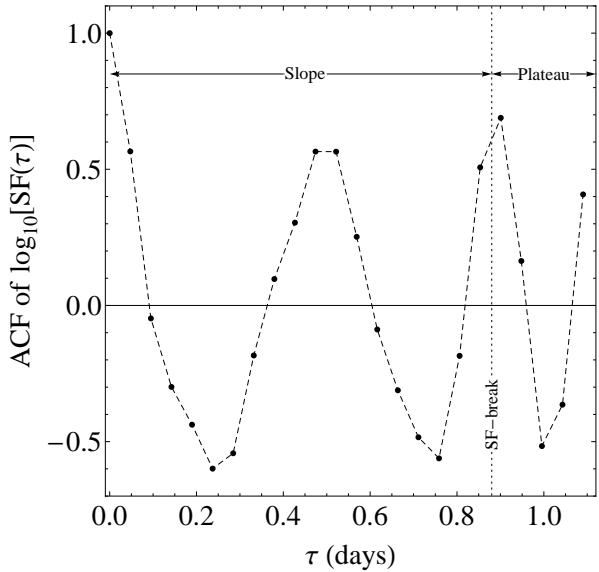}
\caption{The ACF of the first 25 detrended logarithmic SF-estimates of \mrk501 (from Fig. \ref{fig:asca_arti_sf}) showing the degree of linear dependence among them. The vertical dotted line shows the position of the SF-break at 0.88 days, and the arrows indicate the SF-slope and plateau regions. The dashed lines among the points of the ACF are linear interpolations intended to guide the eye.}
\label{fig:acf_sf} 
\end{figure}

Unfortunately the adjacent SF-estimates are far from independent. In Fig. \ref{fig:acf_sf} we show the {\it autocorrelation function} (ACF) of the first 25 logarithmic SF-estimates (out of 137) of \mrk501 (Fig. \ref{fig:asca_arti_sf}) after detrending them by subtracting a fifth degree polynomial. Note here that this detrending is crucial since we want to check the linear relations between the SF-estimates (e.g.\ short-term dependence), and not long-term linear-dependence which is due to the general shape of the SF. In Fig. \ref{fig:acf_sf} we quantify the degree of linear dependence of the SF-estimates for the slope, the break and the plateau regions. It is impossible to have linearly correlated values which are independent\footnote{We should note that sometimes the absence of a linear correlation is confused with statistical independence but this is the case only when we are dealing with Gaussian distributions. Genuine statistical independence requires not only linearly uncorrelated values but also the absence of any functional relation between the variables.}. We can readily see that up to $\tau=0.88$ days (the range which is used in order to derive the break time-scale in Section \ref{ssec:asca_sf}) the degree of linear correlation is greater than 0.5.\par

The assumption of Gaussianity is another issue with the SF fitting procedures. Only if the distribution of the SF-estimates within each $\tau_i$ is Gaussian (with a mean value $SF(\tau_i)$ and standard deviation $\sigma_i$), is it valid to consider that
\eqb
&&P_i\bigl(SF_{\rm M}(\tau_i;C,\tau_{\rm max},\beta_1,\beta_2)\bigr)=\nonumber\\
&&\frac{1}{\sigma_i\sqrt{2\pi}}e^{-\frac{1}{2}\left[\frac{SF(\tau_i)-SF_{\rm M}(\tau_i;C,\tau_{\rm max},\beta_1,\beta_2)}{\sigma_i}\right]^2}
\label{eqe:indi_gaus_prob}
\eqe
and estimate the product in the joint probability (equation (\ref{eqe:independ_prob})) (which implies independence) as
\eqb
&&P(C,\tau_{\rm max},\beta_1,\beta_2)=\nonumber\\
&&\prod_{i=1}^N\left(\frac{1}{\sigma_i\sqrt{2\pi}}\right)e^{-\frac{1}{2}\sum_{i=1}^{N}\left[\frac{SF(\tau_i)-SF_{\rm M}(\tau_i;C,\tau_{\rm max},\beta_1,\beta_2)}{\sigma_i}\right]^2} 
\label{eqe:join_gaus_prob} 
\eqe
According to the maximum likelihood method, in order to find the most probable set of parameters $\left(C,\tau_{\rm max},\beta_1,\beta_2\right)$ we must maximize $P(C,\tau_{\rm max},\beta_1,\beta_2)$ (equation (\ref{eqe:join_gaus_prob})) or equivalently we must minimize the sum-argument $\chi^2$ in the exponential of equation (\ref{eqe:join_gaus_prob})
\eqb
\chi^2=\sum_{i=1}^{N}\left[\frac{SF(\tau_i)-SF_{\rm M}(\tau_i;C,\tau_{\rm max},\beta_1,\beta_2)}{\sigma_i}\right]^2
\label{eqe:chi_square}
\eqe
Thus the quantity $\chi^2$ automatically defines the goodness-of-fit which is actually a measure of the probability of obtaining the observed SF-estimates from a given set of parameters $\left(C,\tau_{\rm max},\beta_1,\beta_2\right)$ but only when we are dealing with independent and Gaussian SF-estimates.\par
\begin{figure*} 
\includegraphics[width=2.2in]{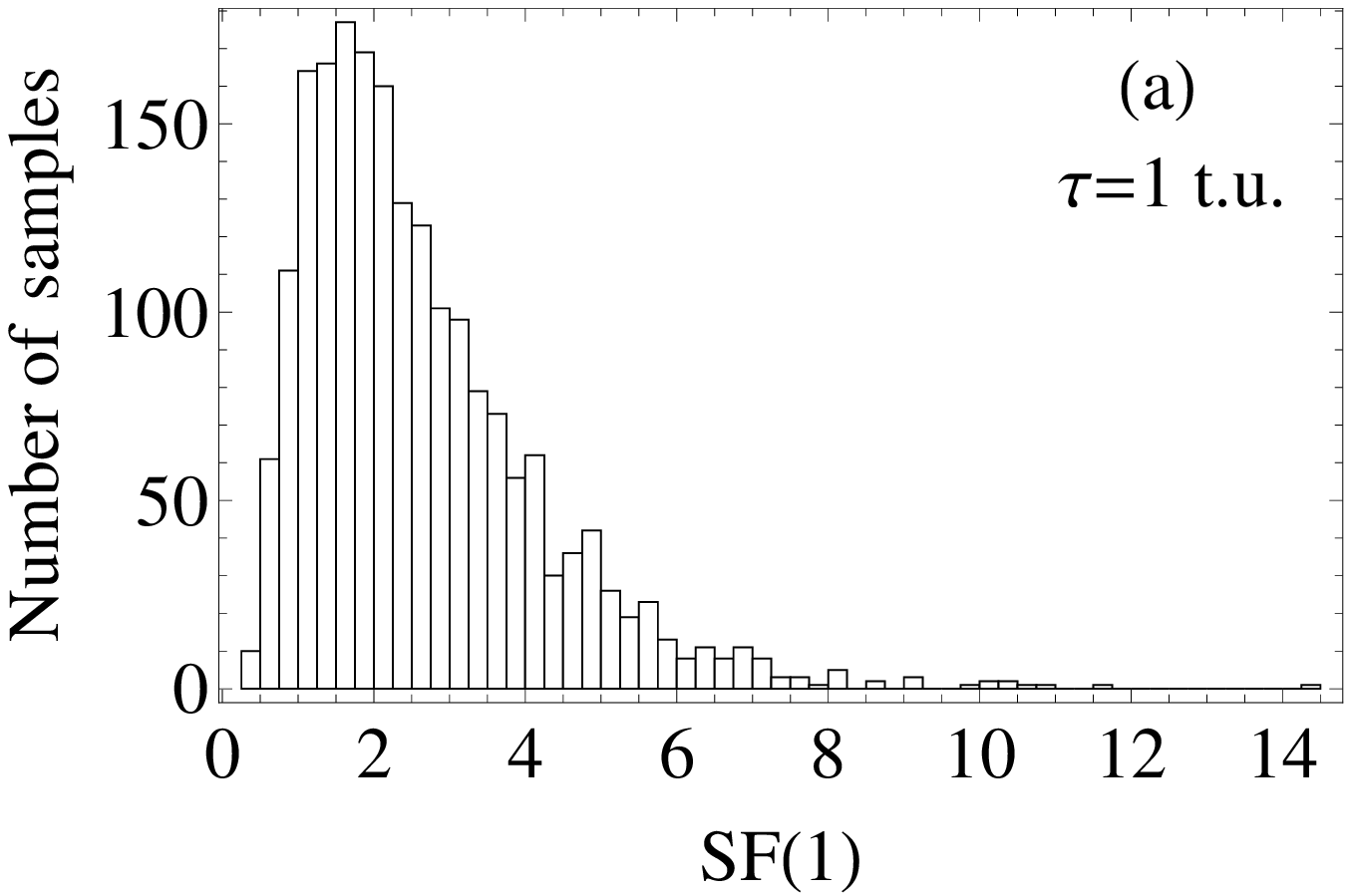}
\includegraphics[width=2.25in]{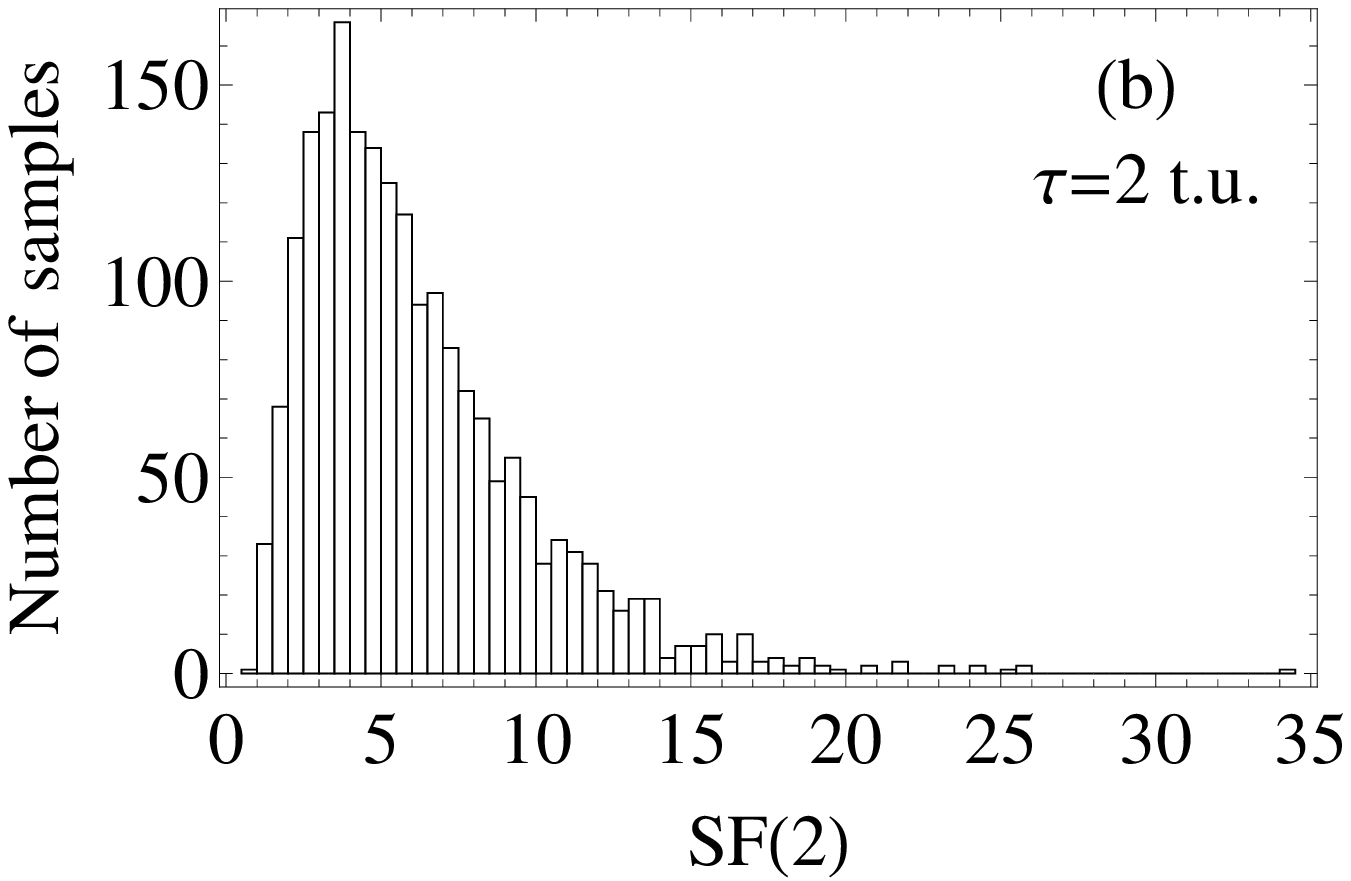}
\includegraphics[width=2.2in]{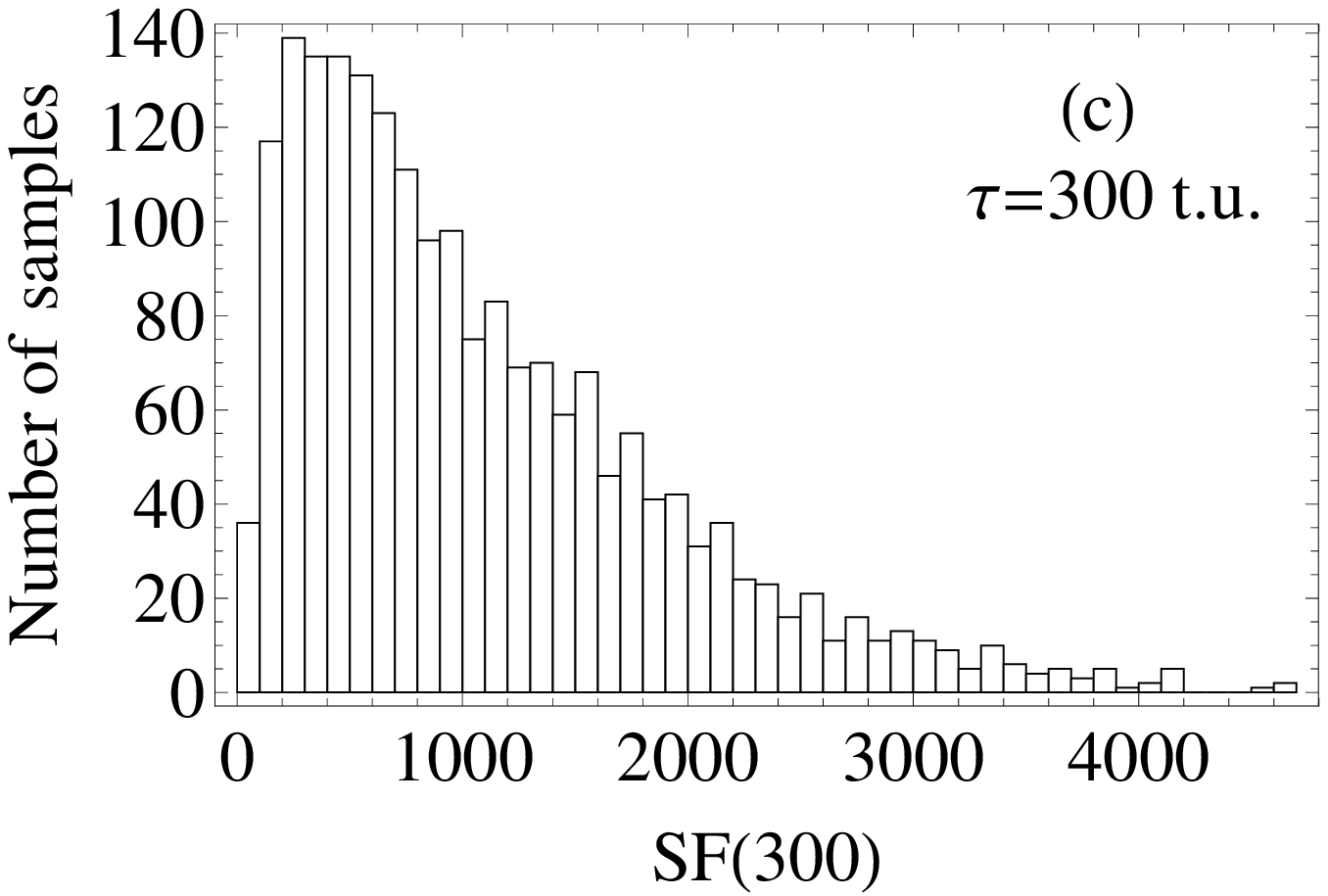}\\
\hspace{0.5em}{\includegraphics[width=2.12in]{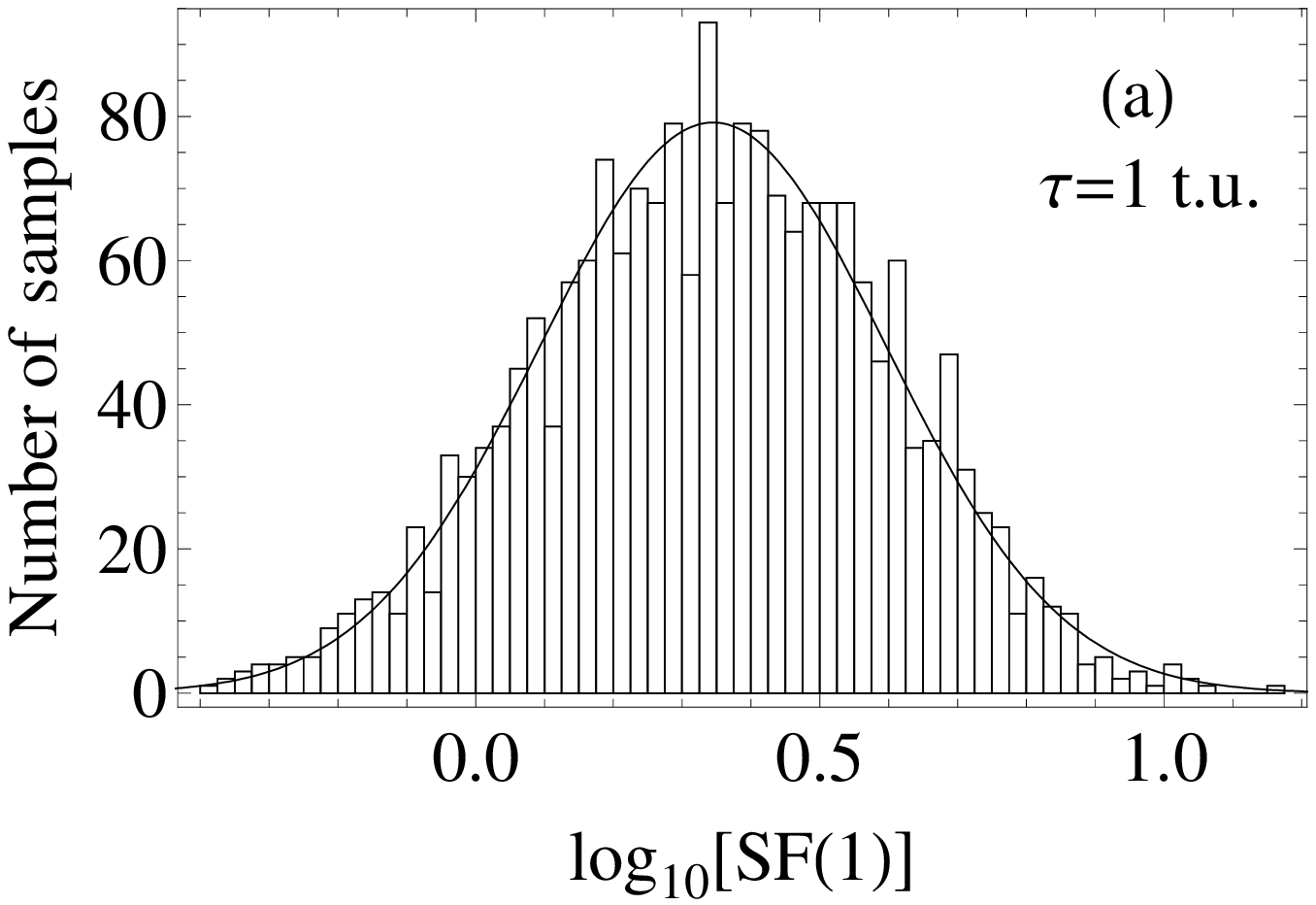}
\hspace{0.8em}\includegraphics[width=2.12in]{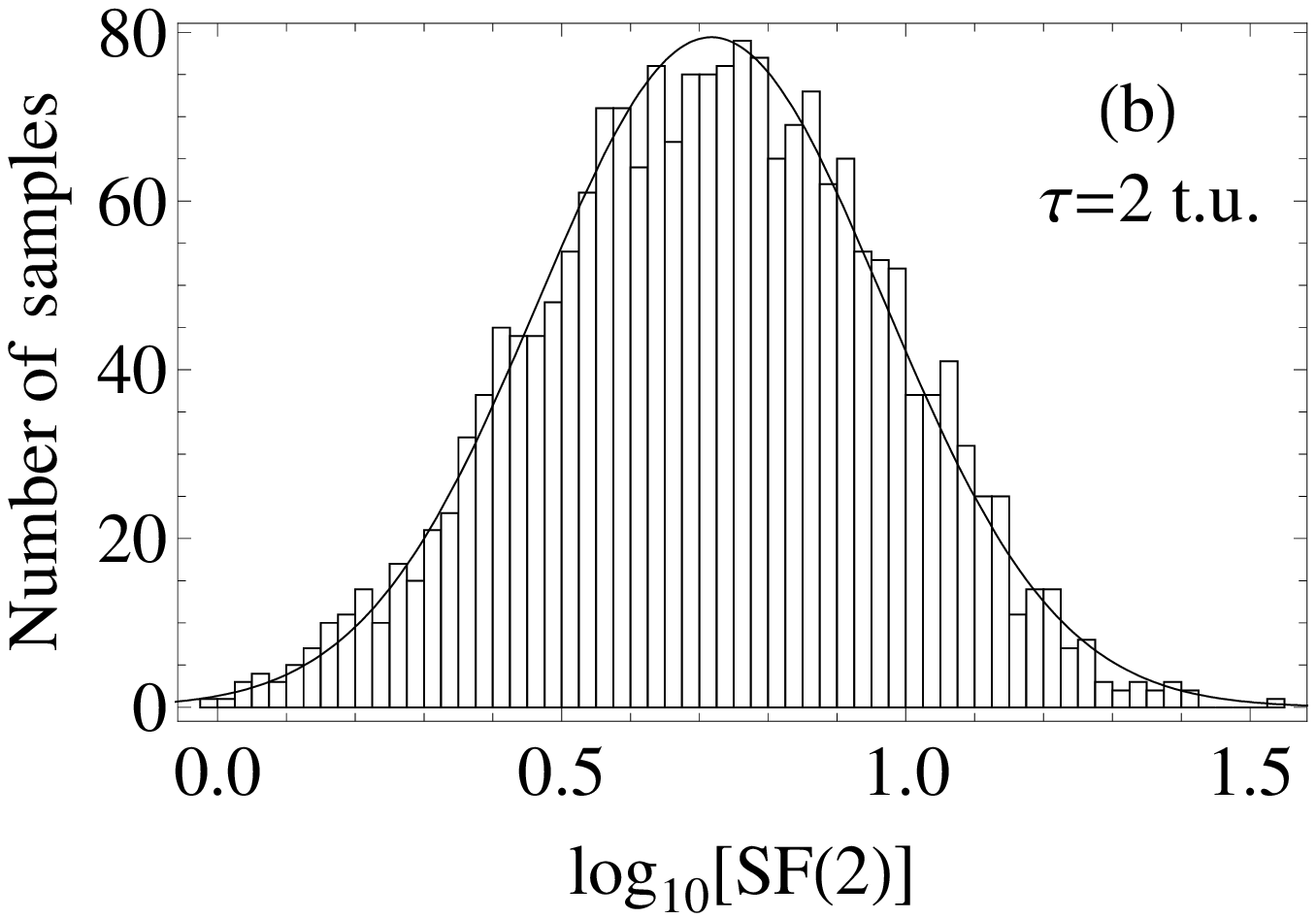}
\hspace{0.6em}\includegraphics[width=2.15in]{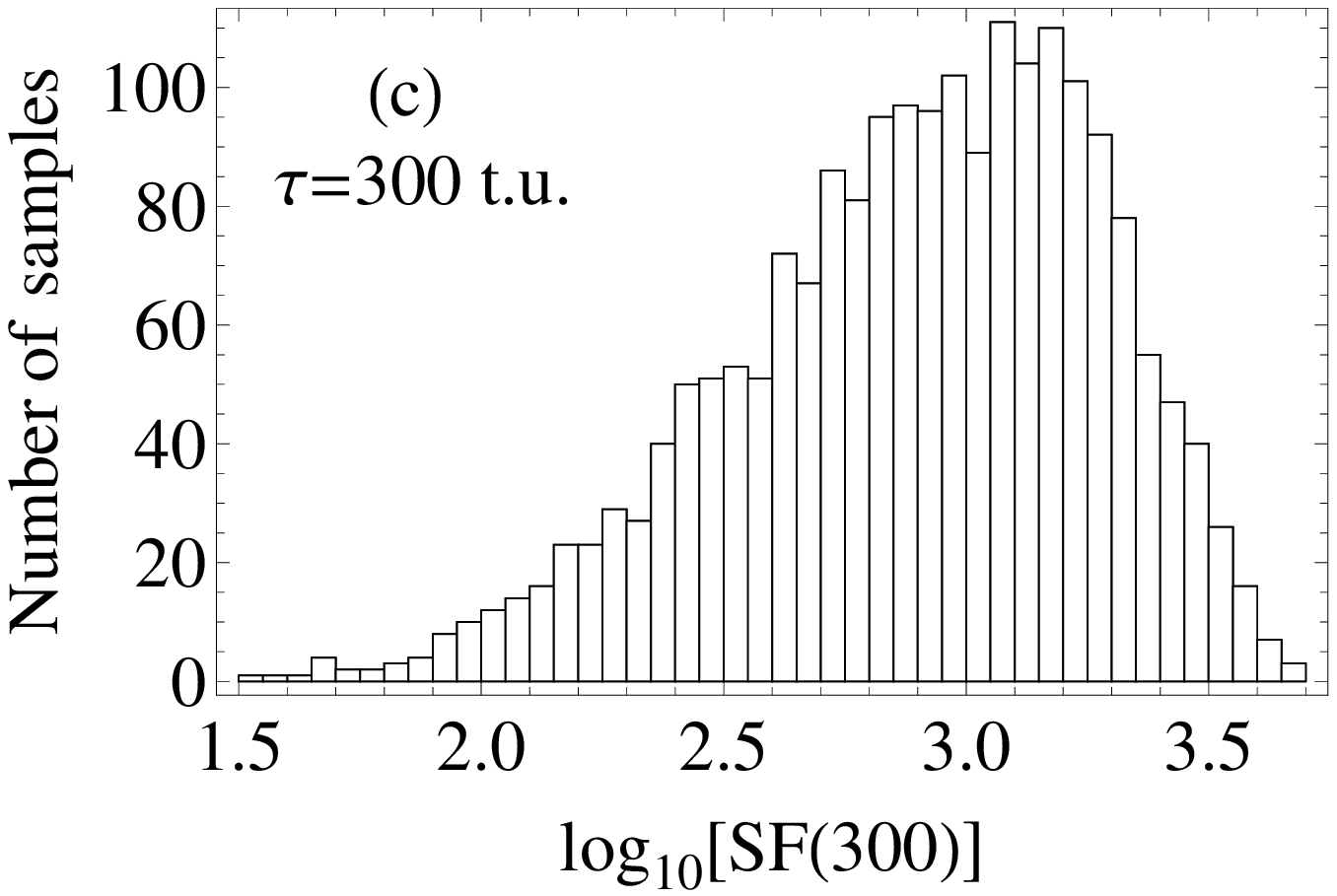}}\\
\caption{The distributions of the ensemble of 2000 SF-estimates, coming from 2000 random data sets 500 t.u. long having a PSD of a power-law form of index -2, for the first, second and three-hundredth time bin respectively.\newline
[Top panels] The distributions of the normal SF-estimates (the histogram bins have a length of 1,1 and 100 respectively) is not Gaussian for all the $\tau$.\newline
[Bottom panels] The distribution of the logarithmic SF-estimates (the histogram bins have a length of 0.025, 0.025 and 0.05 respectively) approximates adequately the Gaussian distribution only for $\tau\apprle50$ t.u.}
\label{fig:sf_three_hist}
\end{figure*}

However, in reality we do not deal with Gaussian SF-estimates. By producing 2000 random data sets 500 t.u. long from a PSD of a simple power-law form having index -2, we form the distribution of the SFs for the first, the second, and the three-hundredth time bins respectively (Fig. \ref{fig:sf_three_hist}, top panels). The histograms clearly have a non-Gaussian form and remains non-Gaussian for all the time bins $\tau$. That means that any fitting relation of the form of equation (\ref{eqe:sf_fit_model}), dealing directly with the SF-estimates, does not provide reliable results since Gaussianity i.e.\ equation (\ref{eqe:indi_gaus_prob}), is not valid.\par
Interestingly, at least for those time bins which are below the fake SF-break, occurring at $\tau_{\rm br}\approx50$ t.u. (Fig. \ref{fig:sfbr_psd_length}), the logarithms of the SF-estimates follow a distribution which is closer to Gaussian (Fig. \ref{fig:sf_three_hist}, bottom panels a and b). For $\tau>\tau_{\rm br}$ even the logarithmic-estimates of the SF deviate significantly from Gaussianity (Fig. \ref{fig:sf_three_hist}, bottom panel c) and of course for all the time bins $\tau$ the measurements continue to be statistically dependent on each other.\par
\citet{kataoka01} define the sum of square differences $\chi^2_{\rm sim}=\sum_k\{\log_{10}[\langle SF(\tau_k)\rangle]-\log_{10}[SF(\tau_k)]\}^2$ , where the angle brackets indicate the arithmetic mean, in order to derive a statistical significance of the goodness of their SF fit, accounting in this way for the non-Gaussianity. They argue that $\chi^2_{\rm sim}$ is different from the traditional $\chi^2$ (i.e.\ equation (\ref{eqe:chi_square})) but the statistical meaning is the same. Since the SF measurements are not independent, a $\chi^2$ function of the form $\sum(S-E)^2$, where $S$ comes from simulations (i.e.\ from a given assumed underlying model) and $E$ are the actual estimates, makes sense only if for every SF bin the distribution of the entries is Gaussian (yielding the $(S-E)$ from the exponent of the Gaussian, equation (\ref{eqe:indi_gaus_prob})) and if all the SF bins are independent of each other (yielding the sum from the joint probability, equation (\ref{eqe:join_gaus_prob})).\par
Concerning the Gaussianity it would be more appropriate for \citet{kataoka01} to use the mean logarithm of the SF, $\langle\log_{10}[SF(\tau_k)]\rangle$ rather than the logarithm of the mean SF, $\log_{10}[\langle SF(\tau_k)\rangle]$, in their pseudo chi-square estimate: $\chi^2_{\rm sim,Gauss}=\sum_k\{\langle\log_{10}[SF(\tau_k)]\rangle-\log_{10}[SF(\tau_k)]\}^2$ where both quantities $\langle\log_{10}[SF(\tau_k)]\rangle$ and $\log_{10}[SF(\tau_k)]$ are distributed Gaussian within a $t_{\rm k}$. However, the main problem of non-independence is not avoided and therefore the derived significances are not statistically meaningful.\par
\subsection{Error underestimation}
\label{ssec:error_underestim}
In most of the cases, when we deal with statistically independent variables under the assumption of Gaussianity, the most probable value (i.e.\ the one that minimizes the $\chi^2$) and its standard deviation are enough to give a full picture of the distribution of the fit parameters.\par
There are several methods, as we are going to see, that are used in the literature claiming robustness and statistically meaningful errors for the SF fit parameters i.e.\ slopes and breaks. In this section we show through simulations that the lack of statistical independence in the SF-estimates yields very small estimates of uncertainty in the fitting parameters, which do not reflect the true underlying distributions of the fitting parameters. The essence of any correct fitting procedure is to yield a statistically meaningful description of these distributions.     
\subsubsection{Fitting the SF-slopes and -breaks}
\begin{figure*} 
\includegraphics[width=2.6in]{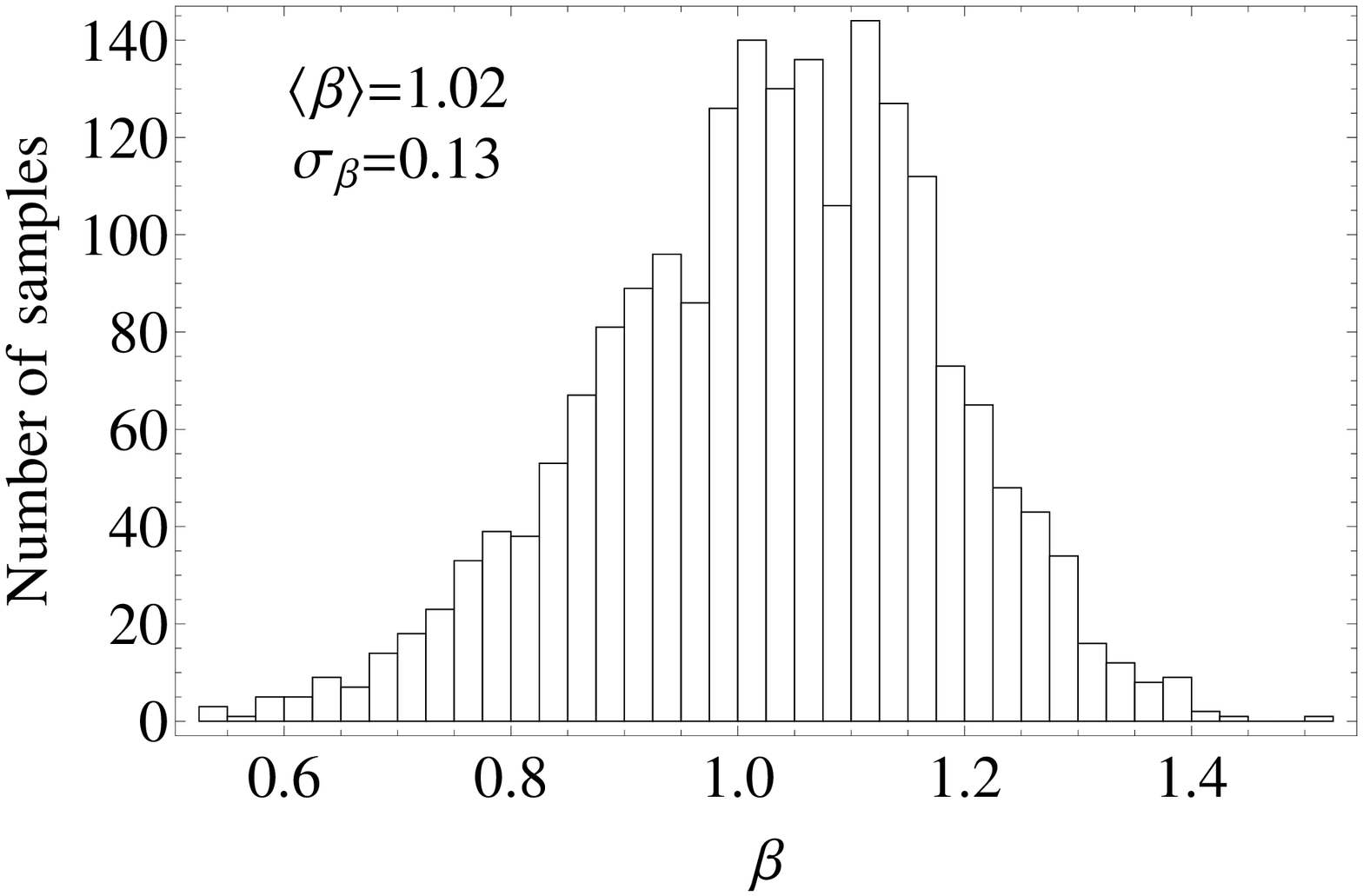}
\hspace*{1.5em}\includegraphics[width=2.6in]{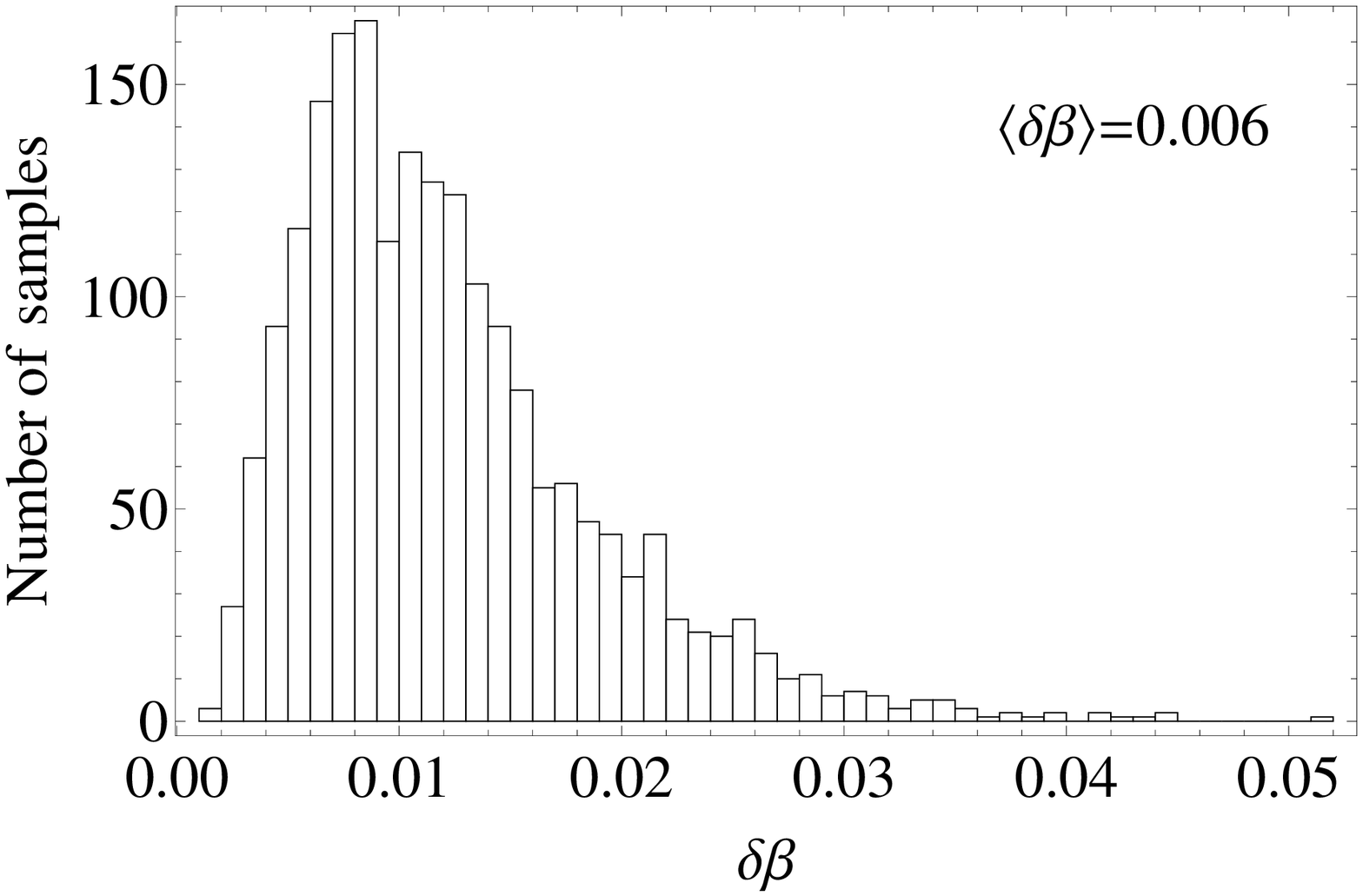}\\
\includegraphics[width=2.6in]{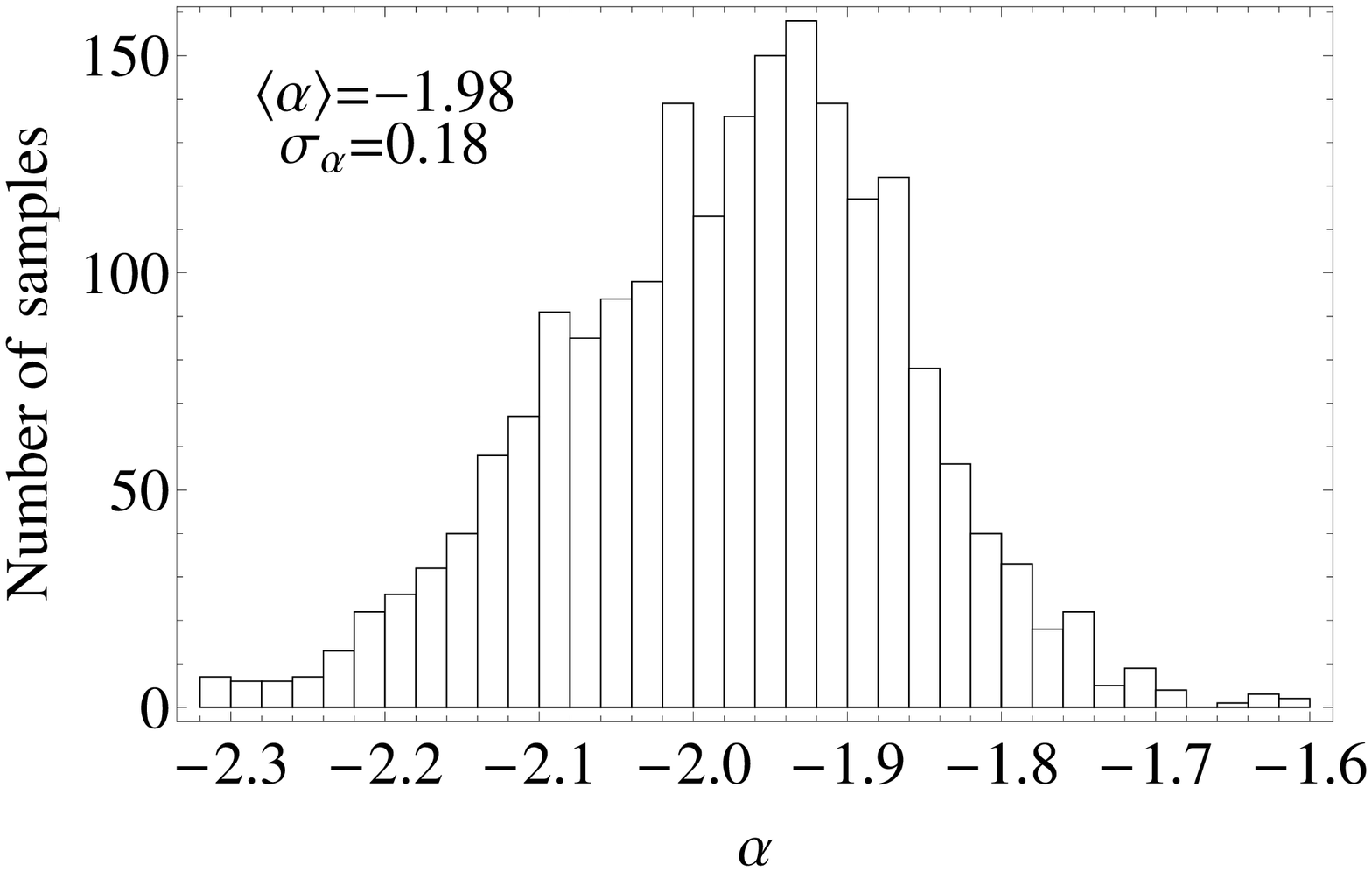}
\hspace*{1.5em}\includegraphics[width=2.6in]{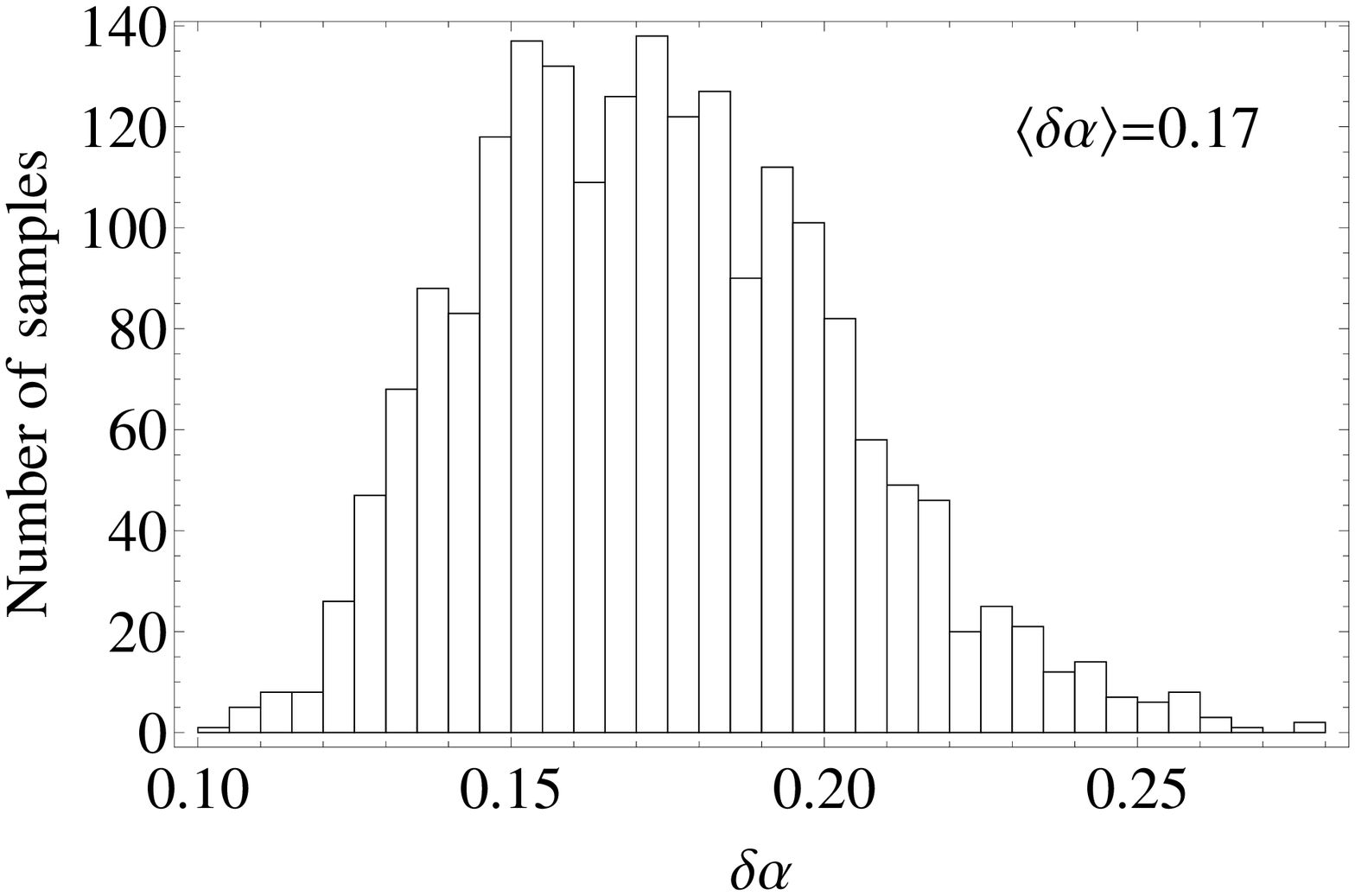}
\caption{The distributions of SF- and PSD-slopes estimated from an ensemble of 2000 artificial light curves having an underlying PSD of a power-law form with slope -2.\newline
[Top panels] (Left) The distribution of the SF slopes $\beta$ has a mean of 1.02 and a standard deviation 0.13 (the histogram bins have a length of 0.025). (Right) The distribution of the errors coming from the fit $\delta\beta$, having a mean of 0.006 (the histogram bins have a length of 0.001).\newline
[Bottom panels] (Left) The distribution of the PSD slopes $\alpha$ has a mean of -1.98 and a standard deviation 0.18 (the histogram bins have a length of 0.02). (Right) The distribution of the errors coming from the fit, $\delta\alpha$, having a mean of $0.17$ (the histogram bins have a length of 0.005).}
\label{fig:sf_psd_fits} 
\end{figure*}

We use the previously generated simulations (2000 light curves, 500 t.u. long from a power-law PSD having an index -2), and for each light curve we calculate the logarithm of the SF-estimates. We attribute to every estimate a standard error based on the error of the sample mean for each time bin, which is one of the most usual methods \citep[e.g.][]{collier01,zhang02a,czerny03}. We then fit, using a least-squares fitting procedure, a simple linear model of the form 
\eqb
\log_{10}[SF(\tau;C,\beta)]=C+\beta\log_{10}(\tau)
\eqe
where $\beta$ represents the slope of the SF. Hence, for each light curve we estimate the value of the SF-slope $\beta$ together with an error $\delta\beta$. The top panels of Fig. \ref{fig:sf_psd_fits} show the distribution of $\beta$ and $\delta\beta$. The errors derived from the fit, $\delta\beta$, are very small in comparison to the actual scatter of the measured $\beta$, depicting clearly the effect of non-independency. From the simulations we expect 68 per cent of the derived slopes to be within the interval $1.02\pm0.13$ but erroneously the $\delta\beta$ distribution has a mean of 0.006, dictating a range for the $\beta$ within $1.020\pm0.006$. The reason for the very small uncertainties in the fitted slopes, $\delta\beta$, is that the errors on the actual SF-estimates are very small themselves due to their non-independent nature. Every SF-estimate follows smoothly the increasing trend, defined from the adjacent points, and thus the errors derived directly from the SF-estimates in this way are always too small.\par 
On the contrary the binned logarithmic periodogram-estimates together with their corresponding errors do provide us with statistically correct information about the behaviour of the slope. For each one of the previous simulated light curves we fit the binned logarithmic periodogram-estimates with a simple linear model and once again we derive the slope $\alpha$ and the error $\delta\alpha$. We can see from the bottom panels of Fig. \ref{fig:sf_psd_fits}, that 68 per cent of the $\alpha$-estimates are expected to fall within $-1.98\pm0.18$, in accordance with the distribution of $\delta\alpha$ (having a mean of 0.17) coming directly from the fits to the binned logarithmic periodograms of the simulated light curves. A meaningful error analysis originating from a fitting procedure, is correct only when its predictive character (i.e.\ in our case for the PSD: 68 per cent of the measurements should be within the range $-1.98\pm0.18$) coincides with the actual fluctuational outcome of the process (i.e.\ $-1.98\pm0.17$). These simulations show us clearly that this is not the case for the SF errors, which are much smaller than the true spread of the SF-estimates for the same variability process.\par
Occasionally authors attempt to take account some of the SF problems. In \citet{agudo06}, for the case of the intra-day variable blazar S5\,0716+71 observed at 86 GHz, the authors use the interpolated version of the SF which takes into account only the errors caused by the interpolation \citep{quirrenbach00}. By repeating this procedure for our simulations we get a mean error in the slope of $0.011$ which is very small compared to the expected 0.13. Another weakness of this method is that several points in the SF have zero uncertainty (i.e.\ the intersection point of the forward and backward interpolations, see figure 4 in \citealp{agudo06} and figures 24--40 in \citealp{quirrenbach00}) therefore they do not really have any statistical meaning.\par
In \citet{fuhrmann08}, again for the case of S5\,0716+71 but observed at radio frequencies, the authors make use of the term \tql variability timescale\tqr which they relate directly to the SF-break. To estimate its uncertainty they use the saturation level $p_0$ and the two fitted parameters of the SF function $\alpha$, $\beta$, $SF(\tau)=a\tau^\beta$, (see equations (B.2) and (B.3) in the appendix of \citealp{fuhrmann08}). We use the simulations of Section \ref{ssec:asca_sf} in order to see how well we can specify the position of these spurious SF-breaks by applying the same methodology, despite the fact that the simulated light curves do not contain any real characteristic time-scale. The distribution of the spurious break time-scales is shown in the left panel of Fig. \ref{fig:sf_hist_break_and_long_dist} (grey area), having a mean value of 0.92 days and a standard deviation of 0.09 days. By applying the method of \citet{fuhrmann08} in the simulations of Section \ref{ssec:asca_sf} we find a mean error for the breaks of 0.006 days which differs significantly from the expected 0.09 days. Obviously with such small (and incorrect) uncertainties every feature in the SF can be considered as a significant time signature of a special source property, as it seen in the right panel of figure 7 in \citet{fuhrmann08} where the authors find two \tql variability time-scales\tqr embedded in their data set.\par
Each of the above cases, show us that timescales derived from the commonly-used methods to interpret and fit the SFs in the blazar literature, should be treated with great caution. Due to the lack of statistical independence, the errors on the fitting parameters tend to be always very small underestimating their true statistical scatter.
\subsection{Resolving physically meaningful time-scales with the SF.}
\label{ssec:phys_meaningful_ts}
True characteristic time-scales embedded in the blazar light curves, should be mapped in the SF as well as in the PSD. In the limit of a single PSD slope in the range -1 to -3, covering a large frequency range and for a long light curve, there is a direct theoretical relation between the PSD and SF slopes (see Appendix \ref{app:sf_psd}). However for the more common broken power-law PSD and with typical astronomical observing constraints, the relationships are less clear and thus, any attempt to relate SF-breaks or -slopes to PSD-breaks or -slopes by one-to-one relationships can lead to misconceptions about the true underlying variability process.\par
\begin{figure} 
\includegraphics[width=3.3in]{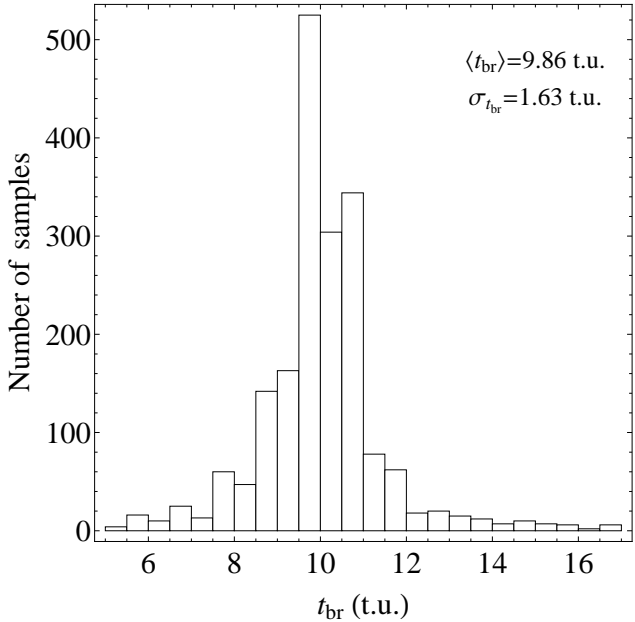}
\caption{The distribution of SF-features (i.e.\ local extremum or inflection point) around 10 t.u., coming from 2000 artificial light curves 500 t.u. long, having a PSD of a broken-power-law shape with a break at $f_{\rm br}=0.1$ (t.u.)$^{-1}$ and with low and high frequency slopes of $\alpha_{\rm l}=-1$ and $\alpha_{\rm h}=-2.5$ respectively. The distribution has a mean value of 9.86 and a standard deviation of 1.63.}
\label{fig:sf_true_breaks_distrib} 
\end{figure}
\begin{figure} 
\includegraphics[width=3.3in]{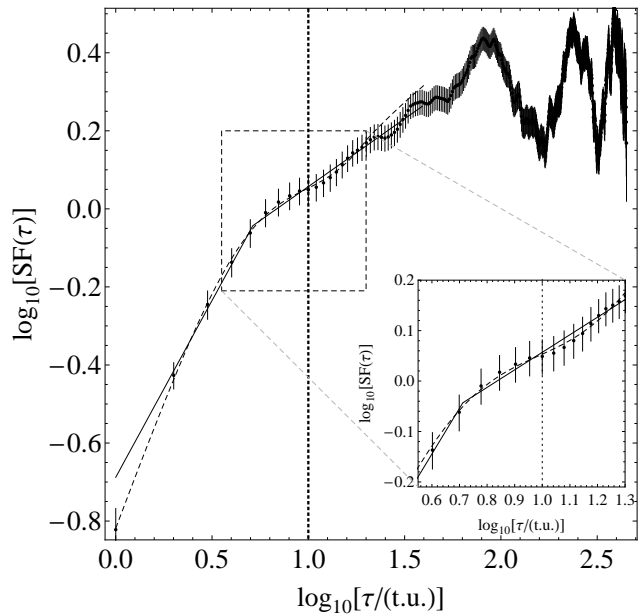}
\caption{The points represent the SF-estimates for a randomly selected light curve, drawn from the ensemble of 2000 data sets 500 t.u. long, having a PSD of a broken-power-law shape with $f_{\rm br}=0.1$ (t.u.)$^{-1}$, $\alpha_{\rm l}=-1$ and $\alpha_{\rm h}=-2.5$ respectively. The thick-dotted vertical line indicates the position of the expected characteristic time-scale of 10 t.u. The specific SF has an inflection point at 10.32 t.u. A physically meaningless break appears at $\tau\sim100$ t.u.\newline
The solid line represents the SF model-fit (equation (\ref{eqe:bkn_logsf_model})) employed to depict $\tau_{\rm br}$ which is constrained between 2--20 t.u., and yields $\tau_{\rm br}=5.15\pm0.43$ t.u. The dashed line represents the SF model-fit (equation (\ref{eqe:db_pow})) employed to depict $\tau_{\rm br}$ which is constrained between 2--20 t.u., yielding $\tau_{\rm br_1}=5.97\pm0.38$ t.u. and $\tau_{\rm br_2}=14.92\pm0.37$ t.u.\newline
The inlay shows the behaviour of the two fitted models around the expected characteristic time-scale of 10.32 t.u.}
\label{fig:sf_from_bknpow} 
\end{figure}

We produce 2000 artificial light curves 500 t.u. long, having a PSD of a broken-power-law shape with a break at $f_{\rm br}=0.1$ (t.u.)$^{-1}$ and with low and high frequency slopes of $\alpha_{\rm l}=-1$ and $\alpha_{\rm h}=-2.5$ respectively. Then, for each light curve we estimate the SF and we produce an interpolated version of it in order to localize any feature around 10 t.u. in the form of either a local extremum or inflection point. We see a wide variety, and sometimes even an absence, of features around 10 t.u., something which places automatically the SF in the category of weak probing methods (Fig. \ref{fig:sf_from_bknpow}). In this particular case we ensure that any feature around 10 t.u. is induced by the existence of the characteristic time-scale since we know {\it a priori} the expected position of it and hence no spurious break is expected at that time-scale. Of course this sort of interesting temporal signature has nothing to do with the previously mentioned (Section \ref{sec:xray_obs}) physically uninteresting SF-break. As we can see from Fig. \ref{fig:sf_from_bknpow} these meaningless breaks, defined by the onset of a plateau on longer time-scales, are also present in our simulated light curves and their positions are simply deduced, as we showed in Section \ref{sec:xray_obs}, from the length of the data set and the underlying PSD of the variability process. The distribution of the physically interesting SF-features, which are expected around 10 t.u., is shown in Fig. \ref{fig:sf_true_breaks_distrib} having a mean value of 9.86 t.u. and a standard deviation of 1.63 t.u. coming from a total of 1789 SFs that exhibit features.\par 
In order to resolve these physically interesting breaks, we employ the commonly used model of a broken-power-law, similar to the form of equation (\ref{eqe:sf_fit_model}), but in its logarithmic version in order to ensure Gaussianity at least for the SF-estimates of the slope (Section \ref{ssect:non_indepen_gauss}). 
\eqb
&&\log_{10}[SF(\tau)]=\nonumber\\
&&\left\{
\begin{array}{c l}
C+\beta_1\log_{10}(\tau) & \tau\leq\tau_{\rm br} \\
C+(\beta_1-\beta_2)\log_{10}(\tau_{\rm br})+\beta_2\log_{10}(\tau) & \tau\geq\tau_{\rm br}
\end{array}
\right.
\label{eqe:bkn_logsf_model}
\eqe
Fitting equation (\ref{eqe:bkn_logsf_model}) to the simulated data typically results in the minimisation routine finding the physically uninteresting SF-breaks, $t_{\rm br}$, occurring on long time-scales ($\approx100$ t.u.) since this is the most prominent break-feature in the SF. Fortunately, and only for the case of the simulations, we know in advance the position of the expected SF feature we constrain our SF fits to the first 50 t.u., corresponding only to the increasing part of the SF, and we estimate the standard errors based on the error of the sample mean as described in Section \ref{ssec:error_underestim}. Moreover we constrain the range of $\tau_{\rm br}$ between 2--20 t.u. in order to force the fitted $\tau_{\rm br}$ parameter to correspond to the true PSD-break (Fig. \ref{fig:sf_from_bknpow}, solid line).\par
The distribution of $\tau_{\rm br}$ from the 2000 simulations, has a mean value of 5.31 t.u. and a standard deviation of 1.46 t.u. The shift of the most probable value from 9.86 t.u. to 5.31 t.u. is clearly a systematic problem of that method which depends on the form of the fitted model (equation (\ref{eqe:bkn_logsf_model})). Nevertheless, the scatter of these breaks, 1.46 t.u., is very close to the expected one, 1.63 t.u. Once again, the errors on $\tau_{\rm br}$ derived directly from the fitting procedure are extremely small, having a mean value of 0.40 t.u. The broken-power-law model is not very representative of the actual shape of the SF which is continuously curved.\par
In order to model the physically interesting SF features we have to perform a much more detailed fitting even in this simple case tested here where we know in advance their approximate position. For this reason we employ a smoothly \tql double-bending\tqr power-law (section 4.2 in \citealp{mchardy04}) of the form
\eqb
SF(\tau)&=&C\tau^{\kappa_1}\left[1+\left(\frac{\tau}{\tau_{\rm br_1}}\right)^{\kappa_2+\kappa_1}\right]^{-1}\times\nonumber\\
&&\left[1+\left(\frac{\tau}{\tau_{\rm br_2}}\right)^{\kappa_3-\kappa_2}\right]
\label{eqe:db_pow}
\eqe
with breaks at $\tau_{\rm br_1}$ and $\tau_{\rm br_2}$ and power-law indices $\kappa_1$ for $\tau<\tau_{\rm br_1}$, $\kappa_2$ for $\tau_{\rm br_1}<\tau<\tau_{\rm br_2}$ and $\kappa_3$ for $\tau>\tau_{\rm br_2}$. Again, we constrain the fitted break time-scales to be between 2--20 t.u (Fig. \ref{fig:sf_from_bknpow}, dashed line).\par
For the 2000 simulated light curves the distribution of the fitted position of the first break, $\tau_{\rm br_1}$, has a mean value of 6.13 t.u. and a standard deviation of 1.32 t.u., and for the second break 15.03 t.u. and 1.45 t.u. respectively. Neither of these breaks depicts the actual position of the characteristic time-scale (10 t.u.). Similarly to the simple broken-power-law fitting model, the errors derived directly from the fit are very small, having a mean value of 0.35 and 0.39 for the two breaks respectively.\par
Despite the fact that breaks in the PSD, representing physically interesting variability time-scales, are mapped to the SF in the form of local maxima or inflection points, it is dangerous to make any statistical statement about them since
\begin{enumerate}
\item it is very difficult to depict the breaks robustly based on a fitting procedure due to the awkward shape of the SF; even if we know their position, based on the PSD, in which case we do not need to use the SF in the first place.
\item the estimated errors in the fitted parameters (derived from a $\Delta\chi^2=1$, under the assumption of Gaussianity) are always very small with respect to the true scatter of the parameters due to the statistical dependence of the SF-estimates; even if we find a well behaved function able to depict the break based on single or multiple fitted parameters .
\end{enumerate}     
In reality, we deal with astronomical data sets for which we want to reveal the possible existence of characteristic time-scales in them. In the case that these time signatures do exist, we do not know in advance their position and how they are mapped on the SF. Moreover, since the SF-estimates are affected by the statistical properties of the observed light curves (i.e.\ mean, variance and the length) and not from the actual statistical properties of the underlying variability process (as is the PSD), even in the absence of characteristic time-scales we expect to see SF-artefact-features similar to the ones induced by real characteristic time-scales.
\section{THE STRUCTURE FUNCTION AND DATA-GAPS}
\label{sect:sf_gaps}
The SF method is considered by several researchers \citep[e.g.][]{takahashi00,collier01,kataoka01,kataoka02,zhang02a} to be the ideal method of studying the time properties of gappy data sets as it is believed to be less distorted by gaps than frequency-domain methods. Here, we use simulations to determine whether this confidence in the robustness to data gaps of the SF method is really justified.\par
\begin{figure*} 
\parbox{0.49\linewidth}{\includegraphics[width=3.1in]{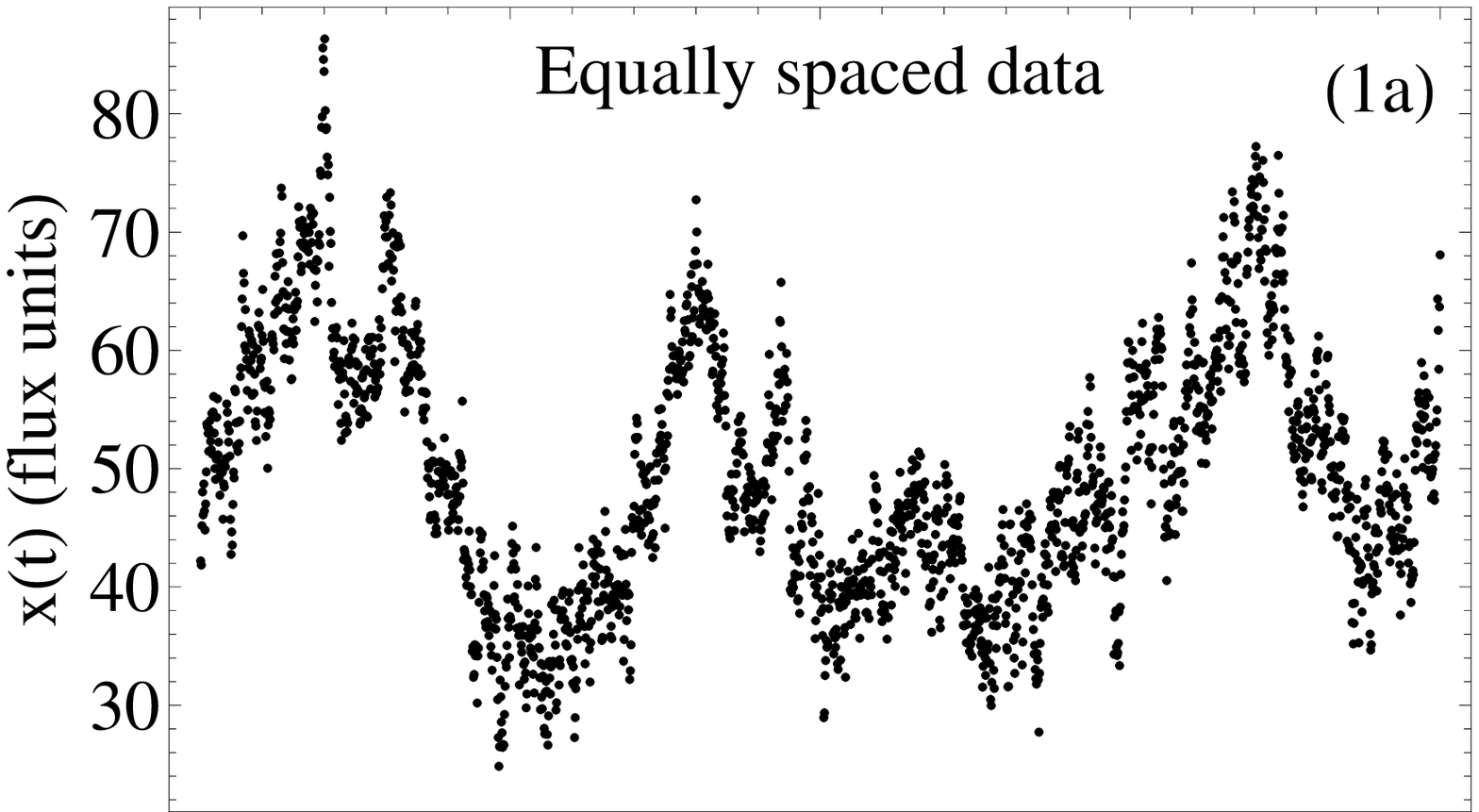}\\[-0.2em]
\includegraphics[width=3.1in]{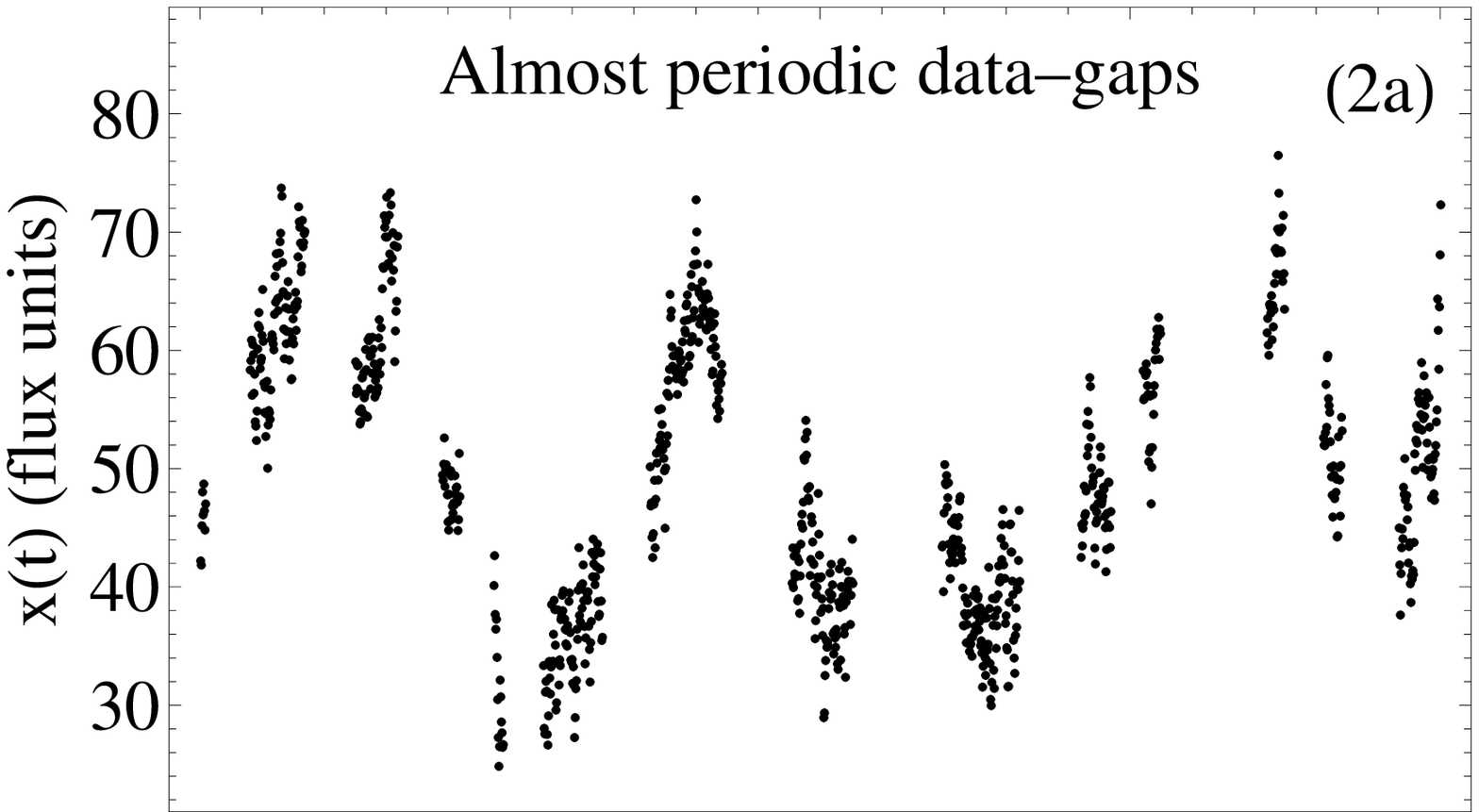}\\[-0.19em]
\includegraphics[width=3.1in]{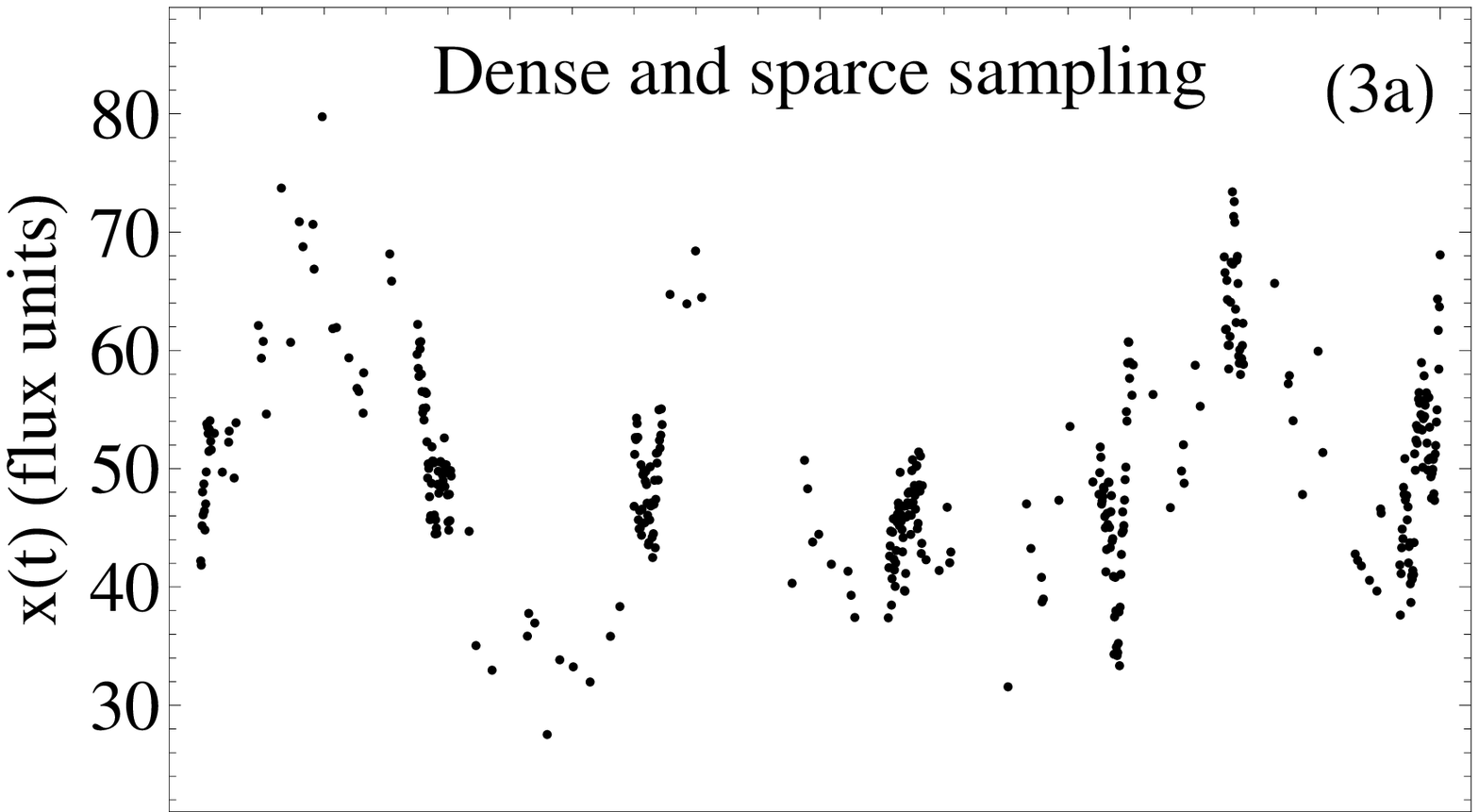}\\[-0.15em]
\hspace*{-0.1em}\includegraphics[width=3.2in]{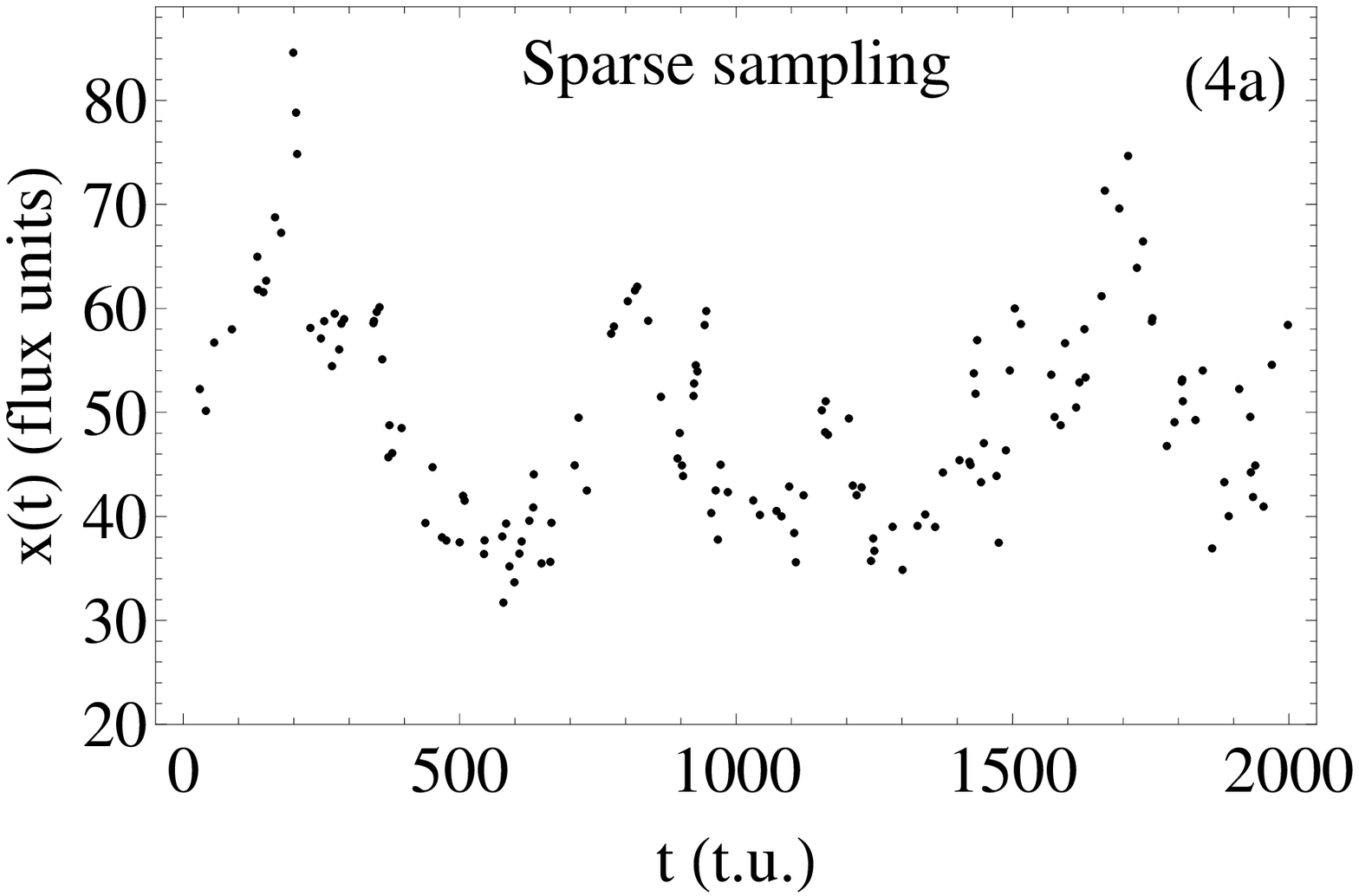}}
\parbox{0.49\linewidth}{\vspace{0.2em}\includegraphics[width=3.15in]{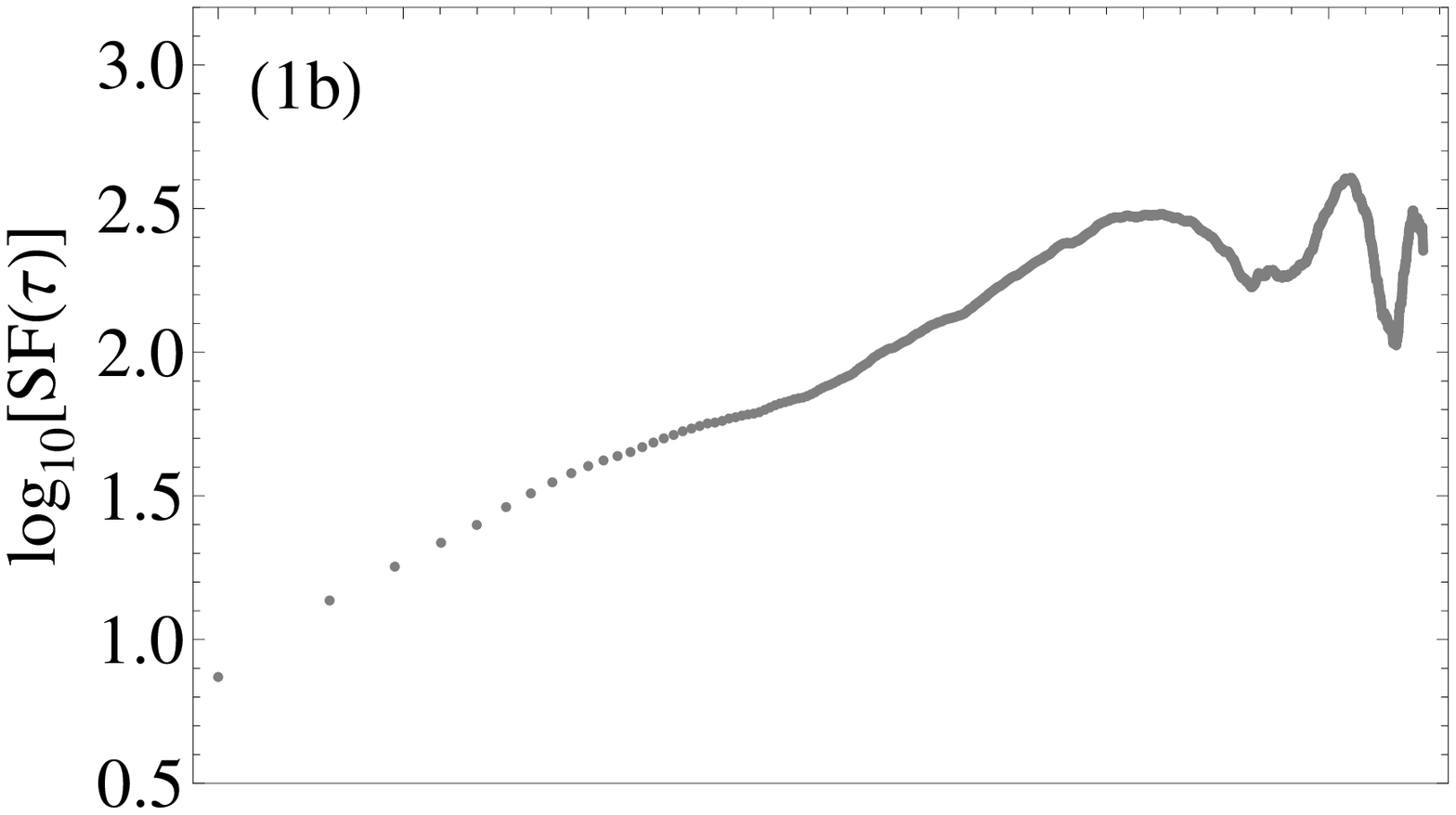}\\[-0.87em]
\includegraphics[width=3.15in]{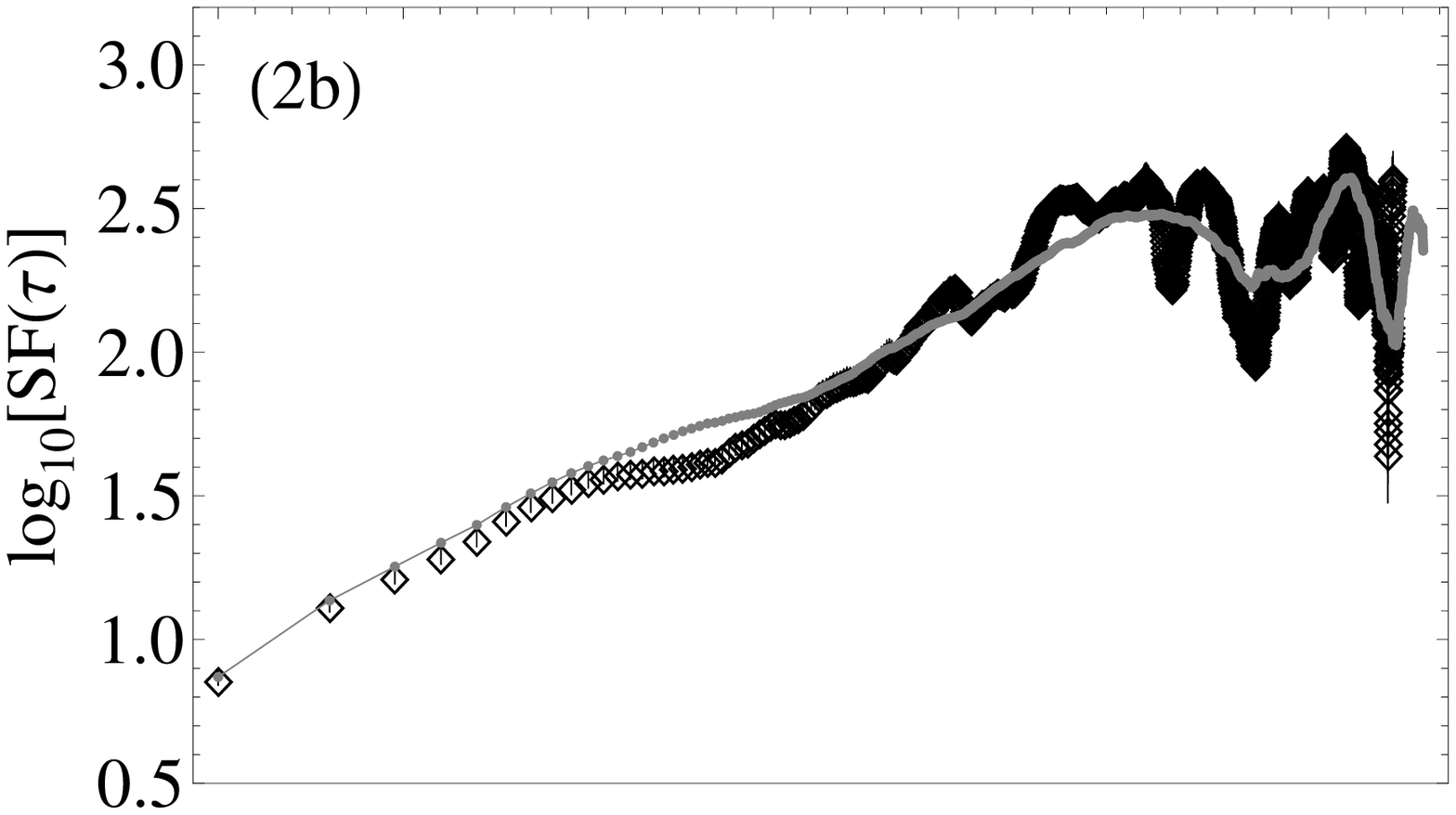}\\[-0.82em]
\includegraphics[width=3.15in]{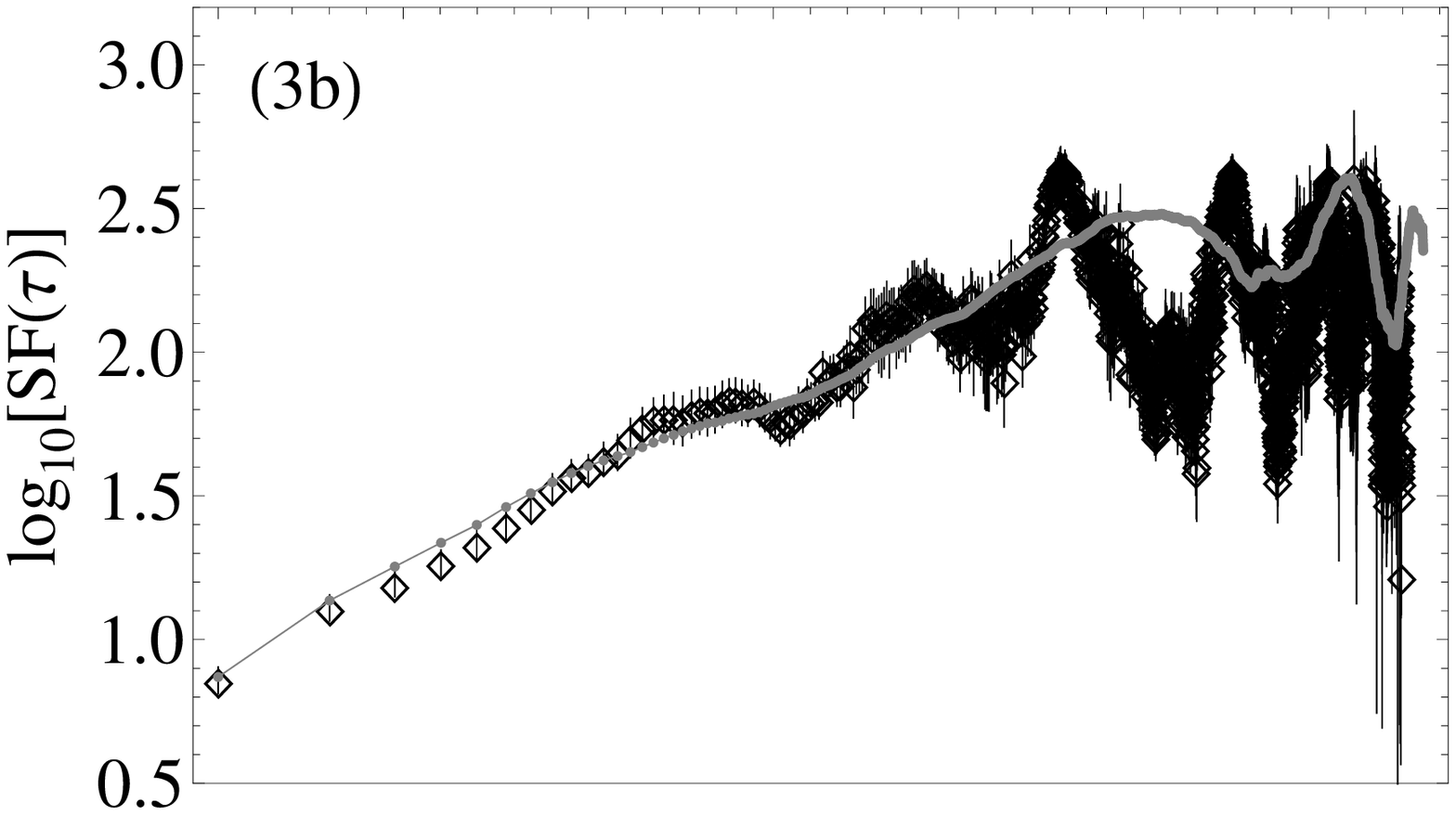}\\[-0.85em]
\hspace{0.2em}\includegraphics[width=3.15in]{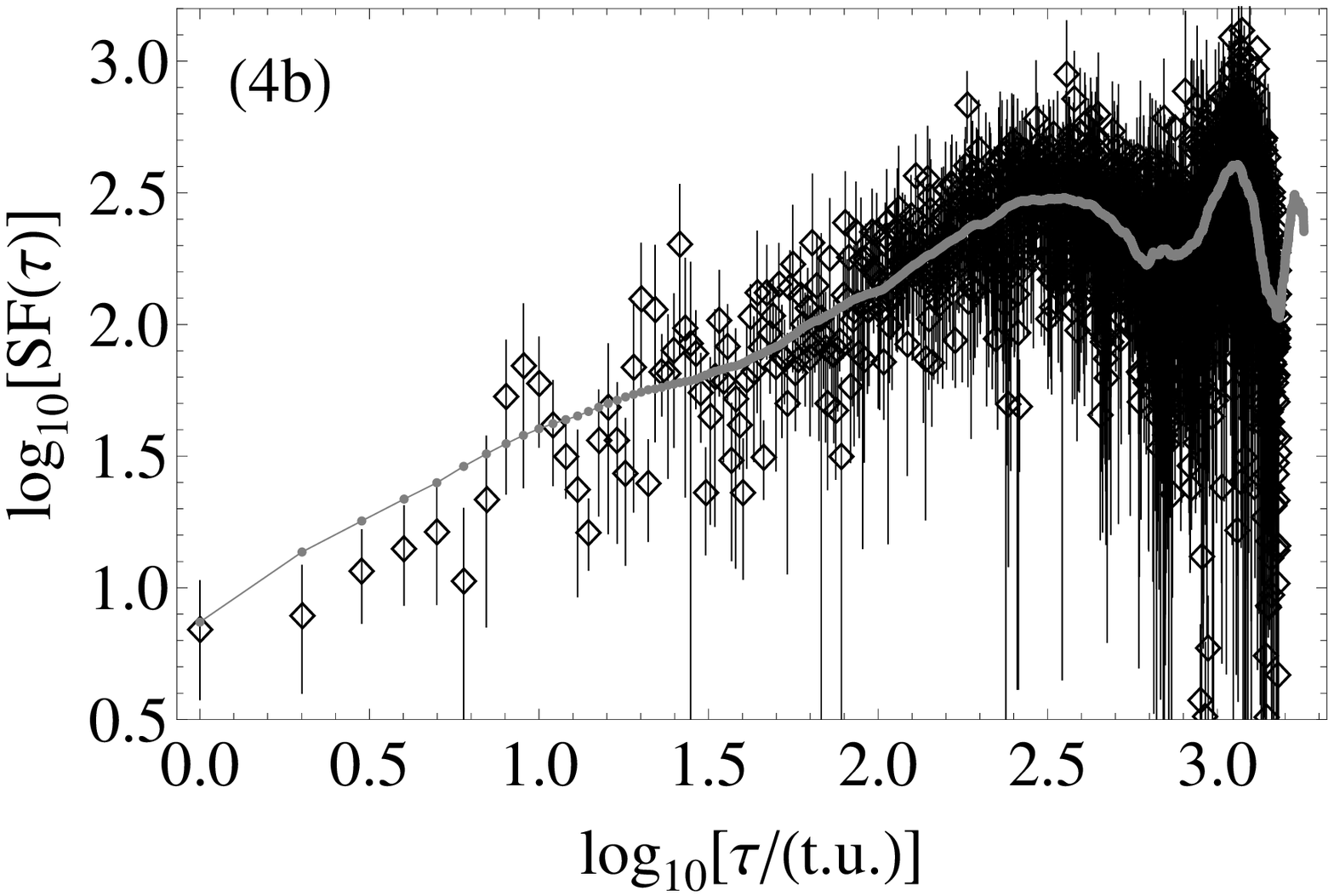}}
\caption{The effects of data-gaps in the SF of a simulated light curve (panel 1a) 2000 t.u. long, having a PSD of a power-law with index -1.5. The left panels show the gappy patterns. The right panels show the corresponding gappy SF-estimates (black points) together with the continuous SF-estimates (grey points connected with grey line), as they are shown in panel 1b.}
\label{fig:gappy_SF} 
\end{figure*}

Initially we produce a single artificial light curve 2000 t.u. long from a featureless PSD with a power-law shape with index -1.5 and we compute its SF (Fig. \ref{fig:gappy_SF}, panels 1a and 1b, respectively). We again note the spurious break at 315 t.u. (as expected from Fig. \ref{fig:sfbr_psd_length}) which does not correspond to any real time property of the underlying variability process.\par
We then add to the simulated light curve three different types of data-gaps and we estimate the SF (Fig. \ref{fig:gappy_SF}, left panels). The first sampling scheme represents the case of {\it almost periodic data-gaps} where 57 per cent of the data are missing in almost periodic epochs, separated by 50--100 time units (Fig. \ref{fig:gappy_SF}, panels 1b). The second sampling scheme consists of {\it dense and sparse sampling} where data are equally spaced for small time periods, and more sparsely sampled for the remaining ones, yielding in total a data set having 83 per cent less data (Fig. \ref{fig:gappy_SF}, panels 1c). The last case that we consider is that of purely sparsely sampled data where 92 per cent of the data are missing (Fig. \ref{fig:gappy_SF}, panels 1d).\par
The effects of the abovementioned data-gaps on the SF are shown in the right panels of Fig. \ref{fig:gappy_SF}. To take into account the uncertainties that are introduced by these gaps for the ensemble of our simulated light curves, we use the {\it bootstrap method}, discussed by \citet{czerny03}. From each gappy light curve we select randomly, allowing repetition, 1000 sub-samples of 2000 t.u. long, and we drop the multiple entries. Then we estimate the SF and from the ensemble of 1000 SFs we calculate for every time bin $\tau$ a standard deviation, including additionally the standard error in the sample mean coming from the gappy light curve.\par
We should emphasise that here we study the differences between the SFs coming from the gappy data sets (empty diamonds, right panels of Fig. \ref{fig:gappy_SF}) and those coming from the uninterrupted data set (grey points connected with the grey line, right panels of Fig. \ref{fig:gappy_SF} which are also displayed in panel (1b) of Fig. \ref{fig:gappy_SF}). That does not mean that the uninterrupted SF is necessarily the \tql true\textquotedblright, \tql original\textquotedblright, or \tql parent\tqr SF of the underlying variability process. This is only one SF, chosen randomly from an ensemble of light curves having a featureless PSD of a power-law form with index -1.5. Our intention is to check whether the observed SF is affected by the data-gaps since this is the method that several authors \citep[e.g.][]{takahashi00,collier01,kataoka01,kataoka02,zhang02a} use to derive physically interesting results. A direct visual comparison of the SFs (Fig. \ref{fig:gappy_SF}) reveals that actually gaps do affect the shape of the SF since they introduce wiggles and bends. The latter can create the wrong impression that the underlying variability process is described by an underlying PSD of a broken-power-law form (as in Section \ref{ssec:phys_meaningful_ts}).\par
\begin{figure} 
\includegraphics[width=3.3in]{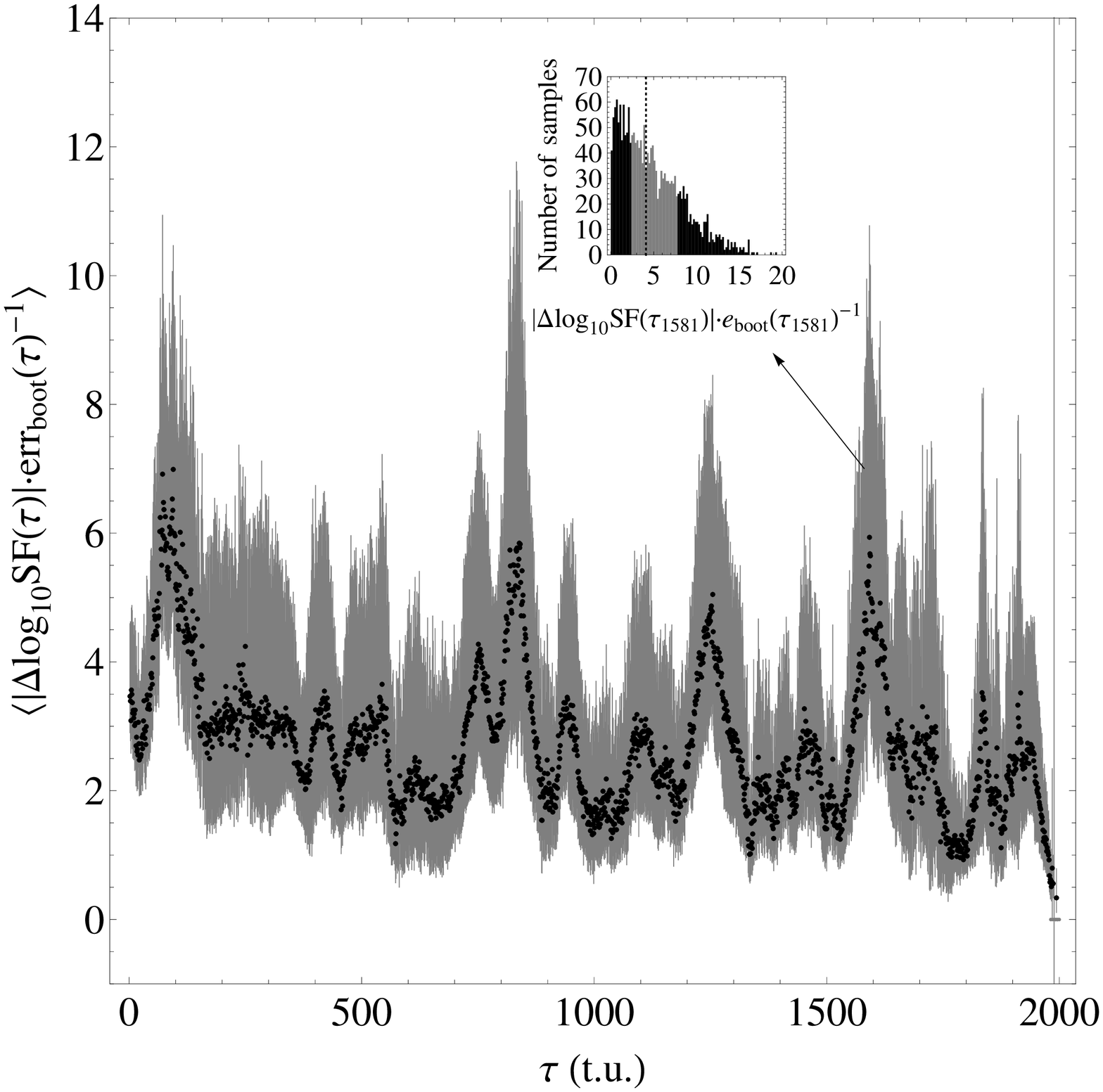}
\caption{The mean difference between gappy and continuous SFs coming from an ensemble of 2000 artificial light curves, having the same underlying PSD of a power-law form with index -1.5. The gappy sampling scheme, applied to the continuous light curves, is shown in panel (3a) of Fig. \ref{fig:gappy_SF}. The errors $err_{\rm boot}(\tau)$ are estimated for each light curve based on the bootstrap method (described in the text). The inset depicts the distribution of the points in the $\tau=1581$ time-bin having a mean value of 4.16 and the standard deviation (i.e\ 34.1\% of the entries) above and below the mean is shown in the grey region.}
\label{fig:sf_gaps_deviat} 
\end{figure} 
At this point it is interesting to check whether or not the SF-errors estimated from the bootstrap method depict correctly the deviations between the SFs of continuously sampled and gappy light curves, $SF_{\rm conti}$ and $SF_{\rm gappy}$ respectively. We produce 2000 artificial light curves 2000 t.u. long, having a power-law PSD of slope -1.5, and we estimate for each one of them the $SF_{\rm conti}$. Then, we apply to the light curves the dense and sparse sampling (example shown in Fig.\ref{fig:gappy_SF}, panel 3a) and we estimate for each one of them the $SF_{\rm gappy}$ as well as an error for each time bin, $err_{{\rm boot,}i}(\tau)$ (i.e.\ the standard deviation of the gappy-SF of the $i^{\rm th}$ simulated light curve at the time bin $\tau$, $SF_{\rm gappy}^i(\tau)$) as it is derived from the abovementioned bootstrap method. For each SF-bin we derive the following quantity
\eqb
\lefteqn{\left<\frac{|\Delta\log_{10}[SF(\tau)]|}{err_{{\rm boot,}i}(\tau)}\right>=}\nonumber\\
&&\frac{1}{2000}\sum_{i=1}^{2000}\frac{|\log_{10}[SF^i_{\rm gappy}(\tau)]-\log_{10}[SF^i_{\rm conti}(\tau)]|}{err_{{\rm boot,}i}(\tau)}
\eqe
\par
As we can see from Fig.\ref{fig:sf_gaps_deviat} the values of $\left<|\Delta\log_{10}[SF(\tau)]|err_{{\rm boot,}i}^{-1}\right>$ differ significantly from unity, having a mean value of $2.79\pm0.91$. That means that the bootstrap method does not yield statistically meaningful errors that reflect the true deviations between the continuous and the gappy SFs. In other words, data gaps introduce {\it systematic deviations} in the structure function which depend on the light curve realization and cannot be accounted for without extensive simulations.
\section{CONCLUSIONS}
We have performed an extensive series of simulations in order to test the properties of the classical \tql running variance\tqr SF-method. Under this fully controlled environment our studies have focused on the apparent SF-breaks, the correspondence of these breaks to the time-scales involved in the shot-noise model, the statistical behaviour of the SF with respect to the commonly used fitting procedures and its response to data-gaps. Our results can be summarized clearly as follows:
\begin{itemize}
\item Strong SF-breaks frequently occur in data sets lacking any sort of characteristic time-scales. The position of these physically uninteresting breaks depends on the length of the observations and the shape of the underlying PSD.
\item SFs derived from blazar observations resemble those coming from the shot-noise model i.e.\ superposition of triangular shots. However in the frequency domain the same model does not show the same behaviour as the observations and thus is not physically realistic.
\item Non-independence and non-Gaussianity mean that it is not possible to derive a meaningful goodness-of-fit from normal fitting procedures. We see, for example, that the derived uncertainties on the SF quantities i.e.\ positions of breaks and slopes, are always much less than the actual scatter of these variables during multiple realizations of the same variability process.
\item Data-gaps affect severely the SF-estimates in an unpredictable way, introducing systematic deviations. The bootstrap method can not yield statistically meaningful errors depicting the deviations between the gappy and the continuous SFs.
\end{itemize}
We finally comment that for blazar variability present implementations merely determine the shape of the SF for the one realisation under consideration. That one realisation is, of course, just one of many possible realisations and so does not define precisely the properties of the true underlying variability process, which is really what we want to know. Current PSD analysis methods \citep[e.g.][]{uttley02} generally do, however, try to determine the shape of the underlying PSD by means of simulation-based modelling that take into account aliasing and other sampling effects. We encourage the blazar community to explore the use of these techniques or develop similar methodologies in the SF-field.

\section*{Acknowledgments}

DE and IMM acknowledge the Science and Technology Facilities Council (STFC) for support under grant ST/G003084/1. PU is supported by an STFC Advanced Fellowship. This research has made use of NASA's Astrophysics Data System Bibliographic Services.


\begin{thebibliography}{70}
\expandafter\ifx\csname natexlab\endcsname\relax\def\natexlab#1{#1}\fi

\bibitem[{{Agudo} {et~al.}(2006){Agudo}, {Krichbaum}, {Ungerechts}, {Kraus},
  {Witzel}, {Angelakis}, {Fuhrmann}, {Bach}, {Britzen}, {Zensus}, {Wagner},
  {Ostorero}, {Ferrero}, {Gracia} \& {Grewing}}]{agudo06}
{Agudo} I., {et~al.}, 2006, \aap, 456, 117

\bibitem[{{Aharonian} {et~al.}(2003){Aharonian}, {Akhperjanian}, {Beilicke},
  {Bernl{\"o}hr}, {B{\"o}rst}, {Bojahr}, {Bolz}, {Coarasa}, {Contreras},
  {Cortina}, {Denninghoff}, {Fonseca}, {Girma}, {Goebel}, {G{\"o}tting},
  {Heinzelmann}, {Hermann}, {Heusler}, {Hofmann}, {Horns}, {Jung}, {Kankanyan},
  {Kestel}, {Kettler}, {Kohnle}, {Konopelko}, {Kranich}, {Krawczynski},
  {Lampeitl}, {L{\'o}pez}, {Lorenz}, {Lucarelli}, {Mang}, {Meyer}, {Mirzoyan},
  {Moralejo}, {O{\~n}a-Wilhelmi}, {Paneque}, {Panter}, {Plyasheshnikov},
  {P{\"u}hlhofer}, {de los Reyes}, {Rhode}, {Ripken}, {Rowell}, {Sahakian},
  {Samorski}, {Schilling}, {Schweizer}, {Sevilla}, {Siems}, {Sobczy{\'n}ska},
  {Stamm}, {Tluczykont}, {Tonello}, {Vitale}, {V{\"o}lk}, {Wagner}, {Wiedner}
  \& {Wittek}}]{aharonian03}
{Aharonian} F., {et~al.}, 2003, \aap, 410, 813

\bibitem[{{Aharonian} {et~al.}(2005{\natexlab{a}}){Aharonian}, {Akhperjanian},
  {Aye}, {Bazer-Bachi}, {Beilicke}, {Benbow}, {Berge}, {Berghaus},
  {Bernl{\"o}hr}, {Boisson}, {Bolz}, {Braun}, {Breitling}, {Brown}, {Bussons
  Gordo}, {Chadwick}, {Chounet}, {Cornils}, {Costamante}, {Degrange},
  {Djannati-Ata{\"i}}, {O'C.~Drury}, {Dubus}, {Emmanoulopoulos}, {Espigat},
  {Feinstein}, {Fleury}, {Fontaine}, {Fuchs}, {Funk}, {Gallant}, {Giebels},
  {Gillessen}, {Glicenstein}, {Goret}, {Hadjichristidis}, {Hauser},
  {Heinzelmann}, {Henri}, {Hermann}, {Hinton}, {Hofmann}, {Holleran}, {Horns},
  {de Jager}, {Kh{\'e}lifi}, {Komin}, {Konopelko}, {Latham}, {Le Gallou},
  {Lemi{\`e}re}, {Lemoine}, {Leroy}, {Lohse}, {Marcowith}, {Masterson},
  {McComb}, {de Naurois}, {Nolan}, {Noutsos}, {Orford}, {Osborne}, {Ouchrif},
  {Panter}, {Pelletier}, {Pita}, {P{\"u}hlhofer}, {Punch}, {Raubenheimer},
  {Raue}, {Raux}, {Rayner}, {Redondo}, {Reimer}, {Reimer}, {Ripken}, {Rob},
  {Rolland}, {Rowell}, {Sahakian}, {Saug{\'e}}, {Schlenker}, {Schlickeiser},
  {Schuster}, {Schwanke}, {Siewert}, {Sol}, {Steenkamp}, {Stegmann},
  {Tavernet}, {Terrier}, {Th{\'e}oret}, {Tluczykont}, {Vasileiadis}, {Venter},
  {Vincent}, {V{\"o}lk} \& {Wagner}}]{aharonian05_mrk421}
---, 2005{\natexlab{a}}, \aap, 437, 95

\bibitem[{{Aharonian} {et~al.}(2005{\natexlab{b}}){Aharonian}, {Akhperjanian},
  {Bazer-Bachi}, {Beilicke}, {Benbow}, {Berge}, {Bernl{\"o}hr}, {Boisson},
  {Bolz}, {Borrel}, {Braun}, {Breitling}, {Brown}, {Chadwick}, {Chounet},
  {Cornils}, {Costamante}, {Degrange}, {Dickinson}, {Djannati-Ata{\"i}},
  {O'C.~Drury}, {Dubus}, {Emmanoulopoulos}, {Espigat}, {Feinstein}, {Fontaine},
  {Fuchs}, {Funk}, {Gallant}, {Giebels}, {Gillessen}, {Glicenstein}, {Goret},
  {Hadjichristidis}, {Hauser}, {Heinzelmann}, {Henri}, {Hermann}, {Hinton},
  {Hofmann}, {Holleran}, {Horns}, {Jacholkowska}, {de Jager}, {Kh{\'e}lifi},
  {Komin}, {Konopelko}, {Latham}, {Le Gallou}, {Lemi{\`e}re},
  {Lemoine-Goumard}, {Leroy}, {Lohse}, {Martin}, {Martineau-Huynh},
  {Marcowith}, {Masterson}, {McComb}, {de Naurois}, {Nolan}, {Noutsos},
  {Orford}, {Osborne}, {Ouchrif}, {Panter}, {Pelletier}, {Pita},
  {P{\"u}hlhofer}, {Punch}, {Raubenheimer}, {Raue}, {Raux}, {Rayner}, {Reimer},
  {Reimer}, {Ripken}, {Rob}, {Rolland}, {Rowell}, {Sahakian}, {Saug{\'e}},
  {Schlenker}, {Schlickeiser}, {Schuster}, {Schwanke}, {Siewert}, {Sol},
  {Spangler}, {Steenkamp}, {Stegmann}, {Tavernet}, {Terrier}, {Th{\'e}oret},
  {Tluczykont}, {Vasileiadis}, {Venter}, {Vincent}, {V{\"o}lk} \&
  {Wagner}}]{aharonian_05B_pks2155}
---, 2005{\natexlab{b}}, \aap, 442, 895

\bibitem[{{Aller} {et~al.}(1999){Aller}, {Aller}, {Hughes} \&
  {Latimer}}]{aller99}
{Aller} M.~F., {Aller} H.~D., {Hughes} P.~A., {Latimer} G.~E., 1999, \apj, 512,
  601

\bibitem[{{Bendat} \& {Piersol}(1986)}]{bendat86}
{Bendat} J.~S., {Piersol} A.~G., 1986, {Random Data: Analysis and Measurement
  Procedures}. New York: John Wiley \& Sons, |c1986, 2nd ed.

\bibitem[{{Bevington} \& {Robinson}(1992)}]{bevington92}
{Bevington} P.~R., {Robinson} D.~K., 1992, {Data reduction and error analysis
  for the physical sciences}. New York: McGraw-Hill, |c1992, 2nd ed.

\bibitem[{{Blandford} \& {Kundic}(1997)}]{blandford97}
{Blandford} R.~D., {Kundic} T., 1997, in The Extragalactic Distance Scale,
  {Livio} M., {Donahue} M., {Panagia} N., eds., pp. 60--75

\bibitem[{{Bregman} {et~al.}(1988){Bregman}, {Glassgold}, {Huggins}, {Kinney},
  {McHardy}, {Webb}, {Pollock}, {Leacock}, {Smith}, {Pica}, {Aller}, {Aller},
  {Hodge}, {Miller}, {Stephens}, {Dent}, {Balonek}, {Barvainis}, {Neugebauer},
  {Impey}, {Soifer}, {Matthews}, {Elias} \& {Wisniewski}}]{bregman88}
{Bregman} J.~N., {et~al.}, 1988, \apj, 331, 746

\bibitem[{{Brinkmann} {et~al.}(2000){Brinkmann}, {Gliozzi}, {Urry}, {Maraschi}
  \& {Sambruna}}]{brinkmann00}
{Brinkmann} W., {Gliozzi} M., {Urry} C.~M., {Maraschi} L., {Sambruna} R., 2000,
  \aap, 362, 105

\bibitem[{{Brinkmann} {et~al.}(2001){Brinkmann}, {Sembay}, {Griffiths},
  {Branduardi-Raymont}, {Gliozzi}, {Boller}, {Tiengo}, {Molendi} \&
  {Zane}}]{brinkmann01}
{Brinkmann} W., {et~al.}, 2001, \aap, 365, L162

\bibitem[{{Burderi} {et~al.}(1997){Burderi}, {Robba}, {La Barbera} \&
  {Guainazzi}}]{burderi97}
{Burderi} L., {Robba} N.~R., {La Barbera} N., {Guainazzi} M., 1997, \apj, 481,
  943

\bibitem[{{Catanese} {et~al.}(1997){Catanese}, {Bradbury}, {Breslin},
  {Buckley}, {Carter-Lewis}, {Cawley}, {Dermer}, {Fegan}, {Finley}, {Gaidos},
  {Hillas}, {Johnson}, {Krennrich}, {Lamb}, {Lessard}, {Macomb}, {McEnery},
  {Moriarty}, {Quinn}, {Rodgers}, {Rose}, {Samuelson}, {Sembroski},
  {Srinivasan}, {Weekes} \& {Zweerink}}]{catanese97}
{Catanese} M., {et~al.}, 1997, \apjl, 487, L143+

\bibitem[{{Coles} \& {Frehlich}(1982)}]{coles82}
{Coles} W.~A., {Frehlich} R.~G., 1982, Journal of the Optical Society of
  America (1917-1983), 72, 1042

\bibitem[{{Collier} \& {Peterson}(2001)}]{collier01}
{Collier} S., {Peterson} B.~M., 2001, \apj, 555, 775

\bibitem[{{Cordes} \& {Downs}(1985)}]{cordes85}
{Cordes} J.~M., {Downs} G.~S., 1985, \apjs, 59, 343

\bibitem[{{Czerny} {et~al.}(2003){Czerny}, {Doroshenko}, {Niko{\l}ajuk},
  {Schwarzenberg-Czerny}, {Loska} \& {Madejski}}]{czerny03}
{Czerny} B., {Doroshenko} V.~T., {Niko{\l}ajuk} M., {Schwarzenberg-Czerny} A.,
  {Loska} Z., {Madejski} G., 2003, \mnras, 342, 1222

\bibitem[{{Done} {et~al.}(1992){Done}, {Madejski}, {Mushotzky}, {Turner},
  {Koyama} \& {Kunieda}}]{done92}
{Done} C., {Madejski} G.~M., {Mushotzky} R.~F., {Turner} T.~J., {Koyama} K.,
  {Kunieda} H., 1992, \apj, 400, 138

\bibitem[{{Fiedler} {et~al.}(1987){Fiedler}, {Waltman}, {Spencer}, {Johnston},
  {Angerhofer}, {Florkowski}, {Josties}, {Klepczynski}, {McCarthy} \&
  {Matsakis}}]{fiedler87}
{Fiedler} R.~L., {et~al.}, 1987, \apjs, 65, 319

\bibitem[{{Fuhrmann} {et~al.}(2008){Fuhrmann}, {Krichbaum}, {Witzel}, {Kraus},
  {Britzen}, {Bernhart}, {Impellizzeri}, {Agudo}, {Klare}, {Sohn}, {Angelakis},
  {Bach}, {Gab{\'a}nyi}, {K{\"o}rding}, {Pagels}, {Zensus}, {Wagner},
  {Ostorero}, {Ungerechts}, {Grewing}, {Tornikoski}, {Apponi},
  {Vila-Vilar{\'o}}, {Ziurys} \& {Strom}}]{fuhrmann08}
{Fuhrmann} L., {et~al.}, 2008, \aap, 490, 1019

\bibitem[{{Gliozzi} {et~al.}(2001){Gliozzi}, {Brinkmann}, {O'Brien}, {Reeves},
  {Pounds}, {Trifoglio} \& {Gianotti}}]{gliozzi01}
{Gliozzi} M., {Brinkmann} W., {O'Brien} P.~T., {Reeves} J.~N., {Pounds} K.~A.,
  {Trifoglio} M., {Gianotti} F., 2001, \aap, 365, L128

\bibitem[{{Heeschen} {et~al.}(1987){Heeschen}, {Krichbaum}, {Schalinski} \&
  {Witzel}}]{heeschen87}
{Heeschen} D.~S., {Krichbaum} T., {Schalinski} C.~J., {Witzel} A., 1987, \aj,
  94, 1493

\bibitem[{{Heidt} \& {Wagner}(1996)}]{heidt96}
{Heidt} J., {Wagner} S.~J., 1996, \aap, 305, 42

\bibitem[{{Hjellming} \& {Narayan}(1986)}]{hjellming86}
{Hjellming} R.~M., {Narayan} R., 1986, \apj, 310, 768

\bibitem[{{Hufnagel} \& {Bregman}(1992)}]{hufnagel92}
{Hufnagel} B.~R., {Bregman} J.~N., 1992, \apj, 386, 473

\bibitem[{{Hughes} {et~al.}(1992){Hughes}, {Aller} \& {Aller}}]{hughes92}
{Hughes} P.~A., {Aller} H.~D., {Aller} M.~F., 1992, \apj, 396, 469

\bibitem[{{Iyomoto} \& {Makishima}(2001)}]{iyomoto01}
{Iyomoto} N., {Makishima} K., 2001, \mnras, 321, 767

\bibitem[{{Kataoka} {et~al.}(2001){Kataoka}, {Takahashi}, {Wagner}, {Iyomoto},
  {Edwards}, {Hayashida}, {Inoue}, {Madejski}, {Takahara}, {Tanihata} \&
  {Kawai}}]{kataoka01}
{Kataoka} J., {et~al.}, 2001, \apj, 560, 659

\bibitem[{{Kataoka} {et~al.}(2002){Kataoka}, {Tanihata}, {Kawai}, {Takahara},
  {Takahashi}, {Edwards} \& {Makino}}]{kataoka02}
{Kataoka} J., {Tanihata} C., {Kawai} N., {Takahara} F., {Takahashi} T.,
  {Edwards} P.~G., {Makino} F., 2002, \mnras, 336, 932

\bibitem[{{Kirk} {et~al.}(1998){Kirk}, {Rieger} \& {Mastichiadis}}]{kirk98}
{Kirk} J.~G., {Rieger} F.~M., {Mastichiadis} A., 1998, \aap, 333, 452

\bibitem[{{Kolmogorov}(1941a)}]{kolmogorov41a}
{Kolmogorov} A.~N., 1941a, Doklady Akad. Nauk CCCP, 30

\bibitem[{{Kolmogorov}(1941b)}]{kolmogorov41b}
---, 1941b, Doklady Akad. Nauk CCCP, 32

\bibitem[{{Lainela} \& {Valtaoja}(1993)}]{lainela93}
{Lainela} M., {Valtaoja} E., 1993, \apj, 416, 485

\bibitem[{{Lewis} \& {Irwin}(1996)}]{lewis96}
{Lewis} G.~F., {Irwin} M.~J., 1996, \mnras, 283, 225

\bibitem[{{Lochner} {et~al.}(1991){Lochner}, {Swank} \&
  {Szymkowiak}}]{lochner91}
{Lochner} J.~C., {Swank} J.~H., {Szymkowiak} A.~E., 1991, \apj, 376, 295

\bibitem[{{Maraschi} {et~al.}(1999){Maraschi}, {Fossati}, {Tavecchio},
  {Chiappetti}, {Celotti}, {Ghisellini}, {Grandi}, {Pian}, {Tagliaferri},
  {Treves}, {Breslin}, {Buckley}, {Carter-Lewis}, {Catanese}, {Cawley},
  {Fegan}, {Fegan}, {Finley}, {Gaidos}, {Hall}, {Hillas}, {Krennrich},
  {Lessard}, {Masterson}, {Moriarty}, {Quinn}, {Rose}, {Samuelson}, {Weekes},
  {Urry} \& {Takahashi}}]{maraschi99}
{Maraschi} L., {et~al.}, 1999, \apjl, 526, L81

\bibitem[{{Markowitz} {et~al.}(2003){Markowitz}, {Edelson}, {Vaughan},
  {Uttley}, {George}, {Griffiths}, {Kaspi}, {Lawrence}, {McHardy}, {Nandra},
  {Pounds}, {Reeves}, {Schurch} \& {Warwick}}]{markowitz03}
{Markowitz} A., {et~al.}, 2003, \apj, 593, 96

\bibitem[{{McHardy} {et~al.}(2007){McHardy}, {Ar{\'e}valo}, {Uttley},
  {Papadakis}, {Summons}, {Brinkmann} \& {Page}}]{mchardy07}
{McHardy} I.~M., {Ar{\'e}valo} P., {Uttley} P., {Papadakis} I.~E., {Summons}
  D.~P., {Brinkmann} W., {Page} M.~J., 2007, \mnras, 382, 985

\bibitem[{{McHardy} {et~al.}(2005){McHardy}, {Gunn}, {Uttley} \&
  {Goad}}]{mchardy05}
{McHardy} I.~M., {Gunn} K.~F., {Uttley} P., {Goad} M.~R., 2005, \mnras, 359,
  1469

\bibitem[{{McHardy} {et~al.}(2006){McHardy}, {Koerding}, {Knigge}, {Uttley} \&
  {Fender}}]{mchardy06}
{McHardy} I.~M., {Koerding} E., {Knigge} C., {Uttley} P., {Fender} R.~P., 2006,
  \nat, 444, 730

\bibitem[{{McHardy} {et~al.}(2004){McHardy}, {Papadakis}, {Uttley}, {Page} \&
  {Mason}}]{mchardy04}
{McHardy} I.~M., {Papadakis} I.~E., {Uttley} P., {Page} M.~J., {Mason} K.~O.,
  2004, \mnras, 348, 783

\bibitem[{{Neugebauer} {et~al.}(1989){Neugebauer}, {Soifer}, {Matthews} \&
  {Elias}}]{neugebauer89}
{Neugebauer} G., {Soifer} B.~T., {Matthews} K., {Elias} J.~H., 1989, \aj, 97,
  957

\bibitem[{{Papadakis} \& {Lawrence}(1993)}]{papadakis93}
{Papadakis} I.~E., {Lawrence} A., 1993, \mnras, 261, 612

\bibitem[{{Press} {et~al.}(1992){Press}, {Teukolsky}, {Vetterling} \&
  {Flannery}}]{press92}
{Press} W.~H., {Teukolsky} S.~A., {Vetterling} W.~T., {Flannery} B.~P., 1992,
  {Numerical recipes in FORTRAN. The art of scientific computing}. Cambridge:
  University Press, |c1992, 2nd ed.

\bibitem[{{Priestley}(1981)}]{priestley81}
{Priestley} M.~B., 1981, {Spectral Analysis and Time Series: Probability and
  Mathematical Statistics}, Vol. 1-2. London: Academic Press, |c1981

\bibitem[{{Prokhorov} {et~al.}(1975){Prokhorov}, {Bunkin}, {Gochelashvily} \&
  {Shishov}}]{prokhorov75}
{Prokhorov} A.~M., {Bunkin} F.~V., {Gochelashvily} K.~S., {Shishov} V.~I.,
  1975, Proceedings of the IEEE, 63, 790

\bibitem[{{Quirrenbach} {et~al.}(2000){Quirrenbach}, {Kraus}, {Witzel},
  {Zensus}, {Peng}, {Risse}, {Krichbaum}, {Wegner} \&
  {Naundorf}}]{quirrenbach00}
{Quirrenbach} A., {et~al.}, 2000, \aaps, 141, 221

\bibitem[{{Quirrenbach} {et~al.}(1992){Quirrenbach}, {Witzel}, {Kirchbaum},
  {Hummel}, {Wegner}, {Schalinski}, {Ott}, {Alberdi} \&
  {Rioja}}]{quirrenbach92}
---, 1992, \aap, 258, 279

\bibitem[{{Rickett} {et~al.}(1984){Rickett}, {Coles} \& {Bourgois}}]{ricket84}
{Rickett} B.~J., {Coles} W.~A., {Bourgois} G., 1984, \aap, 134, 390

\bibitem[{{Romero} {et~al.}(1997){Romero}, {Combi}, {Benaglia}, {Azcarate},
  {Cersosimo} \& {Wilkes}}]{romero97}
{Romero} G.~E., {Combi} J.~A., {Benaglia} P., {Azcarate} I.~N., {Cersosimo}
  J.~C., {Wilkes} L.~M., 1997, \aap, 326, 77

\bibitem[{{Rutman}(1978)}]{rutman78}
{Rutman} J., 1978, Proceedings of the IEEE, 66, 1048

\bibitem[{{Scargle}(1989)}]{scargle89}
{Scargle} J.~D., 1989, \apj, 343, 874

\bibitem[{{Sikora} {et~al.}(2001){Sikora}, {B{\l}a{\.z}ejowski}, {Begelman} \&
  {Moderski}}]{sikora01}
{Sikora} M., {B{\l}a{\.z}ejowski} M., {Begelman} M.~C., {Moderski} R., 2001,
  \apj, 554, 1

\bibitem[{{Simonetti} {et~al.}(1985){Simonetti}, {Cordes} \&
  {Heeschen}}]{simonetti85}
{Simonetti} J.~H., {Cordes} J.~M., {Heeschen} D.~S., 1985, \apj, 296, 46

\bibitem[{{Smith} {et~al.}(1993){Smith}, {Nair}, {Leacock} \&
  {Clements}}]{smith93}
{Smith} A.~G., {Nair} A.~D., {Leacock} R.~J., {Clements} S.~D., 1993, \aj, 105,
  437

\bibitem[{{Spada} {et~al.}(2001){Spada}, {Ghisellini}, {Lazzati} \&
  {Celotti}}]{spada01}
{Spada} M., {Ghisellini} G., {Lazzati} D., {Celotti} A., 2001, \mnras, 325,
  1559

\bibitem[{{Stalin} {et~al.}(2005){Stalin}, {Gupta}, {Gopal-Krishna}, {Wiita} \&
  {Sagar}}]{stalin05}
{Stalin} C.~S., {Gupta} A.~C., {Gopal-Krishna}, {Wiita} P.~J., {Sagar} R.,
  2005, \mnras, 356, 607

\bibitem[{{Subba Rao} {et~al.}(1997){Subba Rao}, {Priestley} \&
  {Lessi}}]{rao97}
{Subba Rao} T., {Priestley} M.~B., {Lessi} O., 1997, {Applications of time
  series analysis in astronomy and meteorology}. London: Chapman and Hall,
  |c1997, 1st ed.

\bibitem[{{Takahashi} {et~al.}(2000){Takahashi}, {Kataoka}, {Madejski},
  {Mattox}, {Urry}, {Wagner}, {Aharonian}, {Catanese}, {Chiappetti}, {Coppi},
  {Degrange}, {Fossati}, {Kubo}, {Krawczynski}, {Makino}, {Marshall},
  {Maraschi}, {Piron}, {Remillard}, {Takahara}, {Tashiro}, {Terasranta} \&
  {Weekes}}]{takahashi00}
{Takahashi} T., {et~al.}, 2000, \apjl, 542, L105

\bibitem[{{Tanihata} {et~al.}(2003){Tanihata}, {Takahashi}, {Kataoka} \&
  {Madejski}}]{tanihata03}
{Tanihata} C., {Takahashi} T., {Kataoka} J., {Madejski} G.~M., 2003, \apj, 584,
  153

\bibitem[{{Tanihata} {et~al.}(2001){Tanihata}, {Urry}, {Takahashi}, {Kataoka},
  {Wagner}, {Madejski}, {Tashiro} \& {Kouda}}]{tanihata01}
{Tanihata} C., {Urry} C.~M., {Takahashi} T., {Kataoka} J., {Wagner} S.~J.,
  {Madejski} G.~M., {Tashiro} M., {Kouda} M., 2001, \apj, 563, 569

\bibitem[{{Ter{\"a}sranta} {et~al.}(2005){Ter{\"a}sranta}, {Wiren}, {Koivisto},
  {Saarinen} \& {Hovatta}}]{terasranta05}
{Ter{\"a}sranta} H., {Wiren} S., {Koivisto} P., {Saarinen} V., {Hovatta} T.,
  2005, \aap, 440, 409

\bibitem[{{Timmer} \& {Koenig}(1995)}]{timmer95}
{Timmer} J., {Koenig} M., 1995, \aap, 300, 707

\bibitem[{{Tosti} {et~al.}(1998){Tosti}, {Fiorucci}, {Luciani}, {Efimov},
  {Shakhovskoy}, {Valtaoja}, {Teraesranta}, {Sillanpaeae}, {Takalo}, {Villata},
  {Raiteri}, {de Francesco} \& {Sobrito}}]{tosti98}
{Tosti} G., {et~al.}, 1998, \aap, 339, 41

\bibitem[{{Uttley} \& {McHardy}(2005)}]{uttley05b}
{Uttley} P., {McHardy} I.~M., 2005, \mnras, 363, 586

\bibitem[{{Uttley} {et~al.}(2002){Uttley}, {McHardy} \& {Papadakis}}]{uttley02}
{Uttley} P., {McHardy} I.~M., {Papadakis} I.~E., 2002, \mnras, 332, 231

\bibitem[{{Vaughan}(2005)}]{vaughan05}
{Vaughan} S., 2005, \aap, 431, 391

\bibitem[{{Vaughan} {et~al.}(2003){Vaughan}, {Edelson}, {Warwick} \&
  {Uttley}}]{vaughan03}
{Vaughan} S., {Edelson} R., {Warwick} R.~S., {Uttley} P., 2003, \mnras, 345,
  1271

\bibitem[{{Yu} {et~al.}(2003){Yu}, {Gilmore}, {Peebles} \& {Rhodes}}]{yu03}
{Yu} C.~X., {Gilmore} M., {Peebles} W.~A., {Rhodes} T.~L., 2003, Physics of
  Plasmas, 10, 2772

\bibitem[{{Zhang} {et~al.}(2002){Zhang}, {Treves}, {Celotti}, {Chiappetti},
  {Fossati}, {Ghisellini}, {Maraschi}, {Pian}, {Tagliaferri} \&
  {Tavecchio}}]{zhang02a}
{Zhang} Y.~H., {et~al.}, 2002, \apj, 572, 762

\end{thebibliography}

\appendix
\section[]{Functional relation between the structure function and the autocorrelation function}
\label{app:sf_acf}
Consider $N$ observations of a stationary real-valued time process $x(t)$: $\left\{x(t_1),x(t_2),\ldots,x(t_N)\right\}$, then there is a direct relation between the SF and the ACF of the process. 
The SF is given in general by
\eqb
SF(\tau)&=&\frac{1}{N}\sum_{i=1}^N\left[x(t_i)-x(t_i+\tau)\right]^2\nonumber\\
&&=\left<\left[x(t)-x(t+\tau)\right]^2\right>
\label{eqe:sf_apend}
\eqe
The autocovariance function of the time process $x(t)$ is given by
\eqb
V_{x,x}(\tau)=\left<\left[x(t)-\overline{x(t)}\right] \left[x(t+\tau)-\overline{x(t)}\right]\right>
\label{eqe:discrete_autocovariance}
\eqe
The variance $S^2$ of the process $x(t)$ is equal to
\eqb
S^2=\frac{1}{N}\sum_{i=1}^N\left[x(t_i)-\overline{x(t)}\right]^2=\left<\left[x(t)-\overline{x(t)}\right]^2\right>
\eqe
note that in the denominator we have $N$ instead of $N-1$ (i.e.\ the biased estimator) due to the fact that theoretically $\overline{x(t)}$ is estimated directly from the parent distribution and not from the data set itself.\newline 
The ACF is given by
\eqb
ACF(\tau)=\frac{V_{x,x}(\tau)}{S^2}
\label{eqe:acf_append}
\eqe
By expanding the terms of equation (\ref{eqe:discrete_autocovariance}) and taking into account the fact that we are dealing with a stationary process i.e.\ $\overline{x(t)}=\overline{x(t+\tau)}=\left<\overline{x(t)}\right>={\rm const.}$ we get  
\eqb
V_{x,x}(\tau)=\left< x(t)x(t+\tau)\right>-\overline{x(t)}^2
\label{eqn:auco}
\eqe
Similarly for the variance
\eqb
S^2=\overline{x(t)^2}-\overline{x(t)}^2=V_{x,x}(0)
\label{eqn:var}
\eqe
From equation (\ref{eqe:sf_apend})
\eqb
SF(\tau)= \overline{x(t)^2}-2\left<x(t)x(t+\tau)\right>+\overline{x(t+\tau)^2}
\eqe
By means of equation (\ref{eqn:var})
\eqb
SF(\tau)= 2\left[S^2+\overline{x(t)^2}-\left<x(t)x(t+\tau)\right>\right]
\eqe
Finally from equation (\ref{eqe:acf_append}) and equation (\ref{eqn:auco}) the SF for a stationary process is given by
\eqb
SF(\tau)&=&2\left[S^2-V_{x,x}(\tau)\right] \nonumber \\
&=&2S^2\left[1-\frac{V_{x,x}(\tau)}{S^2}\right] \nonumber \\
&=&2S^2\left[1-ACF(\tau)\right]
\label{eqn:sfauto}
\eqe
After equation (\ref{eqn:sfauto}) the normalized SF (NSF) is defined as:
\eqb
NSF(\tau)&=&\frac{SF(\tau)}{S^2}\nonumber \\
&=&2\left[1-ACF(\tau)\right]
\label{eqe:normalized_sf}
\eqe
As $t\rightarrow\infty$, $ACF(\tau)\rightarrow 0$ (i.e.\ there is no linear correlation between the various measurements), from equation (\ref{eqn:sfauto}) $SF(\tau)\rightarrow 2S^2$ and from equation (\ref{eqe:normalized_sf}) $NSF(\tau)\rightarrow 2$.

\section[]{Functional relation between the structure function and the power spectral density}
\label{app:sf_psd}
Consider a zero mean stationary real time process $x(t)$ with autocovariance function (equation (\ref{eqn:auco}))
\eqb
V_{x,x}(\tau)=\left<x(t)x(t+\tau)\right>=\left<x(t)x^*(t+\tau)\right> 
\label{eqn:auco_zeromean}
\eqe
where the asterisk denotes complex conjugation. By rewriting the terms inside the mean as a function of their 
Fourier transform, $Y(f)$, for $-\infty<f<\infty$, equation (\ref{eqn:auco_zeromean}) reads
\eqb
V_{x,x}(\tau)&=&\left< \int_{-\infty}^{+\infty}Y(f)e^{-2\pi i f t}df 
\int_{-\infty}^{+\infty}Y^*(f)e^{2\pi i f (t+\tau)}df \right> \nonumber \\
&=&\left< \int_{-\infty}^{+\infty}Y(f)Y^*(f) e^{2\pi i f \tau}df \right> \nonumber \\
&=&\left<\int_{-\infty}^{\infty}\bigr|Y(f)\bigr|^2 e^{2\pi i f \tau}df \right>
\label{eqn:auco_fourier}
\eqe
and due to stationarity, none of the terms is varying as a function of $t$ (that is, independency of time translations e.g. \citet{bendat86}) 
\eqb
V_{x,x}(\tau)= \int_{-\infty}^{\infty}\bigr|Y(f)\bigr|^2 e^{2\pi i f \tau}df 
\eqe
According to the first equality of equation (\ref{eqn:sfauto}) and equation (\ref{eqn:var}) for $-\infty<f<\infty$ 
\eqb
SF(\tau)&=& 2\left[V_{x,x}(0)-V_{x,x}(\tau)\right] \nonumber \\
&=&2\left[\int_{-\infty}^{\infty}\bigr|Y(f)\bigr|^2df-
\int_{-\infty}^{\infty}\bigr|Y(f)\bigr|^2e^{2\pi i f \tau}df\right] \nonumber \\
&=& 2\int_{-\infty}^{\infty}(1- e^{2\pi i f \tau})\bigr|Y(f)\bigr|^2 df  \nonumber \\
\eqe
Based on the Euler's formula, $e^{i \kappa} =\cos(\kappa)+i \sin(\kappa)$, and ignoring the imaginary part dealing only with the phases
\eqb
SF(\tau)=2\int_{-\infty}^{\infty}\left[1- \cos(2\pi f \tau)\right]\bigr|Y(f)\bigr|^2 df 
\label{eqe:sf_integration_amplitude}
\eqe
In order to estimate the variability power contained in the frequency interval between $f$ and $f+df$, we can omit the distinction between positive and negative frequencies and regard $f$ as varying between 0 and $\infty$ \citep{press92}. By considering the sum of the modulus-squared of the sinusoidal amplitudes $Y(f)$ and $Y(-f)$, we estimate the {\it one-sided} PSD of $x(t)$ \citep{press92} as 
\eqb
\mathcal{P}(f)=|Y(f)|^2+|Y(-f)|^2\;\;\mathrm{for}\;\;0\leq f<\infty
\label{eqe:conti_psd}
\eqe
Since $x(t)\in\mathbb{R}$,
\eqb
|Y(f)|^2=|Y(-f)|^2\;\;{\mathrm{(symmetry\,about\,the\;y\!-\!axis)}}
\label{eqe:symmet_prop}
\eqe
and thus,
\eqb
\mathcal{P}(f)=2|Y(f)|^2\;\;\mathrm{for}\;\;0\leq f<\infty\;\; 
\label{eqe:pds_amplit}
\eqe 
From the symmetric property (equation (\ref{eqe:symmet_prop})) and due to the fact that $\cos(f)=\cos(-f)$ for $0\leq f<\infty$, equation (\ref{eqe:sf_integration_amplitude}) reads
\eqb
SF(\tau)=4\int_{0}^{\infty}\left[1- \cos(2\pi f \tau)\right]\bigr|Y(f)\bigr|^2
\eqe
yielding from equation (\ref{eqe:pds_amplit})
\eqb
SF(\tau)=2\int_{0}^{\infty}\left[1- \cos(2\pi f \tau)\right]\mathcal{P}(f) df\;\;\mathrm{for}\;\;0\leq f<\infty\;\; 
\label{eqe:sf_integration_PDS}
\eqe
For a PSD of a power-law form $\mathcal{P}(f)=\kappa f^{-\lambda}$, with $1<\lambda<3$ and $\kappa$ a positive constant the last integration (equation (\ref{eqe:sf_integration_PDS})) has an analytical solution 
\eqb
SF(\tau)=-2^{\lambda}\kappa\pi^{\lambda-1}\Gamma(1-\lambda)\sin\left(\frac{\lambda\pi}{2}\right)\tau^{\lambda-1}
\label{eqe:relation_sf_PDS}
\eqe
where $\Gamma(x)$ is the (complete) Gamma function.
That means that only when the following requirements
\begin{enumerate}
\item stationarity (also called {\it weakly stationarity} i.e.\ mean value and autocovariance function independent of time translations, \citealp[see e.g.][]{bendat86}).
\item zero mean data set.
\item The frequency range $f$ should vary from 0 to $\infty$.
\item The PSD should be given from a power-law form with index $1<\lambda<3$.
\end{enumerate}
are fulfilled then there is a direct relation between SF and PSD connecting the slopes of the two quantities, $\beta$ and $\lambda$ respectively, $\beta=\lambda-1$.
\label{lastpage}

\section[]{The general power spectral density of a single triangular shot}
\label{app:psd_triang}
\begin{figure} 
\includegraphics[width=3.3in]{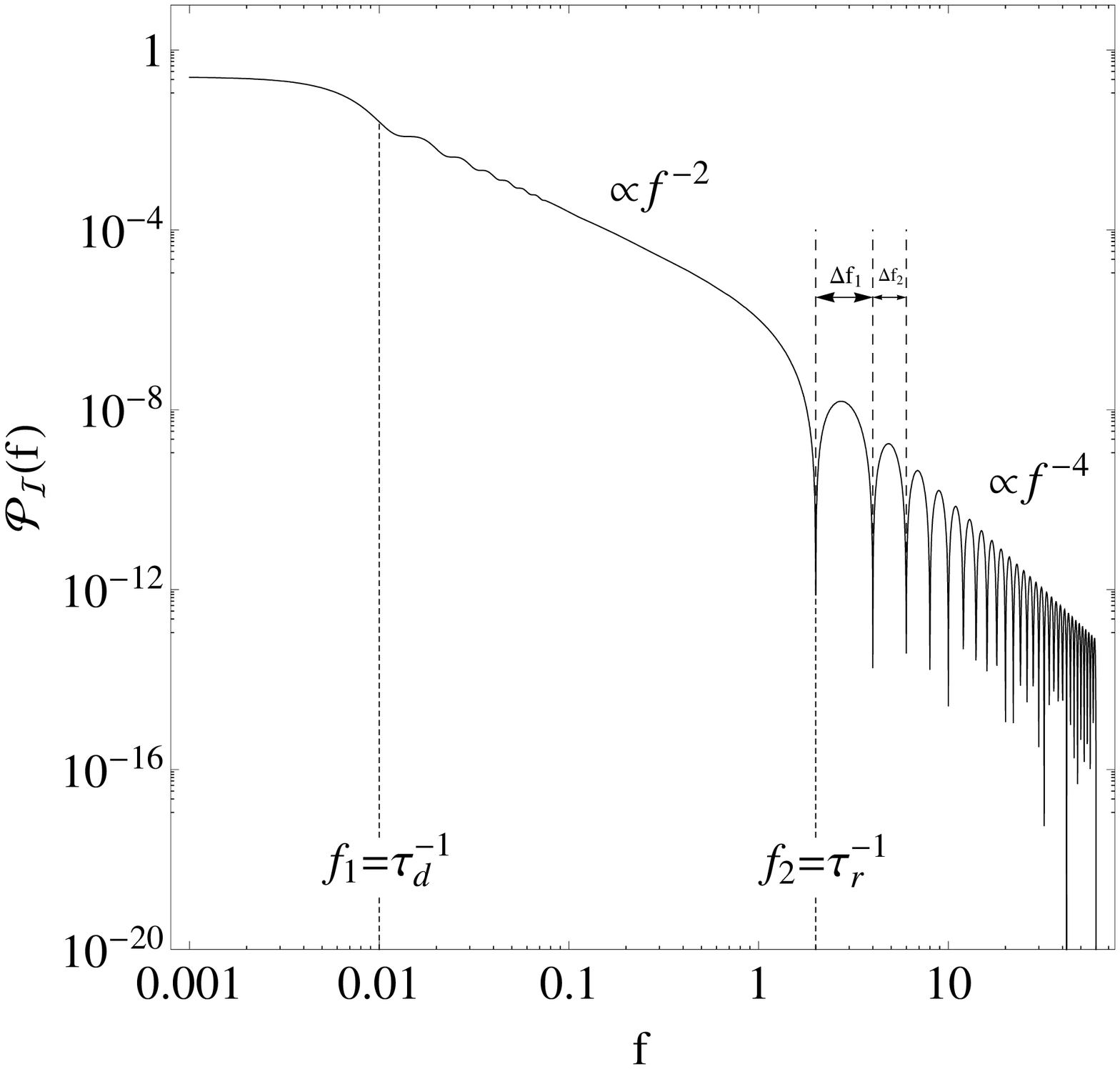}
\caption{The general PSD function of a triangular-shaped shot (equation (\ref{eqe:triang_shot})) (in logarithmic scale). The shot parameters are $\tau_{\rm r}=0.5$ t.u., $\tau_{\rm d}=100$ t.u. and $I_0=0.04$ flux units. The shape of the PSD is characterized by two breaks at $f_1=\tau_{\rm d}^{-1}$ and $f_2=\tau_{\rm r}^{-1}$. For $f<f_1$ it is flat and for $f_1<f<f_2$ it becomes steep with a slope of -2. For $f>f_2$ the PSD becomes even steeper with a slope of -4 and beat frequencies separated by $\Delta f_1=\Delta f_2=\ldots=\Delta f_n=\tau_{\rm r}^{-1}$ start to appear in the general trend.}
\label{fig:anal_psd_triang} 
\end{figure}

The general PSD function of a triangular shot of the form of equation (\ref{eqe:triang_shot}) is given by
\eqb
\mathcal{P}_{\tiny\mathcal{I}}(f)&=&\frac{I_0^2}{8\pi^4 t_{\rm d}^2 t_{\rm r}^2 f^4}\{t_{\rm d}^2+t_{\rm d}t_{\rm r}+t_{\rm r}^2-\nonumber\\
&&(t_{\rm d}+t_{\rm r})\left[t_{\rm r}\cos(2 \pi t_{\rm d} f)+t_{\rm d}\cos(2 \pi t_{\rm r} f)\right]+\nonumber\\
&&t_{\rm d}t_{\rm r}\cos( 2 \pi (t_{\rm r}+t_{\rm d}) f )\}\,\,\,\mathrm{for}\, t_{\rm r}\neq0\;\&\,t_{\rm d}\neq0
\label{eqe:general_psd_triang}
\eqe
Figure \ref{fig:anal_psd_triang} shows the form of this PSD function for the case of $\tau_{\rm r}=0.5$ t.u., $\tau_{\rm d}=100$ t.u. and $I_0=0.04$ flux units. It is characterized by two distinct breaks at $f_1=\tau_{\rm d}^{-1}$ and $f_2=\tau_{\rm r}^{-1}$ and the slope changes from 0 to -2 as we pass from $f<f_1$ to $f_1<f<f_2$ and becomes even steeper for $f>f_2$ with a slope of -4. For the latter region beat frequencies appear which are separated by $\Delta f=\tau_{\rm r}^{-1}$.\par

For instantaneous rise $t_{\rm r}=0$ the PSD is equal to
\eqb
\mathcal{P}_{\tiny\mathcal{I}}(f)&=&\frac{I_0^2}{8\pi^4 t_{\rm d}^2 f^4}[1+2\pi^2t_{\rm d}^2 f^2-\cos(2 \pi t_{\rm d} f)- \nonumber\\ && 2\pi t_{\rm d} f \sin\left(2 \pi t_{\rm d} f\right)]
\label{eqe:psd_triang_rise0}
\eqe
for instantaneous decay the PSD is given again from the last relation (equation (\ref{eqe:psd_triang_rise0})) having the $t_{\rm d}$ replaced with $t_{\rm r}$.\par
Finally, for the case of symmetric shots $t_{\rm r}=t_{\rm d}=\tau$ and equation (\ref{eqe:general_psd_triang}) reads
\eqb
\mathcal{P}_{\tiny\mathcal{I}}(f)=\frac{I_0^2}{\pi^4\tau^2 f^4}\sin^4(\pi\tau f)
\label{eqe:psd_symm_triang}
\eqe
\end{document}